\documentclass[11pt]{article}
\usepackage{cite}
\usepackage{color}

\usepackage{amsmath,amsfonts,amssymb}
\usepackage[small,bf,hang]{caption}
\usepackage{slashed}
\usepackage{mathabx}
\usepackage{latexsym,epsfig}

\def\hybrid{
        \topmargin -20pt
        \oddsidemargin 0pt
        \headheight 0pt \headsep 0pt
        \textwidth 6.25in 
        \textheight 9.5in 
        \marginparwidth .875in
        \parskip 5pt plus 1pt \jot = 1.5ex}

\hybrid

 \csname
@addtoreset\endcsname{equation}{section}

\def\moth{\mathsurround=0pt}
\newdimen\zo \zo=0pt

\def\tick{\leaders\hrule height 0.5ex depth 0pt \hskip 0.5pt}
\def\upboxfill{$\moth \setbox\zo\hbox{\tick}%
  \hskip 3pt\hbox to 0pt{$\tick$\hss}\hrulefill \hbox to 7.5pt{$\tick$\hss}$}

\def\dtick{\leaders\hrule height .34pt depth 0.5ex \hskip 0.5pt}
\def\downboxfill{$\moth \setbox\zo\hbox{\dtick}%
  \hskip 2pt\hbox to 0pt{$\dtick$\hss}\hrulefill \hbox to 2pt{$\dtick$\hss}$}

\def\bec{\begin{center}}
\def\ec{\end{center}}

\def\cB{{\cal B}}
\def\cG{{\cal G}}

\def\cD{{\cal D}}
\def\cF{{\cal F}}

\def\cA{{\cal A}}

\def\cM{{\cal M}}

\def\cR{{\cal R}}

\def\cP{{\cal G}}

\def\cV{{\cal V}}

\def\cT{{\cal T}}

\def\cA{{\cal A}}

\def\cQ{{\cal Q}}
\def\cP{{\cal P}}

\def\bfa{{\boldsymbol{\alpha}}}

\def\MQ{{\mathbb{Q}}}

\def\nn{\nonumber}

\def\be{\begin{equation}}
\def\ee{\end{equation}}
\def\bea{\begin{eqnarray}}
\def\eea{\end{eqnarray}}
\def\ba{\begin{array}}
\def\ea{\end{array}}

\begin{document}

\begin{titlepage}
\rightline{}
\rightline{\tt  DAMTP-2014-34}
\rightline{\tt  MIT-CTP 4556}
\rightline{\tt  AEI-2014-015}
\rightline{June 12, 2014}
\begin{center}
\vskip .6cm
{\Large \bf {Supersymmetric E$_{7(7)}$ Exceptional Field Theory}}\\
\vskip .6cm
{\large {Hadi Godazgar${}^1$, Mahdi Godazgar${}^1$, Olaf Hohm${}^2$, \\[0.5ex]
Hermann Nicolai${}^3$ and Henning Samtleben${}^4$}}
\vskip .6cm
{\it {${}^1$DAMTP, Centre for Mathematical Sciences}}\\
{\it {University of Cambridge}}\\
{\it {Wilberforce Road, Cambridge, CB3 0WA, UK}}\\
H.M.Godazgar@damtp.cam.ac.uk
\vspace{-3mm}

\hspace{-2mm} M.M.Godazgar@damtp.cam.ac.uk
\vskip 0.2cm
{\it {${}^2$Center for Theoretical Physics}}\\
{\it {Massachusetts Institute of Technology}}\\
{\it {Cambridge, MA 02139, USA}}\\
ohohm@mit.edu
\vskip 0.2cm
{\it {${}^3$Max-Planck-Institut f\"ur Gravitationsphsyik}}\\
{\it { Albert-Einstein-Institut (AEI)}}\\
{\it {M\"uhlenberg 1, D-14476 Potsdam, Germany}}\\
Hermann.Nicolai@aei.mpg.de
\vskip 0.2cm
{\it {${}^4$Universit\'e de Lyon, Laboratoire de Physique, UMR 5672, CNRS}}\\
{\it {\'Ecole Normale Sup\'erieure de Lyon}}\\
{\it {46, all\'ee d'Italie, F-69364 Lyon cedex 07, France}}\\
henning.samtleben@ens-lyon.fr

\vskip .5cm
{\bf Abstract}
\end{center}

\vskip 0.2cm

\noindent
\begin{narrower}
{\footnotesize
We give the supersymmetric extension of exceptional field theory 
for E$_{7(7)}$, which is based on a $(4+56)$-dimensional 
generalized spacetime subject to a covariant constraint. 
The fermions are tensors under the local Lorentz 
group ${\rm SO}(1,3)\times {\rm SU}(8)$ and transform as 
scalar densities under the E$_{7(7)}$ (internal) generalized diffeomorphisms. 
The supersymmetry transformations are manifestly 
covariant under these symmetries and close, in particular, 
into the generalized diffeomorphisms of the 56-dimensional space.
We give the fermionic field equations and prove supersymmetric 
invariance.  We establish the consistency of these results
with the recently constructed generalized geometric formulation of $D=11$ supergravity.
}

\end{narrower}

\end{titlepage}

\newpage

\tableofcontents

\newpage

\section{Introduction}

Ever since the discovery of `hidden' exceptional symmetries in maximal $N=8$ supergravity
\cite{Cremmer:1978ds} a recurring theme has been the question of whether these symmetries 
are specifically tied to dimensional reduction on tori, or whether they reflect more general 
properties of the underlying {\em uncompactified} maximal theories, possibly even providing clues
towards a better understanding of M-theory. Starting from $D=11$ supergravity \cite{Cremmer:1978km}
clear evidence for the existence of hidden structures beyond those of standard differential geometry was already given in the early work of Refs.~\cite{deWit:1986mz,Nicolai:1986jk}, a line of development which was continued in \cite{Koepsell:2000xg} and taken up again 
in \cite{deWit:2013ija,Godazgar:2013dma,Godazgar:2014sla}. Somewhat independently 
of these developments, an important 
insight has been the emergence of generalized geometric concepts in string and 
M-theory, which enable a duality-covariant formulation of the low-energy effective 
spacetime theories, as  manifested in double field theory  
\cite{Siegel:1993th, Hull:2009mi, Hull:2009zb, Hohm:2010jy, Hohm:2010pp},
and  in the recently constructed `exceptional field theory' (EFT)
\cite{Hohm:2013pua,Hohm:2013uia}. See also Refs.~\cite{Hull:2007zu, Pacheco:2008ps, Coimbra:2011ky,Coimbra:2012af} for a generalized geometric approach in the sense of Ref.~\cite{Hitchin:2004ut, Gualtieri:2003dx}. The purpose of this paper, then, is to bring
together these strands of development: first we complete the  construction of the 
E$_{7(7)}$ EFT by giving the fully supersymmetric 
extension by fermions; second, we relate the resulting theory to the formulation
of \cite{deWit:1986mz,deWit:2013ija,Godazgar:2013dma,Godazgar:2014sla}. 
As one of our main results we will demonstrate the compatibility of these two formulations, and explain 
the subtleties involved in making a detailed comparison.

The approach of \cite{deWit:1986mz}, which has been extended and completed in 
\cite{Godazgar:2013dma, Godazgar:2014sla} to also take into account aspects of
the E$_{7(7)}$-based exceptional geometry, takes $D=11$ supergravity as the 
starting point and reformulates it in order to make a local SO$(1,3)\times {\rm SU}(8)$ 
tangent space symmetry manifest. To this end  the fields and coordinates 
are decomposed in a $(4+7)$ splitting, as in Kaluza-Klein compactifications, 
but keeping the full coordinate dependence of all fields (however, unlike in EFT, 
no extra coordinates beyond those of the original theory are introduced).
The fermions transform under the local SU$(8)$ subgroup, and their 
supersymmetry transformations, already given in \cite{deWit:1986mz}, 
are manifestly SU$(8)$ covariant. Moreover, those parts of the bosonic sector
which lead to scalar and vector fields in the dimensionally reduced maximal supergravity 
can then be assembled into E$_{7(7)}$ objects, namely a 56-bein encoding 
the internal field components and a 56-plet of vectors combining the 28 electric and
28 magnetic vectors of $N=8$ supergravity; their supersymmetry
transformations can be shown to take the precise form of the four-dimensional maximal gauged 
supergravity. While in this approach the fermions are included 
from the beginning (with the supersymmetry variations constituting the starting point of the analysis) and the
on-shell equivalence with $D=11$ supergravity is thus guaranteed 
at each step of the construction, a proper understanding of the role of E$_{7(7)}$ 
in {\em eleven dimensions} (as well as of the E$_{7(7)}$-covariant 
dynamics of the bosonic sector) was lacking in the original work of \cite{deWit:1986mz},
and has only emerged with the recent advances.  Nevertheless it is remarkable 
that the combinations of SU(8) connections in the supersymmetry variations of the 
fermions found `empirically' in Ref.~\cite{deWit:1986mz} are precisely the ones 
required by E$_{7(7)}$-covariance as identified here.

The results of Ref.~\cite{Koepsell:2000xg} suggest that a formulation that is 
properly covariant under the exceptional groups should include extended coordinates 
transforming under this group, an idea that also 
appears in the proposal of  Ref.~\cite{West:2003fc}. Such an extended spacetime has 
later been implemented for E$_{7(7)}$ in a particular 
truncation of $D=11$ supergravity that retains only 
the internal coordinates and field components of the $(4+7)$ splitting \cite{Hillmann:2009ci}. 
More recently, similar reformulations of $D=11$ supergravity have been given for the analogous truncations, 
casting the theory and their residual gauge transformations into a covariant form \cite{Berman:2011jh,Coimbra:2011ky,Coimbra:2012af}.
In contrast to the original approach of Ref.~\cite{deWit:1986mz}, however, these formulations 
are not immediately applicable to the untruncated $D=11$ supergravity. By contrast, 
the construction of Refs.~\cite{Godazgar:2013dma,Godazgar:2014sla}, the recent construction of complete EFTs in Refs.~\cite{Hohm:2013pua,Hohm:2013uia} and finally, the present 
work  extend the formulation of Ref.~\cite{deWit:1986mz} to a fully E$_{7(7)}$-covariant theory. 

The E$_{7(7)}$ EFT, which is a natural extension of double field theory, is based on a 
4+56-dimensional generalized spacetime, with fields in E$_{7(7)}$ representations initially depending on all coordinates $x^\mu$ and $Y^M$ (with fundamental indices $M=1,\ldots, 56$).
The theory is given by an action along with non-abelian twisted self-duality equations for the 56 vector fields. The fields transform appropriately under E$_{7(7)}$-generalized diffeomorphisms. Crucially, the theory is subject to an E$_{7(7)}$-covariant section condition \cite{Coimbra:2011ky} that implies that the fields depend only on a subset of 
coordinates. In order to compare with the usual $D=11$ supergravity, and thus with the results of Ref.~\cite{deWit:1986mz, Godazgar:2013dma}, 
one has to pick a particular solution of this constraint, which reduces the spacetime to 4+7 dimensions.
After solving the section constraint, the various components of the generalized diffeomorphisms can be interpreted as conventional diffeomorphisms and tensor gauge transformations. 
In addition, and in analogy to type II double field theory \cite{Hohm:2011zr,Hohm:2011dv}, 
the section constraint has two inequivalent solutions: $D=11$ supergravity and type IIB supergravity. 
After solving the section constraint, 
the E$_{7(7)}$ EFT also encodes, as 7 components among the 56 gauge vectors, dual gravity 
degrees of freedom. This description is consistent by virtue of a covariantly constrained 
compensating two-form gauge field ${\cal B}_{\mu\nu M}$ \cite{Hohm:2013jma,Hohm:2013uia}.
The status of this field may appear somewhat mysterious, but its appearance is already 
implied by consistency of the EFT gauge symmetries. 
In this paper we will give further credibility 
to this field by showing that it has consistent supersymmetry variations. 

In this paper we introduce the fermions of the 
E$_{7(7)}$ EFT and give the supersymmetry variations of all fields in a manifestly 
E$_{7(7)}\times {\rm SU}(8)$-covariant form, showing that they close, in particular, into the 
external and internal generalized diffeomorphisms.  This is in analogy with the supersymmetrization of DFT \cite{Coimbra:2011nw, Hohm:2011nu, Jeon:2011sq}.
Importantly, we find that the supersymmetry transformations of all fields can be written solely in terms of the fields of EFT, 
in particular the 56-bein, without recourse to the $D=11$ fields that can be thought of as parametrising these structures in a GL(7) decomposition.
Furthermore, we determine the fermionic field equations and verify supersymmetric 
on-shell invariance. To this end we have to further develop the generalized exceptional geometry 
underlying the E$_{7(7)}$ 
covariant formulation by introducing connections and invariant curvatures generalizing the geometry of
double field theory~\cite{Siegel:1993th,Hohm:2010xe,Hohm:2011si,Hohm:2012mf,Jeon:2011cn}. 
For the internal, $56$-dimensional sub-sector, such a geometry is to a large extent already contained in the literature \cite{Coimbra:2011ky,Coimbra:2012af,Aldazabal:2013mya,Cederwall:2013naa}. In particular, Refs.~\cite{Coimbra:2011ky,Coimbra:2012af} give the full dynamics and supersymmetry transformation
rules for the truncated theory, where the fields and parameters are independent of the four-dimensional external coordinates, in terms of such geometrical objects. We use the opportunity to give a complete and self-contained presentation of this geometry. 
We give compact and E$_{7(7)}$-covariant expressions for the internal connections in terms 
of the 56-bein and other covariant objects. 
One of the main results of this paper then is the formulation including 
external and internal connection components
${\cal Q}_{\mu}$ and ${\cal Q}_M$  for the local SU$(8)$, 
respectively, and similarly external and internal connection components $\omega_{\mu}$ and $\omega_{M}$ 
for the local SO$(1,3)$, with all geometric objects being also covariant under E$_{7(7)}$-generalized diffeomorphisms. 
The various connection components are summarized in the following scheme 
\bea
\begin{tabular}{c|c}
$\begin{array}{c} \boxed{{\omega_\mu}} \\  
\scalebox{0.7}{$\Gamma_{[\mu\nu]}{}^\rho = 0$}
\end{array} $ & 
$\begin{array}{c} \boxed{{\cal Q}_\mu} \\  
\scalebox{0.7}{$ {\cal D}_\mu{\cal V}_M{}^{AB} \equiv {\cal P}_\mu{}^{ABCD} \,{\cal V}_{M CD}  $}
\end{array} $
\\ \hline\\[-2ex]
$\begin{array}{c} \boxed{\omega_M} \\  
\scalebox{0.7}{${\cal D}_M e_\mu{}^{\alpha}  \equiv \pi_M{}^{\alpha \beta}   e_\mu{}_{\beta}$}
\end{array} $
 & 
 $\begin{array}{c}\boxed{ {\cal Q}_M} \\  
\scalebox{0.7}{$\Gamma_{MN}{}^K |_{\bf 912} = 0$}
\end{array} $
\end{tabular}
\quad\quad\;.
\label{full_spinIntro}
\eea
Here we also indicate the corresponding covariant torsion-type constraints satisfied by the connections.  
The precise definitions of the various tensors and our conventions will be given in the main text. 
The formulation  is manifestly covariant under all gauge symmetries except 
for the external diffeomorphisms of $x^{\mu}$ that depend also on the `internal' E$_{7(7)}$ coordinates. 
The structure of the various diagonal and off-diagonal connection components in (\ref{full_spinIntro}) hints at a larger geometrical framework in which they would emerge from a single `master connection', whose introduction would finally render all gauge 
symmetries manifest.

A distinctive feature of generalized geometries is that, in contrast to conventional geometry, the 
connections are not completely determined by imposing covariant constraints, necessarily featuring
undetermined connections that are not given in terms of the physical fields, 
as first discussed in the geometry of double field theory 
\cite{Siegel:1993th,Hohm:2010xe, Coimbra:2011nw, Hohm:2011si,Hohm:2012mf} and 
later extended to exceptional groups \cite{Coimbra:2011ky,Aldazabal:2013mya,Cederwall:2013naa}. 
As in double field theory, however, this is consistent with the final form of the (two-derivative) 
theory depending only on the physical fields, 
as the undetermined connections drop out of the action and all (supersymmetry) variations, as shown in \cite{Coimbra:2012af} for the truncated theory. 
We also clarify the relation of these geometrical structures to the 
formulation of \cite{deWit:1986mz,Godazgar:2013dma,Godazgar:2014sla}, 
in which connections carry `non-metricities' that can be absorbed, as we will show, 
into SU$(8)$ connections once we include components along 
the E$_{7(7)}$-extended directions.

One obvious question concerns the precise significance of the term `symmetry' in the present
context. The E$_{7(7)}$ identified here is analogous to the GL($D$) that appears
in general relativity, and is `spontaneously broken' when one  picks a particular
non-trivial solution to the section constraint $(t_\bfa){}^{MN} \partial_M\otimes \partial_N =0$.\,\footnote{It is an old idea to interpret the graviton as a Goldstone boson
of spontaneously broken GL(4) symmetry \cite{Ogievetsky:1973ik, bo, West:2001as}, but the present scheme should not be viewed as a realization of this idea.} However, the new structures exhibited 
here do {\em not} imply that $D=11$ supergravity or IIB supergravity have
any new local symmetries beyond the ones already known.\,\footnote{The only new
  local symmetry would be the one associated with the seven `dual' internal diffeomorphisms,
  but the corresponding transformation parameters `miraculously' drop 
  out in all relevant formulae, as shown in Ref.~\cite{Godazgar:2014sla}. 
  In the formulation of Ref.~\cite{Hohm:2013uia} this fact is explained by the `St\"uckelberg-like' gauge invariance
  associated with the two-form field ${\cal B}_{\mu\nu M}$. } Nevertheless it is remarkable 
and significant that the internal diffeomorphisms can be combined with the tensor gauge transformations of the form fields and their duals in an E$_{7(7)}$-covariant form. 
Evidently, the true advantage of the reformulation would only become fully apparent 
if solutions of the section constraint, besides those corresponding to $D=11$ or IIB supergravity, exist. Such solutions would give genuinely {\em new} theories (but see below). Although such solutions are 
somewhat unlikely to exist for the case at hand, the situation may become more interesting 
when one considers infinite dimensional extensions of the E-series.

A second question concerns the utility of the supersymmetric EFT constructed 
here in a more general  perspective. Here we see two main possible applications 
and extensions. The first application concerns the non-linear consistency of 
Kaluza-Klein compactifications other than torus compactifications. These can be investigated 
along the lines of \cite{deWit:1986iy, Nicolai:2011cy,Godazgar:2013pfa}, exploiting the 
present formalism and the fact that it casts the higher-dimensional theory in a form
adapted to (gauged) lower dimensional supergravity. Indeed, the full non-linear 
Kaluza-Klein ans\"atze for those higher-dimensional fields (including dual fields) 
yielding scalar or vector fields in the compactification have already been obtained
in this way for the AdS$_4 \times S^7$ compactification
\cite{deWit:1984nz, deWit:2013ija, Godazgar:2013nma, Godazgar:2013pfa},
as well as for general Scherk-Schwarz compactifications with fluxes \cite{Godazgar:2013oba}.\,\footnote{See also Ref.~\cite{Lee:2014mla}, where uplift ans\"atze for sphere reductions of the $D=11$ and type IIB theories are conjectured using similar ideas.}
Apart from the non-linear ans\"atze for higher rank tensors, which can now also
be deduced in a straightforward fashion, and beyond the extension to 
other non-trivial compactifications of $D=11$ supergravity, the main outstanding 
problem here is to extend these results to the compactification
of IIB supergravity on AdS$_5 \times S^5$, for which either the supersymmetric
extension of E$_{6(6)}$ EFT \cite{Hohm:2013vpa} or the present version with the IIB solution 
of the section constraint might be employed. Indeed, a study of the ambiguities inherent in defining generalized connections and how the supersymmetry transformations (and hence the theory) remain invariant under such redefinitions in this paper has lead to an understanding of the hook-type ambiguities observed in the $D=11$ theory in Ref.~\cite{Nicolai:2011cy}.

Secondly, the fact that the supersymmetric EFT has a structure very similar to four-dimensional maximal gauged supergravity \cite{deWit:2007mt} may lead to a higher-dimensional understanding of the 
new SO(8) gauged supergravities of Ref.~\cite{Dall'Agata:2012bb}, obtained by performing an electromagnetic U(1) rotation of the 56 electric and magnetic vectors, which is {\em not}
in E$_{7(7)}$. Partial evidence presented in Refs.~\cite{deWit:2013ija,Godazgar:2013nma}, as well as a more explicit argument based on the higher-dimensional embedding tensor in Ref.~\cite{Godazgar:2014sla}, show 
that these gaugings cannot originate from the $D=11$ supergravity of Ref.~\cite{Cremmer:1978km}. Specifically,
the deformed theories can be obtained from the standard SO(8) gauged supergravity
by `twisting' the 56-bein relative to the vectors \cite{deWit:2013ija}, that is, by making 
the replacement 
\begin{equation}\label{omegadef}
\cV (x)\, \rightarrow \, \cV(x;\omega) \equiv
\left(\begin{matrix} \cos\omega & \sin\omega \\[2mm]
-\sin\omega & \cos\omega \end{matrix}\right)\, \cV(x) 
\end{equation}
in all formulae, where each element of the U(1) rotation matrix acts on a 28$\times$28 subblock 
of the 56$\times$56 matrix $\cV$, in precise analogy with the deformation of the four-dimensional theory \cite{Dall'Agata:2012bb}.\,\footnote{In fact, in the context of four-dimensional maximal gauged theories, the U(1) rotation above is to be understood as part of a more general SL$(2, \mathbb{R})$ symplectic deformation \cite{Dall'Agata:2014ita}.} The present reformulation naturally suggests that a 
higher-dimensional ancestor of the deformed SO(8) gauged supergravities might thus be obtained 
by performing an analogous `twist' of the 56-bein of EFT (see also Ref.~\cite{Godazgar:2013oba}), $\cV(x,Y)\rightarrow \cV(x,Y;\omega)$,
relative to all vectors and tensors, where the 56-bein is now taken to also depend 
on the 56 extra coordinates $Y^M$.
Because of the inequivalence of the corresponding gauged SO(8) supergravities in
four dimensions, it is clear that such a theory would  no longer be on-shell 
equivalent to the $D=11$ supergravity of Ref.~\cite{Cremmer:1978km}, and hence would correspond to a 
non-trivial deformation of that theory. In fact, this would be the first example of a  
genuinely new maximal supergravity in the maximal space-time dimension $D=11$
since the discovery of Ref.~\cite{Cremmer:1978km} in 1978, and it would be a remarkable vindication 
of the present scheme if such a theory could be shown to exist. Equally important, 
there would be no way to reconcile this deformed theory with $D=11$ diffeomorphism 
and Lorentz invariance; in other words, the four-dimensional $\omega$-deformation 
of Ref.~\cite{Dall'Agata:2012bb} would lift to an analogous deformation of $D=11$ 
supergravity that is encoded in a suitably generalized geometric framework 
transcending conventional supergravity.  

The outline of the paper is as follows.  In section~\ref{sec:eftrev} we review the bosonic 
E$_{7(7)}$-covariant exceptional field theory, of Refs.~\cite{Hohm:2013pua,Hohm:2013uia};
in section~\ref{sec:e7covgeo} we construct its supersymmetric completion
upon introducing the proper fermion connections and working out the
supersymmetry algebra.
In section~\ref{sec:d11}, we discuss how this theory relates to the 
reformulation \cite{deWit:1986mz,Godazgar:2013dma,Godazgar:2014sla}
of the full (untruncated) $D=11$ supergravity after an explicit solution of the section constraint is chosen.

We refer the reader to appendix \ref{app:notations} for a summary of index notations and conventions.


\section{Bosonic E$_{7(7)}$ exceptional field theory}
\label{sec:eftrev}


In this section we give a brief review of the bosonic sector of the E$_{7(7)}$-covariant exceptional field theory,
constructed in Refs.~\cite{Hohm:2013pua,Hohm:2013uia} (to which we refer for details)
and translate it into the variables appropriate for the coupling to fermions,
in particular the 56-bein parametrizing the coset space ${\rm E}_{7(7)}/{\rm SU}(8)$\,. 
To begin with, all fields in this theory depend on the four external variables $x^\mu$, $\mu=0, 1, \dots, 3$, and the 
56 internal variables $Y^M$, $M=1, \dots, 56$, transforming in
the fundamental representation of ${\rm E}_{7(7)}$,
however the latter dependence is strongly restricted by the section condition
\be
 \begin{split}
  (t_\bfa)^{MN}\,\partial_M \partial_N A \ &= \ 0\;, \qquad   (t_\bfa)^{MN}\,\partial_MA\, \partial_N B \ = \ 0 \,,\\
  &\Omega^{MN}\,\partial_MA\, \partial_N B \ = \ 0 \,, 
 \end{split}
 \label{sectioncondition}
 \ee  
for any fields or gauge parameters $A,B$. 
Here, $\Omega^{MN}$ is the symplectic invariant matrix which we use for lowering and raising
of fundamental indices according to $X^M=\Omega^{MN}X_N$, $X_N=X^M\Omega_{MN}$\,.
The tensor $(t_\bfa)_M{}^{N}$ is the representation matrix of ${\rm E}_{7(7)}$ in the fundamental representation.
These constraints admit (at least) two inequivalent solutions, in which the fields
depend on a subset of seven or six of the internal variables, respectively, according to the decompositions
\begin{subequations}
\begin{align}
 {\bf 56} &~\longrightarrow~
\boxed{{7}_{+3}}+ { 21}'_{+1}+{21}_{-1}+ {7}'_{-3}\;,
\label{decIIA}\\
{\bf 56} &~\longrightarrow~
\boxed{({6},1)_{+2}}+(6',2)_{+1}+(20,1)_0 +(6,2)_{-1}+ (6',1)_{-2}
\;,
\label{decIIB}
\end{align}
\end{subequations}
of the fundamental representation of ${\rm E}_{7(7)}$ with respect to 
the maximal subgroups ${\rm GL}(7)$ and ${\rm GL}(6)\times {\rm SL}(2)$, 
respectively. The resulting theories are the full $D=11$ supergravity and the type IIB theory, respectively.
The bosonic field content of the E$_{7(7)}$-covariant exceptional field theory is given by 
\bea
\left\{e_\mu{}^{\alpha}\,,\; {\cal V}_{M}{}^{AB},\; \cA_\mu{}^M\,,\; \cB_{\mu\nu\,\bfa}\,,\; 
\cB_{\mu\nu\,M} \right\}
\;,
\label{fieldcontent}
\eea
which we describe in the following.
The field $e_\mu{}^{\alpha}$ is the vierbein, from which the external (four-dimensional) 
metric is obtained as $g_{\mu\nu}=e_\mu{}^{\alpha} e_{\nu \alpha}$.
Its analogue in the internal sector is the complex 56-bein 
\bea
{\cal V}_M{}^{\underline{N}} &=& \{{\cal V}_{M}{}^{AB}, {\cal V}_{M}{}_{AB}\}\;,
\eea
satisfying
\bea
{\cal V}_{M}{}^{AB}={\cal V}_{M}{}^{[AB]}
\;,\qquad
{\cal V}_{M}{}_{AB} &=& ({\cal V}_{M}{}^{AB})^*\;,
\eea
with ${\rm SU}(8)$ indices $A, B, \dots=1, \dots, 8,$ in the fundamental ${\bf 8}$ representation 
and collective index $\underline{N}$ labelling the ${\bf 28}+\bar{\bf 28}$.\,\footnote{While
 the SU(8) indices were taken to be $i,j,k, \dots$ in Ref.~\cite{Hohm:2013uia}, we here revert
 to the notation of Ref.~\cite{deWit:1986mz}, also employed in Refs.~\cite{Godazgar:2013dma,Godazgar:2014sla}, where SU(8) indices are
 denoted by the letters $A,B,C,\dots$. The reason is 
 that, when considering non-trivial compactifications, one must distinguish between the
 SU(8) indices $A,B,\dots$ in eleven dimensions, and the SU(8) indices $i,j,\dots$
 in the four-dimensional compactified theory. These are only the same for the torus
 compactification. Any other compactification involves Killing spinors 
 as `conversion matrices' (hence the distinction between `curved' and `flat' SU(8) indices 
 in Ref.~\cite{deWit:1986iy}).
 However, in accord with previous conventions, fundamental ${\rm SU}(8)$ indices 
 are raised and lowered by complex conjugation.}
The fact that the 56-bein is an ${\rm E}_{7(7)}$ group-valued matrix is most efficiently encoded in 
the structure of its infinitesimal variation, 
\bea
\delta {\cal V}_M{}^{AB} &=& 
-\delta q_C{}^{[A} \,{\cal V}_M{}^{B]C} + \delta p{}^{ABCD}\,{\cal V}_M{}_{CD} 
\;,
\label{varV}
\eea
with
\bea
\delta q_A{}^{B} &=& - \delta q^{B}{}_A\;,\qquad
\delta p{}^{ABCD} ~=~ \frac1{24}\,\epsilon^{ABCDEFGH}\,\delta p{}_{EFGH}
\;.
\eea
This is equivalent to
\begin{eqnarray}
   {\cal V}_{M AB} \,\delta{\cal V}_N{}^{CD} \,\Omega^{MN}   &=&{}
  \frac{2}{3}\,\delta_{[A}{}^{[C} \; {\cal V}_{M B]E} \,\delta{\cal V}_N{}^{D]E}
  \,\Omega^{MN}  \,, \nn\\
  {\cal V}_{M AB} \,\delta{\cal V}_{N CD} \,\Omega^{MN}   &=&{}
  {\cal V}_{M [AB} \,\delta{\cal V}_{N CD]} \,\Omega^{MN}  \,, \nn\\
  {\cal V}_{M}{}^{AB} \,\delta{\cal V}_N{}^{CD} \,\Omega^{MN} &=& - \frac1{24}
  \varepsilon^{ABCDEFGH} \,   {\cal V}_{M \, EF} \,\delta{\cal V}_{N \, GH}
  \,\Omega^{MN}  \,. 
\end{eqnarray}

A particular consequence of the group property is
\begin{eqnarray}
  {\cal V}_M{}^{AB} \,{\cal V}_{N\,AB} - {\cal V}_{M\,AB}\, {\cal V}_N{}^{AB}  &=&  
  \mathrm{i}\,\Omega_{MN}\,, \nonumber\\ 
  \Omega^{MN} \,{\cal V}_M{}^{AB} \,{\cal V}_{N\,CD} &=& 
  \mathrm{i}\,\delta^{AB}_{CD}\,, \nonumber\\  
  \Omega^{MN} \,{\cal V}_M{}^{AB} \, {\cal V}_N{}^{CD} &=& 0\,. 
  \label{VOM}
\end{eqnarray}
The analogue of the external metric $g_{\mu\nu}$ in the internal sector is the positive definite symmetric 
real matrix
\bea
{\cal M}_{MN} &\equiv& {\cal V}_{M\,AB} {\cal V}_N{}^{AB} + {\cal V}_{N\,AB} {\cal V}_M{}^{AB}
\;,
\label{defM}
\eea
 in terms of which the bosonic sector in Ref.~\cite{Hohm:2013uia}
 has been constructed.

The 56 gauge fields $\cA_\mu{}^M$ in (\ref{fieldcontent}) are subject to the first order duality equations 
given by\,\footnote{\label{convfootnote}
We use the space-time conventions of Ref.~\cite{deWit:2007mt}, such that our tensor density 
$\varepsilon_{\mu\nu\rho\sigma}$  is related to the one employed in Ref.~\cite{Hohm:2013uia} by
$ \varepsilon^{\rm [0705.2101]}_{\mu\nu\rho\sigma}=i\varepsilon^{\rm [1312.4542]}_{\mu\nu\rho\sigma}$\,.
}
\bea
{\cal F}^-_{\mu\nu\,AB} &\equiv& \frac12\,{\cal F}_{\mu\nu\,AB}
-\frac14\,e\,\varepsilon_{\mu\nu\rho\sigma}\, {\cal F}^{\rho\sigma}{}_{AB}
~=~0
\;.
\label{dualityV}
\eea
Here, the 56 non-abelian field strengths are defined as
\bea
{\cal F}_{\mu\nu\,AB} &\equiv& {\cal F}_{\mu\nu}{}^M \, {\cal V}_{M\,AB}\;,
\eea
\bea
{\cal F}_{\mu\nu}{}^M &\equiv&  2 \partial_{[\mu} \cA_{\nu]}{}^M 
-2\,\cA_{[\mu}{}^N \partial_N \cA_{\nu]}{}^M 
-\frac1{2}\left(24\, (t_\bfa)^{MN} (t^\bfa)_{KL}
-\Omega^{MN}\Omega_{KL}\right)\,\cA_{[\mu}{}^K\,\partial_N \cA_{\nu]}{}^L
\nonumber\\
&&{}
- 12 \,  (t^\bfa)^{MN} \,\partial_N \cB_{\mu\nu\,\bfa}
-\frac12\,\Omega^{MN}\,\cB_{\mu\nu\,N}
\;,
\label{YM}
\eea
with the 2-forms $\cB_{\mu\nu\,\bfa}$, $\cB_{\mu\nu\,N}$ from (\ref{fieldcontent}), transforming 
in the adjoint and the fundamental representation of ${\rm E}_{7(7)}$, respectively.
The latter form is a covariantly constrained tensor field,
i.e.\ it is constrained by algebraic equations analogous to (\ref{sectioncondition})
\be
 \begin{split}
  (t_\bfa)^{MN}\,\cB_{M }\cB_{N} \ &= \ 0\;, \qquad 
  (t_\bfa)^{MN}\,\cB_M\,\partial_NA \ = \ 0 \,, \qquad (t_\bfa)^{MN}\,\partial_M\,\cB_N \ = \ 0 \,, \\
  &\Omega^{MN}\,\cB_M\,\cB_N\ = \ 0 \,, \qquad \Omega^{MN}\,\cB_M\,\partial_NA\ = \ 0 \,.
 \end{split}
 \label{sectionconditionB}
 \ee  
Its presence is necessary for consistency of the hierarchy of non-abelian 
gauge transformations and can be inferred directly from the properties of
the Jacobiator of generalized diffeomorphisms~\cite{Hohm:2013uia}.
In turn, after solving the section constraint it ensures the correct and duality covariant
description of those degrees of freedom that
are on-shell dual to the 11-dimensional gravitational degrees of freedom. 

Using (\ref{VOM}) and (\ref{defM}), equations (\ref{dualityV}) take the form 
of the  twisted self-duality equations\,\footnote{See footnote \ref{convfootnote}.
}
\bea
{\cal F}_{\mu\nu}{}^M &=&  
\frac12\,i\,e\varepsilon_{\mu\nu\rho\sigma}\,\Omega^{MN} {\cal M}_{NK}\,{\cal F}^{\rho\sigma}{}^K
\;.
\label{dualityM}
\eea

The bosonic exceptional field theory is invariant under generalized diffeomorphisms in the internal
coordinates, acting via~\cite{Coimbra:2011ky,Berman:2012vc}
\bea\label{genLie}
\mathbb{L}_{\Lambda} U^M 
\ \equiv \ \Lambda^K \partial_K U^M - 12\, \mathbb{P}^M{}_N{}^K{}_L\,\partial_K \Lambda^L\,U^N
+\lambda(U)\,\partial_P \Lambda^P\,U^M
\;,
\eea
on a fundamental vector $U^M$ of weight $\lambda(U)$. The projector on the adjoint representation
\begin{equation} \label{adjproj}
\mathbb{P}^K{}_M{}^L{}_N\equiv
(t_\bfa)_M{}^K (t^\bfa)_N{}^L 
=
\frac1{24}\,\delta_M^K\delta_N^L
+\frac1{12}\,\delta_M^L\delta_N^K
+(t_\bfa)_{MN} (t^\bfa)^{KL}
-\frac1{24} \,\Omega_{MN} \Omega^{KL}
\;, 
\end{equation}
ensures that the action (\ref{genLie}) is compatible with the ${\rm E}_{7(7)}$ group structure. The generalized diffeomorphisms also give rise to the
definition of covariant derivatives 
\bea
{D}_\mu=\partial_\mu - \mathbb{L}_{\cA_\mu}
\;,
\label{covD}
\eea
whose commutator precisely
closes into the field strength (\ref{YM}). The full bosonic theory is invariant under the vector and tensor gauge symmetries
\bea
 \delta _{\Lambda} e_\mu{}^{\alpha} &=& \mathbb{L}_{\Lambda} e_\mu{}^{\alpha}\;,
 \nonumber\\
   \delta _{\Lambda}{\cal V}_M{}^{AB} &=&  \mathbb{L}_{\Lambda} {\cal V}_M{}^{AB}   \;,
\nonumber\\
   \delta_{\Lambda,\Xi} \cA_\mu{}^M &=&  {D}_\mu \Lambda^M  
   +  12\,(t^\bfa)^{MN}\,\partial_N\Xi_{\mu\,\bfa}
   +\frac12\,\Omega^{MN}\,\Xi_{\mu\,N} \;,
   \nonumber\\
      \delta_{\Lambda,\Xi} \cB_{\mu\nu \, \bfa} &=&   2\,{ D}_{[\mu}\Xi_{\nu]\bfa} +
   (t_\bfa)_{KL}\, \Lambda^K{\cal F}_{\mu\nu}{}^{L}-(t_\bfa)_{KL}\,\cA_{[\mu}{}^K\, \delta \cA_{\nu]}{}^L\;, 
     \nonumber\\
     \delta_{\Lambda,\Xi}  \cB_{\mu\nu M} &=& 
      2\,{D}_{[\mu}\Xi_{\nu]M} 
           +48\,(t^\bfa)_L{}^K  \left(\partial_K \partial_M \cA_{[\mu}{}^L\right) \Xi_{\nu]\bfa}
     \nonumber\\
     &&{}
   +  \Omega_{KL}\left(\cA_{[\mu}{}^K \partial_M \delta \cA_{\nu]}{}^L
-\partial_M \cA_{[\mu}{}^K \delta \cA_{\nu]}{}^L
-{\cal F}_{\mu\nu}{}^K\partial_M \Lambda^L
+ \partial_M {\cal F}_{\mu\nu}{}^K   \Lambda^L \right)
     ,\;\;\;\;\; \quad
    \label{gaugeLX}
 \eea
with parameters $\Lambda^M$, $\Xi_{\mu \, \bfa}$, $\Xi_{\mu \, M}$, the latter constrained according to (\ref{sectionconditionB}).
The $\Lambda$-weights of the various bosonic fields and parameters are collected in table~\ref{tab:weights},
where we have also included the $\Lambda$-weights of the fermionic fields to be introduced later.
Note that $\cB_{\mu\nu\,\bfa}$ and $\cB_{\mu\nu\,M}$ appear in the field strength (\ref{YM})
only via the combination/projection
\bea
- 12 \,  (t^\bfa)^{MN} \,\partial_N \cB_{\mu\nu\,\bfa}
-\frac12\,\Omega^{MK}\,\cB_{\mu\nu\,K}
\;.
\label{projB}
\eea
As a result, we observe the following additional gauge transformations that leave the field strengths
invariant
\bea
\delta_\Omega \cB_{\mu\nu\,\bfa} &=& \partial_M \Omega_{\mu\nu}{}^M{}_\bfa + (t_\bfa)_M{}^N \Omega_{\mu\nu}{}_N{}^M\;,
\nonumber\\
\delta_\Omega \cB_{\mu\nu\,M} &=& 
- \partial_M \Omega_{\mu\nu}{}_N{}^N - 2 \,\partial_N \Omega_{\mu\nu}{}_M{}^N
\;,
\label{shiftB}
\eea
where $\Omega_{\mu\nu}{}^M{}_\bfa$ is a parameter living 
in the ${\bf 912}$ of ${\rm E}_{7(7)}$, i.e.
\bea
(t^\bfa){}^{(KL} \Omega_{\mu\nu}{}^{M)}{}_\bfa &=& 0
\;,
\label{912Omega}
\eea
and $\Omega_{\mu\nu}{}_N{}^M$ is a parameter constrained in the index $_N$ 
just as the $N$ index in partial derivatives $\partial_N$, see equations (\ref{sectioncondition}), 
and the two-form $\cB_{\mu\nu\,N}$, see equations (\ref{sectionconditionB}).
The shift transformations (\ref{shiftB}) should be understood as the tensor gauge transformations
of the three-form gauge potentials of the theory (which we have not explicitly introduced)
that also act on the two-forms due to the St\"uckelberg couplings of
their field strengths. They precisely drop out in the projection (\ref{projB}) which is the one appearing in 
the vector field strengths.

\begin{table}[tb]
\begin{center}
\begin{tabular}{|c||c|c|c|c|c||c|c|}\hline
{\rm field} & $e_{\mu}{}^{\alpha}$ & ${\cal V}_{M}{}^{AB}$  & $\cA_\mu{}^M$, $\Lambda^M$ &
$\cB_{\mu\nu\,\bfa}$\,, $\Xi_{\mu\,\bfa}$ &
$\cB_{\mu\nu\,M}$, $\Xi_{\mu\,M}$ & $\chi_{ABC}$ & $\psi_{\mu}^A$\,, $\epsilon^A$
\\ \hline
$\lambda$ & $\frac12$ & 0 & $\frac12$ & $1$ & $\frac12$ & $-\frac14$ & $\frac14$ \\ \hline
\end{tabular}
\caption{$\Lambda$-weights for the bosonic and fermionic fields and parameters.}
\label{tab:weights}
\end{center}
\end{table}

Other than the first-order duality equations (\ref{dualityV}), the remaining equations of motion of the bosonic theory
are most compactly described by a Lagrangian\,\footnote{
Due to the self-duality (\ref{dualityM}) of the vector fields, this is understood as a ``pseudo-Lagrangian"
in the sense of a democratic action~\cite{Bergshoeff:2001pv}
such that the duality equations (\ref{dualityM}) are to be imposed {\em after}
varying the Lagrangian.}
\bea
\label{finalaction}
 {\cal L}_{\rm EFT}&=&  e\,  \widehat{R}
 +\frac{1}{48}\,e\,g^{\mu\nu}\,{D}_{\mu}{\cal M}^{MN}\,{ D}_{\nu}{\cal M}_{MN}
  -\frac{1}{8}\,e\,{\cal M}_{MN}\,{\cal F}^{\mu\nu M}{\cal F}_{\mu\nu}{}^N
 \nonumber\\&&{}
+{\cal L}_{\rm top}-e\,V({\cal M}_{MN},g_{\mu\nu}) \,.
\eea
Let us present the different terms. The modified Einstein Hilbert term carries the Ricci scalar $\widehat{R}$ obtained
from contracting the modified Riemann tensor
  \be
  \widehat{R}_{\mu\nu}{}^{\alpha \beta} \ \equiv \  R_{\mu\nu}{}^{\alpha \beta}[\omega]+{\cal F}_{\mu\nu}{}^{M}
  e^{\alpha}{}^{\rho}\partial_M e_{\rho}{}^{\beta}\,,
  \label{improvedRE7}
 \ee
with the spin connection $\omega_\mu{}^{\alpha \beta}$ obtained from the covariantized vanishing torsion condition
\bea
0 &=& {\cal D}_{[\mu} e_{\nu]}{}^{\alpha} ~\equiv~
\partial_{[\mu} e_{\nu]}{}^{\alpha} -\cA_{[\mu}{}^K \partial_K  e_{\nu]}{}^{\alpha} -\frac12\,\partial_K \cA_{[\mu}{}^K\,e_{\nu]}{}^{\alpha}
+ \omega_{[\mu}{}^{\alpha \beta}\,e_{\nu] \beta}
\;.
\label{omega}
\eea
The scalar kinetic term can be equivalently expressed as
\bea
\frac{1}{48}\,{D}^\mu {\cal M}^{KL} D_\mu {\cal M}_{KL} 
&=&
-\frac{1}{6}\,
 {\cal P}^\mu{}_{ABCD} \, {\cal P}_\mu{}^{ABCD}  
\;,
  \label{kinS}
\eea
where we have introduced the coset currents ${\cal P}_\mu{}^{ABCD}$
as follows
\bea
 {\cal D}_\mu{\cal V}_M{}^{AB} &\equiv& 
 D_\mu{\cal V}_M{}^{AB} +  {\cal Q}_{\mu \,C}{}^{[A} \,{\cal V}_M{}^{B]C} 
 ~=~  {\cal P}_\mu{}^{ABCD} \,{\cal V}_{M CD} \;
\label{PQ}
\eea
according to the decomposition (\ref{varV}) and where $D_\mu$ refers to the covariant derivative defined in equation (\ref{covD}).
This moreover defines the composite ${\rm SU}(8)$ connection
\begin{equation}
  {\cal Q}_\mu{}_A{}^B ~=~
\frac{2i}{3}\,{\cal V}^N{}^{BC} \,D_\mu {\cal V}_N{}_{CA}
\;,
\label{defQ}
\end{equation}
indicating that the 56-bein transforms under local ${\rm SU}(8)$ transformations. Thus, we will in the following use 
${\cal D}_{\mu}\equiv D_\mu + {\cal Q}_\mu$ to denote the resulting ${\rm SU}(8)$-covariant derivatives. 
The vector kinetic term in (\ref{finalaction}) 
\bea
  -\frac{1}{8}\,e\,{\cal M}_{MN}\,{\cal F}^{\mu\nu M}{\cal F}_{\mu\nu}{}^N &=&
  -\frac14\,e\, {\cal F}_{\mu\nu}{}^{AB} {\cal F}^{\mu\nu}{}_{AB}
  \;,
\eea
simply contracts the non-abelian field strengths (\ref{YM})
with the internal metric (\ref{defM}), while the topological term is most compactly given as the boundary 
contribution of a five-dimensional bulk integral
 \bea
 \label{Ltop}
\int_{\partial\Sigma_5} d^4x\,  \int d^{56}Y\, {\cal L}_{\rm top} 
&=& \frac{i}{24}\,\int_{\Sigma_5} d^5x \,\int d^{56}Y
\varepsilon^{\mu\nu\rho\sigma\tau}\,{\cal F}_{\mu\nu}{}^M\,
{\cal D}_{\rho} {\cal F}_{\sigma\tau}{}_M
\;.
 \eea
Finally, the last term in (\ref{finalaction}) is given by
\bea
  V({\cal M}_{MN},g_{\mu\nu})  &=& 
  -\frac{1}{48}{\cal M}^{MN}\partial_M{\cal M}^{KL}\,\partial_N{\cal M}_{KL}+\frac{1}{2} {\cal M}^{MN}\partial_M{\cal M}^{KL}\partial_L{\cal M}_{NK}\label{fullpotential}\\
  &&{}-\frac{1}{2}g^{-1}\partial_Mg\,\partial_N{\cal M}^{MN}-\frac{1}{4}  {\cal M}^{MN}g^{-1}\partial_Mg\,g^{-1}\partial_Ng
  -\frac{1}{4}{\cal M}^{MN}\partial_Mg^{\mu\nu}\partial_N g_{\mu\nu}\;,
\nonumber 
\eea 
in terms of the internal and external metric. For later use, we note that
in terms of the 56-bein and modulo a total derivative 
$e^{-1} \partial_M (e K^M)$, the potential takes the form
\bea
  V({\cal V}_M{}^{AB},g_{\mu\nu})  &=& 
 4\,{\cal V}^{M}{}_{[AB} {\cal V}^N{}_{CD]}\,\left(
 \partial_M p_{N}{}^{ABCD} - \frac12\,q_{M\,E}{}^{A}\,p_{N}{}^{EBCD}\right)
 \nonumber\\
 &&{}
 +\frac16\,
 {\cal M}^{MN}\,p_{M}{}^{ABCD}p_{N\,ABCD} 
 +4\,
 {\cal V}^{M}{}_{AB} {\cal V}^N{\,}^{CD}\,p_{M}{}^{ABEF}p_{N\,CDEF} 
  \nonumber\\
 &&{}
 -\frac{1}{4}  {\cal M}^{MN}g^{-1}\partial_Mg\,g^{-1}\partial_Ng
  -\frac{1}{4}{\cal M}^{MN}\partial_Mg^{\mu\nu}\partial_N g_{\mu\nu}\;,
\label{fullpotentialV} 
\eea 
expressed via the standard decomposition of the Cartan form $\cV^{-1}\partial_M\cV$ along the compact and non-compact parts of the 
E$_{7(7)}$ Lie algebra
\be\label{Cartan0}
{q}_M{}_A{}^B \equiv
\frac{2i}{3}\,{\cal V}^N{}^{BC} \,\partial_M{\cal V}_N{}_{CA}\;, \quad
{p}_{M}{}^{ABCD} \equiv i{\cal V}^{N}{}^{AB} \,\partial_M {\cal V}_{N}{}^{CD} 
\;.
\ee
Written in the form of~(\ref{fullpotentialV}), it is easy to observe that
the first two lines of the potential reproduce the corresponding terms in equation (7.5) 
of Ref.~\cite{deWit:1986mz}.

All five terms in (\ref{finalaction}) are separately gauge invariant under generalized diffeomorphisms (\ref{gaugeLX})
in the internal coordinates.
In addition, the full set of equations of motion is invariant under generalized diffeomorphisms in the external
coordinates acting as
 \bea
 \delta_\xi e_{\mu}{}^{\alpha} &=& \xi^{\nu}{D}_{\nu}e_{\mu}{}^{\alpha}
 + {D}_{\mu}\xi^{\nu} e_{\nu}{}^{\alpha}\;, \label{diffx}\\
\delta_\xi {\cal M}_{MN} &=& \xi^\mu \,{D}_\mu {\cal M}_{MN}\;,\nonumber\\
\delta_\xi \cA_{\mu}{}^M &=& \xi^\nu\,{\cal F}_{\nu\mu}{}^M + {\cal M}^{MN}\,g_{\mu\nu} \,\partial_N \xi^\nu
\;,
\nonumber\\
\delta_\xi \cB_{\mu\nu\,\bfa} &=& \xi^\rho\,{\cal H}_{\mu\nu\rho\,\bfa}
-(t_\bfa)_{KL}\,\cA_{[\mu}{}^K\, \delta_\xi {\cal A}_{\nu]}{}^L
\;,\nonumber\\
\delta_\xi \cB_{\mu\nu\,M} &=& \xi^\rho\,{\cal H}_{\mu\nu\rho\,M}
-2ie\,{\varepsilon}_{\mu\nu\rho\sigma} g^{\sigma\tau} {D}^\rho\left(g_{\tau\lambda}\partial_M \xi^\lambda\right) 
-\left(\cA_{[\mu}{}^K \partial_M \delta_\xi \cA_{\nu]}{}_K
-\partial_M \cA_{[\mu}{}^K \delta_\xi \cA_{\nu]}{}_K\right)
\;.
\nonumber
\eea
When $\partial_M = 0$, this reduces to the 
action of standard four-dimensional diffeomorphisms.
Remarkably, the invariance of the theory under (\ref{diffx}) fixes all relative coefficients in 
(\ref{finalaction}) and thus uniquely determines all equations of motion.

Variation of (\ref{finalaction}) gives the field equations for the scalar fields parametrizing 
${\cal M}_{MN}$ and the Einstein field equations for $g_{\mu\nu}$. Variation with respect to the
two-forms $\cB_{\mu\nu\,\bfa}$ and $\cB_{\mu\nu\,M}$ yields projections of the first-order
vector field equations (\ref{dualityM}). Finally, the variation of the action with respect to the vector fields
leads to second order field equations 
\bea
{D}_\nu\left(e\, {\cal M}_{MN}\,{\cal F}^{\mu\nu}{}^N\right) &=& 
e\left(\widehat{J}^\mu{}_{M} +{\cal J}^\mu{}_M \right)
\label{DF}
\eea
after combining with the derivative of (\ref{dualityM}), and where
the gravitational and matter currents are defined by the respective contributions
from the Einstein-Hilbert and the scalar kinetic term
\bea\label{STep}
\widehat{J}^\mu{}_{M} &\equiv&
-2  e_{\alpha}{}^\mu e_{\beta}{}^\nu \left(
\partial_M \omega_{\nu}{}^{\alpha \beta} 
-  {\cD}_{\nu} \left(   e^{\rho[\alpha} \partial_M e_{\rho}{}^{\beta]} \right) \right)
\;,\nonumber\\
 {\cal J}^\mu{}_{M} &\equiv& 
 2i\,e^{-1}\,\partial_N\left(e\, {\cal P}^\mu{\,}^{ABCD} {\cal V}^N{}_{AB}  {\cal V}_{M}{}_{CD} 
~-~{\rm c.c.} \right)
 -\frac{1}{24}\,{\cD}^\mu {\cal M}^{KL} \partial_M {\cal M}_{KL}
 \;.
\eea
Equation \eqref{DF} may be compared to the second order field
equations obtained from combining the derivative of (\ref{dualityM}) with the
Bianchi identities
\bea
  3\, {\cal D}_{[\mu}{\cal F}_{\nu\rho]}{}^M &=& 
 - 12\, (t^\bfa)^{MN}\partial_N{\cal H}_{\mu\nu\rho\,\bfa} - \frac12\,\Omega^{MN}\,
  {\cal H}_{\mu\nu\rho\,N}
  \;,
  \label{Bianchi}
 \eea
where ${\cal H}_{\mu\nu\rho\,\bfa}$ and ${\cal H}_{\mu\nu\rho\,M}$ denote the non-abelian
field strengths of the two-forms
\bea
 {\cal H}_{\mu\nu\rho\,\bfa} &=&
3\,{\cal D}_{[\mu} \cB_{\nu\rho]\,\bfa}
-3\,(t_\bfa)_{KL}\,\cA_{[\mu}{}^K\partial_{\nu\vphantom]} \cA_{\rho]}{}^L+ \dots
\nonumber\\
 {\cal H}_{\mu\nu\rho\,M} &=&
3\,{\cal D}_{[\mu} \cB_{\nu\rho]\,M}
-3\left(\cA_{[\mu}{}^N \partial_M \partial_{\nu\vphantom]} \cA_{\rho]}{}_N-
 \partial_M \cA_{[\mu}{}^N  \partial_{\nu\vphantom]} \cA_{\rho]}{}_N
\right) +\dots
\;.
\eea
Combining (\ref{dualityM}), (\ref{DF}), and (\ref{Bianchi}) 
gives rise to the first-order duality equations describing the dynamics of the
two-forms
\bea
i\widehat{J}^\mu{}_{M} +
\frac{1}{3}\, {\cal D}^\mu
 {\cal V}^{N}{}^{AB} \,\partial_M {\cal V}_{N}{}_{AB}
&=&
\frac1{12}\,e^{-1}\varepsilon^{\mu\nu\rho\sigma} \,{\cal H}_{\nu\rho\sigma\,M}
\;,\nonumber\\
(t_\bfa)_N{}^M\,
\left({\cal P}^\mu{\,}^{ABCD} {\cal V}^N{}_{AB}  {\cal V}_{M}{}_{CD} 
-{\cal P}^\mu{}_{ABCD} {\cal V}^N{}^{AB}  {\cal V}_{M}{}^{CD}
\right)
&=&
e^{-1}\varepsilon^{\mu\nu\rho\sigma} \,
 {\cal H}_{\nu\rho\sigma\,\bfa} 
 \;.
 \label{dualH}
\eea
Strictly speaking, the second equation only holds under projection with $(t^\bfa)^{KL}\partial_L$.
The first-order equations (\ref{dualH}) show that the two-form fields 
do not bring in additional degrees of freedom to the theory.


\section{SU$(8)\times {\rm E}_{7(7)}$ exceptional geometry} 
\label{sec:e7covgeo}


\subsection{Connections}  \label{subsec:e7covgeo}

In this section we set up the ${\rm E}_{7(7)}$-covariant geometrical formalism for defining derivatives that  are simultaneously covariant with respect to generalized internal 
diffeomorphisms, local SU$(8)$, and ${\rm SO}(1,3)$ Lorentz transformations. This will 
allow us to couple the bosonic ${\rm E}_{7(7)}$-covariant exceptional field 
theory to fermions and to establish the link with the `ground up' approach to
be described in the next section. From the representation content of maximal
$N=8$ supergravity, or equivalently from an appropriate decomposition of the 
$D=11$ gravitino, it follows that the fermionic fields of the theory are ${\rm SO}(1,3)$ spinors, 
and transform in  the $\bf{8}$ (the gravitini $\psi_\mu^A$) and in the $\bf{56}$
(the matter fermions $\chi^{ABC}$) of ${\rm SU}(8)$, respectively.\,\footnote{
We use spinor conventions from Ref.~\cite{deWit:2007mt}, i.e.\ in particular 
$\gamma^{\mu\nu\rho\sigma} = e^{-1} \epsilon^{\mu\nu\rho\sigma}\,\gamma^5$ and 
$\gamma^5 \epsilon_A = -\epsilon_A$\,.} The main new feature is that,
like the bosonic fields (\ref{fieldcontent}), the fermions are here taken to  
depend on $4+56$ coordinates modulo the section condition~(\ref{sectioncondition}).
Under `internal' generalized diffeomorphisms (\ref{genLie}) they transform as
scalar densities with weights as given in table~\ref{tab:weights}.

For the external derivatives, the relevant connections have been introduced in the previous section.  On a spinorial object in the fundamental representation of E$_{7(7)}\times {\rm SU}(8)$, the covariant derivative is defined as
\bea
{\cal D}_{\mu} X_{A\,N} &=&
D_{\mu} X_{A \,N} + \frac14\,\omega_{\mu}{}^{\alpha \beta}\gamma_{\alpha \beta} X_{A\,N} + \frac{1}{2}{\cal Q}_{\mu\,A}{}^{B}X_{BN}\;,
\label{covDmu}
\eea
with the  E$_{7(7)}$-covariant derivative $D_\mu$ from (\ref{covD}), and the spin- and SU$(8)$-connections 
defined by (\ref{omega}) and (\ref{defQ}), respectively. By construction, these connections ensure covariance
of ${\cal D}_{\mu} X_{A\, N}$.
As usual, for covariant derivatives on four-dimensional space-time 
tensors we may also introduce the covariant derivative $\nabla_\mu$ which in addition to
(\ref{covDmu}) carries the Christoffel connection defined by the  standard 
(though covariantized) vierbein postulate
\bea
{\cal D}_{\mu} e_\nu{}^{\alpha} - \Gamma_{\mu\nu}{}^\rho\,e_\rho{}^{\alpha} &=&0 
\;.
\label{VP1}
\eea

For the internal sector, we similarly define a covariant derivative in the internal variables 
$Y^M$. The most general such derivative (denoted by $\nabla_M$) 
acts on Lorentz indices, SU(8) indices and E$_{7(7)}$ indices, and has the form
 \bea
{\nabla}_{M} X_{A \, N} \ &=& \ \partial_M X_{A \, N}+ \frac14\,\omega_{M}{}^{\alpha \beta}\gamma_{\alpha \beta} X_{A \, N}  \nn\\[1mm]
&& +\, \frac{1}{2}{\cal Q}_{M \,A}{}^{B}X_{B \,N}-\Gamma_{MN}{}^{K}\,X_{A\,K}
  -\frac23\,\lambda(X)\,\Gamma_{KM}{}^{K} X_{A \, N}\;, 
 \label{full_nabla}
 \eea
if $X$ is a generalized tensor of weight $\lambda(X)$ under generalized 
diffeomorphisms (\ref{genLie}).
Likewise, we use 
 \be
{\cal D}_{M} X_{A \, N} \ = \ \partial_M X_{A \, N}+ \frac14\,\omega_{M}{}^{\alpha \beta}\gamma_{\alpha \beta} X_{A \, N}
+\frac{1}{2}{\cal Q}_{M \,A}{}^{B}X_{B \, N}\;, 
 \label{full_DM}
 \ee
for the derivative without the Christoffel connection $\Gamma_{MN}{}^{K}$.
The required transformation rules for the connections are determined by covariance. 
Under generalized diffeomorphisms (\ref{genLie}), 
the non-covariant variation of the first term in (\ref{full_nabla}) is given by 
 \be
  \Delta^{\rm nc}_{\Lambda}\big(\partial_M X_{A \, N}\big) \ = \ 
  12 \,\mathbb{P}^{K}{}_{N}{}^{P}{}_{Q}\,\partial_M\partial_P\Lambda^Q\,X_{A \,K}\;,
 \ee
where we recall that the covariant terms carry a weight of $-\frac{1}{2}$~\cite{Hohm:2013uia}.     
Thus, $\Gamma_{MN}{}^P$ also carries a weight of $-\frac{1}{2}$ and has the inhomogeneous transformation 
 \be\label{delinhomo}
  \delta_{\Lambda}\Gamma_{MK}{}^{N} \ = \ \mathbb{L}_{\Lambda}\Gamma_{MK}{}^{N}
  +12 \,\mathbb{P}^{N}{}_{K}{}^{P}{}_{Q}\,\partial_M\partial_P\Lambda^Q\;.
 \ee
This implies in particular, 
 \be\label{traceVAR}
    \delta_{\Lambda}\Gamma_{MK}{}^{M} \ = \ \mathbb{L}_{\Lambda}\Gamma_{MK}{}^{M}
  +\frac{3}{2} \partial_K\partial_P\Lambda^P\;, 
 \ee
explaining the factor $\frac{2}{3}$ in the last term of (\ref{full_nabla}). In the following, we will discuss the
definition of the internal spin- and ${\rm SU}(8)$ connection.

The internal spin connection $\omega_{M}{}^{\alpha \beta}$ is defined by analogy with (\ref{PQ}) 
by demanding that
\bea
{\cal D}_M e_\mu{}^{\alpha} &=& \pi_M{}^{\alpha \beta}\, e_{\mu \beta}\;,
\label{VtorM}
\label{omegaM}
\eea
with $\pi_M{}^{\alpha \beta}=\pi_M{}^{(\alpha \beta)}$ living on the coset ${\rm GL}(4)/{\rm SO}(1,3)$\,.
As a consequence,
\begin{equation}
 {\omega}_M{}^{\alpha \beta} ~=~ e^{\mu[\alpha} \partial_M e_\mu{}^{\beta]}\;,
\end{equation}
and
\bea
e^{\mu[\alpha} {\cal D}_M e_\mu{}^{\beta]} &=& 
0~=~ e_{\alpha[\mu}  {\cal D}_M e_{\nu]}{}^{\alpha} 
\;.
\label{vTo}
\eea
Later, it will turn out to be convenient to also introduce a modified 
spin connection $\widehat{\omega}_M{}^{\alpha \beta}$
\bea
\widehat{\omega}_M{}^{\alpha \beta} &\equiv& {\omega}_M{}^{\alpha \beta}
-\frac1{4}\,{\cal M}_{MN}\,{\cal F}_{\mu\nu}{}^N\,e^{ \mu \, \alpha} e^{\nu\,\beta }
\;,
\label{omegaMhat}
\eea
including the non-abelian field strengths ${\cal F}_{\mu\nu}{}^N$ in a fashion reminiscent of Kaluza-Klein theory, whereby we view fields 
$e_\mu{}^{\alpha}$, ${\cal V}_{M}{}^{AB}$, and $A_\mu{}^M$ as parts of a single big vielbein.
We will denote the corresponding covariant derivatives by $\widehat{\cal D}$ and $\widehat{\nabla}$,
respectively.

In order to discuss the remaining connections in (\ref{full_nabla}), let us first require that 
the internal ${\rm SU}(8)$ connection and the Christoffel connection are related by
a generalized vielbein postulate (or `GVP', for short)
\bea
0 &\equiv&
{\nabla}_{M} {\cal V}_N{}^{AB} ~=~ \partial_M {\cal V}_N{}^{AB}
+{\cal Q}_{M \,C}{}^{[A}{\cal V}_N{}^{B]C} -\Gamma_{MN}{}^{K}\,{\cal V}_K{}^{AB}
\;, 
\label{VP2}
\eea
which is the analogue of (\ref{VP1}) for the internal sector.  In analogy with 
standard differential geometry one would now like to solve this relation 
for both the SU(8) connection $\cQ_{M \,A}{}^B$ and the generalized affine 
connection $\Gamma_{MN}{}^P$ in terms of the 56-bein $\cV$ and its 
derivatives $\partial_M \cV$. While in ordinary differential geometry, a unique such
answer can be obtained by imposing vanishing torsion, here there
remain further ambiguities. In addition one would like the resulting expressions
to satisfy all requisite covariance properties, to wit: $Q_{M \, A}{}^B$ should transform as 
a proper connection under local SU(8) and as a generalized vector under generalized
diffeomorphisms, while $\Gamma_{MN}{}^P$ should transform as a generalized
affine connection under generalized diffeomorphisms and as a singlet under local
SU(8). However, parallel to DFT it is not possible to express a connection satisfying these combined
covariance requirements as a function of only $\cV$ and $\partial_M\cV$ in a covariant way,
as we will also confirm in terms of a simplified example in appendix~\ref{app:gl7}, and in 
terms of an explicit calculation for the SU(8) connection in appendix~\ref{app:su8connection}.

The first step in reducing the ambiguities is to constrain the connections by requiring 
the generalized torsion to vanish; this amounts to the constraint~\cite{Coimbra:2011ky}
\be\label{GENtorsion}
  {\cal T}(V,W)^M \ = \ {\cal T}^{M}{}_{NK}V^NW^K \ \equiv \ \mathbb{L}^{\nabla}_VW^M-\mathbb{L}_V W^M \ \equiv \ 0
\ee
for vectors $V,W$ {of weight $\frac{1}{2}$} where
$\mathbb{L}^{\nabla}$ denotes the generalized Lie derivative with
all partial derivatives replaced by covariant derivatives.  Explicit evaluation of this condition yields
 \be\label{FULLtau}
  {\cal T}_{NK}{}^{M} \ = \ \Gamma_{NK}{}^{M}-12\,\mathbb{P}^M{}_{K}{}^{P}{}_{Q}\,\Gamma_{PN}{}^{Q}
  +4\,\mathbb{P}^M{}_{K}{}^{P}{}_{N}\Gamma_{QP}{}^{Q}
  \;, 
 \ee
with $\mathbb{P}$ the adjoint projector defined in equation (\ref{adjproj}). Indeed, it is a straightforward computation to show that this combination transforms covariantly under generalized diffeomorphisms.  From (\ref{delinhomo}) and using the cubic identity (A.3) of Ref.~\cite{Hohm:2013uia}
 \be
 \begin{split}
    \Delta^{\rm nc}_{\Lambda}\big(\Gamma_{PM}{}^{N}-12\,\mathbb{P}^N{}_{M}{}^{K}{}_{L}\,\Gamma_{KP}{}^{L}\big)
  \ &= \  -6\,(t^{\bfa})_{P}{}^R (t_{\bfa})_M{}^N \partial_R\partial_K\Lambda^K\\
  \ &= \ -4\,\mathbb{P}^N{}_{M}{}^{R}{}_{P}\,\Delta^{\rm nc}_{\Lambda}\Gamma_{KR}{}^{K}\;, 
 \end{split}
 \ee
where we have used equation (\ref{traceVAR}) and the fact that all other terms in (A.3) vanish by the section constraint. The last term is of the form of the non-covariant variation of the final term in (\ref{FULLtau}), with the opposite sign. Hence, the generalized torsion transforms as a generalized tensor.  The fact that the generalized torsion is gauge covariant means that it can be set consistently to zero. 

From equation (\ref{VP2}), the last two indices in
the generalized Christoffel connection $(\Gamma_{M})_{N}{}^K$ take values in the adjoint of ${\rm E}_{7(7)}$. Hence, the generalized connection  lives in the ${\rm E}_{7(7)}$ representations
\bea
{\bf 56} \otimes {\bf 133} &=& {\bf 56}+{\bf 912}+{\bf 6480}\;.
\eea
Using the explicit form of the corresponding projectors given in ref.~\cite{deWit:2002vt}, one can verify 
that the vanishing torsion constraint (\ref{GENtorsion}) translates 
into~\cite{Coimbra:2011ky,Aldazabal:2013mya,Cederwall:2013naa}
\bea
\Gamma_{MN}{}^K\Big|_{\bf 912} &=& 0\;.
\label{notorsion912}
\eea
In addition, requiring density compatibility of the internal derivatives
according to
\be \label{viercomp}
   {\nabla}_{M}e \ \equiv \ 0\;,
\ee
fixes
\be\label{traceconn}
   \frac{3}{4}\,e^{-1}\partial_M e ~=~ \Gamma_{KM}{}^{K} 
   ~=~-\Omega_{MN} \Omega^{PQ} \Gamma_{PQ}{}^N
   \;,
\ee
where the second equality is obtained from contraction of (\ref{notorsion912}).
As we will explain below, this trace must drop out in all relevant expressions involving
the fermions.

Next, we work out the most general SU(8) connection compatible with 
vanishing generalized torsion. Using equation (\ref{VP2}), the condition (\ref{notorsion912}) 
is equivalent to the following conditions on the internal ${\rm SU}(8)$
connection ${\cal Q}_M{}$:
\bea
{\cal V}^K{\,}^{AB}\, {\cal D}_P{\cal V}_K{}^{CD}
&=&
6\,{\cal V}^{K}{\,}^{[AB}{}  \left( {\cal D}_K{\cal V}_P{}^{CD]} \right) 
-\frac14\,
\epsilon^{ABCDEFGH}\,{\cal V}^{K}{}_{EF}{}  \left( {\cal D}_K{\cal V}_P{\,}_{GH} \right)
\nonumber\\
&&{}
-2\,\Gamma_{QK}{}^Q\left(
{\cal V}^{K}{\,}^{[AB}{}  {\cal V}_P{}^{CD]} 
-\frac1{24}\,\epsilon^{ABCDEFGH}\,{\cal V}^{K}{}_{EF}{} {\cal V}_P{\,}_{GH} \right),
\label{inter1}\\[2ex]
{\cal V}^K{\,}_{AC}\, {\cal D}_P{\cal V}_K{}^{BC}
&=&
6\left({\cal V}^{K}{}_{AC}{}  \, {\cal D}_K{\cal V}_P{}^{BC} +
{\cal V}^{K}{}^{BC}{} \,{\cal D}_K{\cal V}_P{}_{AC} \right) 
\nonumber\\
&&{}
-\frac34\,\delta_A^B\left( {\cal V}^{K}{}_{CD}{} \, {\cal D}_K{\cal V}_P{}^{CD} 
+{\cal V}^{K}{}^{CD}{}  \,{\cal D}_K{\cal V}_P{}_{CD} \right)
\nonumber\\
&&{}
-2\,\Gamma_{QK}{}^Q\left(
{\cal V}^{K}{}_{AC}{}  \,{\cal V}_P{}^{BC} +
{\cal V}^{K}{}^{BC}{} {\cal V}_P{}_{AC} -\frac18\,\delta_A^B\,{\cal M}_P{}^K\right) 
\;,
\label{inter2}
\eea
which constitute the analogue of (\ref{omega}) in the internal sector. Unlike in the external sector
and standard geometry, the vanishing torsion conditions (\ref{inter1}), (\ref{inter2})
are not sufficient to fully determine the internal ${\rm SU}(8)$
connection \cite{Coimbra:2011ky,Cederwall:2013naa}, but rather constrain it to the following form
\bea
{\cal Q}_M{}_A{}^B &=&
q_M{}_A{}^B + R_M{}_A{}^B + U_M{}_A{}^B + W_M{}_A{}^B \;.
\label{internalQ}
\eea
Here 
\be\label{Cartan}
{q}_M{}_A{}^B \equiv
\frac{2i}{3}\,{\cal V}^N{}^{BC} \,\partial_M{\cal V}_N{}_{CA}\;, \quad
{p}_{M}{}^{ABCD} \equiv i{\cal V}^{N}{}^{AB} \,\partial_M {\cal V}_{N}{}^{CD} 
\ee
are obtained in the standard way from the decomposition of the Cartan
form $\cV^{-1}\partial_M\cV$ along the compact and non-compact parts of the 
E$_{7(7)}$ Lie algebra. We note that $q_{M \, A}{}^B$ transforms properly as a connection
while $p_M{}^{ABCD}$ transforms covariantly under local SU(8), but neither transforms
as a vector under generalized diffeomorphisms.
The remaining pieces in (\ref{internalQ}) are given by 
\bea
R_{M \,A}{}^B &\equiv& 
\frac{4i}3
\left({\cal V}^N{}^{BC} {\cal V}_M{}^{DE}\, p_N{}_{ACDE}
+{\cal V}^N{}_{AC} {\cal V}_M{}_{DE}\, p_N{}^{BCDE}\right)
\nonumber\\
&&{}
+\frac{20i}{27}
\left({\cal V}^N{}^{DE} {\cal V}_M{}^{BC}\, p_N{}_{ACDE}
+{\cal V}^N{}_{DE} {\cal V}_M{}_{AC}\, p_N{}^{BCDE}\right)
\nonumber\\
&&{}
-\frac{7i}{27}\,\delta_A^{B}
\left({\cal V}^N{}^{CD} {\cal V}_M{}^{EF}\, p_N{}_{CDEF}
+{\cal V}^N{}_{CD} {\cal V}_M{}_{EF}\, p_N{}^{CDEF}\right)
\;,
\nonumber\\[2mm]
W_M{}_A{}^B &\equiv&
\frac{8i}{27} \,\Big({\cal V}_M{}_{AC} {\cal V}^{N}{}^{BC}+ {\cal V}_{M}{}^{BC} {\cal V}^N{}_{AC}
-\frac18\,\delta_A^B\,{\cal M}_{MK} \Omega^{NK}\Big) \, \Gamma_{LN}{}^L
\;,
\label{QQUW}
\eea
and by
\be
U_M{}_A{}^B = {\cal V}_{M\,CD}\,u^{CD ,B}{}_A - {\cal V}_{M}{}^{CD}\,u_{CD,A}{}^B
\;,
\label{UU}
\ee
where the SU(8) tensor $u_{CD,A}{}^B$ satisfies
\be
u^{[CD,B]}{}_A \equiv0\;,\quad
u^{CA,B}{}_C \equiv0\;,
\ee
and thus belongs to the $\bf{1280}$ of SU(8).
It is now straightforward to check that $u_{CD,A}{}^B$ drops out of the vanishing 
torsion conditions (\ref{inter1}), (\ref{inter2}) and thus remains undetermined.
An explicit form of ${\cal Q}_M{}_A{}^B$ in terms of the 
${\rm GL}(7)$ components of ${\cal V}_M{}^{AB}$
has been given in Ref.~\cite{Coimbra:2011ky}. 
With $\cQ_{M \, A}{}^B$ given by (\ref{internalQ}), it is now straightforward to
solve (\ref{VP2}) for the affine connection
\be\label{AffCon1}
\Gamma_{MN}{}^P\big( \cV,\partial\cV, \cQ\big) =
i\Big( \cV^{P \,AB} \cD_M(\cQ) \cV_{N\, AB} - \cV^P{}_{AB} \cD_M(\cQ) \cV_N{}^{AB}\Big)
\ee
using (\ref{VOM}). This, then, is the most general expression for a {\em torsion-free} affine connection, 
where the part $U_M{}_A{}^B$ 
of the connection (\ref{internalQ}) corresponding to the ${\bf 1280}$ representation 
of ${\rm SU}(8)$ represents the irremovable ambiguity that remains even after imposition
of the zero torsion constraint \cite{Coimbra:2011ky,Cederwall:2013naa}. 
In appendix~\ref{app:su8connection}
we will derive the unique expression for $U_{M \, A}{}^B$ in terms of only $\cV$ and 
$\partial_M\cV$  that makes $\cQ_{M \, A}{}^B$ a generalized vector, but the resulting 
connection will no longer transform as a proper SU(8) connection, and as a 
consequence the affine connection would no longer be an SU(8) singlet.\,\footnote{By contrast, 
the connections to be derived directly from $D=11$ supergravity in the following section 
do satisfy the required covariance properties, but the corresponding $U_{MA}{}^B$ can
then no longer be expressed in a covariant way in terms of $\cV$ and $\partial_M\cV$ alone.} 

In view of these subtleties it is therefore all the more remarkable how the 
supersymmetric theory manages to sidestep these difficulties and ambiguities.
Namely, in all relevant expressions the internal covariant derivatives 
${\cal D}_M$ appear only in combinations in which the undetermined part 
$U_M{}_A{}^B$ of the connection is projected out 
and for which the covariance under generalized diffeomorphisms is manifest.
We illustrate this with a number of explicit expressions that will be useful in the following.  Using the explicit expression for ${\cal Q}_M{}_A{}^B$, equation (\ref{internalQ}), in equation (\ref{full_DM}), we have, for example\,\footnote{Such projections onto the \textbf{8} and \textbf{56} of SU(8) were shown to be insensitive to the ambiguity $U_M{}_A{}^B$ in Ref.~\cite{Coimbra:2011ky}.}
\bea
{\cal V}^M{}^{AB}\,{\cal D}_M \Xi_B &=&  {\cal V}^M{}^{AB}\,\partial_M \Xi_B 
+ \frac12\,{\cal V}^M{}^{AB}\,{q}_M{}_B{}^C\,\Xi_C
+ \frac12\,{\cal V}^M{}_{CD}\,{p}_M{}^{ABCD}\,\Xi_B 
\nonumber\\
&&{}
+\frac12\,\Gamma_{KM}{}^K\, {\cal V}^M{}^{AB}\, \Xi_B
\;,\nonumber
\\[1ex]
{\cal V}^M{}^{[AB}\,{\cal D}_M\Xi^{C]} &=&
{\cal V}^M{}^{[AB}\,\partial_M\Xi^{C]}-
\frac12\,{\cal V}^M{}^{[AB}\,{q}_M{}_D{}^{C]} \Xi^D
-\frac23\,{\cal V}^M{}_{ED}\,{p}_M{}^{ABCD} \,\Xi^E
\nonumber\\
&&{}
+\frac12\,{\cal V}^M{}_{DE}\,{p}_M{}^{DE[AB} \,\Xi^{C]}
+\frac16 \,\Gamma_{KM}{}^K\,  {\cal V}^M{}^{[AB}\,\Xi^{C]}
\;,
\label{covcom}
\eea
where the piece involving the trace of the affine connection comes
from $W_{M \, A}{}^B$ (we have ignored the possible appearance of the internal 
spin connection $\omega_M{}^{\alpha \beta}$). Indeed, $U_M{}_A{}^B$ does not 
survive in any of these combinations, as can be explicitly verified using 
equations (\ref{UU}). In other words, despite the non-covariance of the Cartan
form, and thus of $q_M$ and $p_M$, under generalized diffeomorphisms, the
above combinations {\em are} covariant under generalized diffeomorphisms
because under generalized diffeomorphisms all terms with second derivatives 
of $\Lambda^M$ cancel out. Modulo density contributions resulting from the non-vanishing 
weights of the fermions (see below), the
particular contractions (\ref{covcom}) of covariant derivatives with the 56-bein
turn out to be precisely those appearing in the supersymmetry transformation rules
and fermionic field equations. More specifically,
now also allowing for a non-trivial weight $\lambda$, and with fully covariant derivatives,
we have
\bea
{\cal V}^M{}^{AB}\,{\nabla}_M \Xi_B &=&  {\cal V}^M{}^{AB}\,\partial_M \Xi_B 
+ \frac12\,{\cal V}^M{}^{AB}\,{q}_M{}_B{}^C\,\Xi_C 
+ \frac12\,{\cal V}^M{}_{CD}\,{p}_M{}^{ABCD}\,\Xi_B \nn \\
&& \qquad\quad
+ \left(\frac12 - \frac23 \lambda(\Xi)\right)\,\Gamma_{KM}{}^K\, {\cal V}^M{}^{AB}\, \Xi_B 
\;,\nonumber
\\[1ex]
{\cal V}^M{}^{[AB}\,{\nabla}_M\Xi^{C]} &=&
{\cal V}^M{}^{[AB}\,\partial_M\Xi^{C]}-
\frac12\,{\cal V}^M{}^{[AB}\,{q}_M{}_D{}^{C]} \Xi^D 
-\frac23\,{\cal V}^M{}_{ED}\,{p}_M{}^{ABCD} \,\Xi^E
\nonumber\\
&&{} \!\!\!\!\!\!\!\!\!\!\!\!\!\!\!\!\!\!\!\!
+\, \frac12\,{\cal V}^M{}_{DE}\,{p}_M{}^{DE[AB} \,\Xi^{C]}
+\left( \frac16 -\frac23 \lambda(\Xi)\right)\,\Gamma_{KM}{}^K\,  {\cal V}^M{}^{[AB}\,\Xi^{C]}
\;.
\label{covcom1}
\eea
As we will see in the following section, and as originally shown in Ref.~\cite{deWit:1986mz},
there is no term proportional to $e^{-1}\partial_Me$ (cf.~(\ref{traceconn})) in the supersymmetry 
variations of the fermions. Consequently, the density terms proportional to $\Gamma_{KM}{}^K$
must cancel. This fixes the weight of the corresponding spinors in (\ref{covcom1})
uniquely, and in agreement with the weight assignments given in
the table. 
In summary, the above expressions are indeed fully covariant under 
both local SU(8) and generalized diffeomorphisms. We will furthermore show in the following section 
that these expressions do agree with the ones already obtained in Ref.\cite{deWit:1986mz}, 
upon imposition of the section constraint.

Similar `miracles' occur in the bosonic sector. For instance, in the bosonic field equations, 
we find after some computation that the scalar contribution to the vector
field equations from (\ref{STep}) can be expressed as
\bea 
{\cal J}^\mu{}_{M} &=&
 -\frac{1}{24}\,{\cal D}^\mu {\cal M}^{KL} \partial_M {\cal M}_{KL} 
+  2i\,e^{-1}\,\partial_N\left(e\, {\cal P}^\mu{\,}^{ABCD} {\cal V}^N{}_{AB}  {\cal V}_{M}{}_{CD} 
~-~{\rm c.c.} \right)
 \nonumber\\
 &=&
 -2i\,{\cal V}_M{}^{AB}\,{\cal V}^{N\,CD} \,\nabla_N\left( g^{\mu\nu}  {\cal P}_{\nu}{\,}_{ABCD} \right)
~+{\rm c.c.} 
 \;,
 \label{JMM}
\eea
with the undetermined connection $U_M{}_A{}^B$ again 
dropping out from this contraction of covariant derivatives.

\medskip

We summarize the structure and definitions
of the various components (external and internal, ${\rm SO}(1,3)$ and ${\rm SU}(8)$) 
of the full spin connection as follows
\bea
\begin{tabular}{c|c}
$\begin{array}{c} \boxed{{\omega_\mu}} \\  
\scalebox{0.7}{$\Gamma_{[\mu\nu]}{}^\rho = 0$}
\end{array} $ & 
$\begin{array}{c} \boxed{{\cal Q}_\mu} \\  
\scalebox{0.7}{$ {\cal D}_\mu{\cal V}_M{}^{AB} \equiv {\cal P}_\mu{}^{ABCD} \,{\cal V}_{M CD}  $}
\end{array} $
\\ \hline\\[-2ex]
$\begin{array}{c} \boxed{\omega_M} \\  
\scalebox{0.7}{${\cal D}_M e_\mu{}^{\alpha}  \equiv \pi_M{}^{\alpha \beta}   e_\mu{}_{\beta}$}
\end{array} $
 & 
 $\begin{array}{c}\boxed{ {\cal Q}_M} \\  
\scalebox{0.7}{$\Gamma_{MN}{}^K |_{\bf 912} = 0$}
\end{array} $
\end{tabular}
\quad
\;.
\label{full_spin}
\eea
The various components of its generalized curvature contain the building blocks for
the bosonic field equations (\ref{dualityM}), (\ref{finalaction})
as we shall discuss in section~\ref{subsec:susy_equations} below.

\subsection{The supersymmetry algebra}
\label{subsec:susy}

A nice illustration of the properties of the full spin connection (\ref{full_spin}) is the algebra of
supersymmetry transformations. In particular, the closure of the algebra on the 56-bein hinges on the 
vanishing of the generalized torsion (\ref{GENtorsion}) in the very same way as the closure on the vierbein
requires the vanishing of the external torsion (\ref{omega}).
The supersymmetry transformations of the bosonic fields (\ref{fieldcontent})
take the same structural form as in the four-dimensional theory
\bea
  \delta_{\epsilon} e_{\mu}{}^{\alpha}&=&
  \bar\epsilon^{A}\gamma^{\alpha}\psi_{\mu A} ~+~
  \bar\epsilon_{A}\gamma^{\alpha}\psi_{\mu}{}^A \;, \nonumber\\[1ex]
  \delta_{\epsilon}{\cal V}_M{}^{AB} &=&  2\sqrt{2}\,{\cal V}_{M CD} \, \Big(
  \bar\epsilon^{[A}\chi^{BCD]}+\frac1{24}\varepsilon^{ABCDEFGH}\,
  \bar\epsilon_{E}\chi_{FGH}\Big)   \,,  \nonumber \\[1ex]
    \delta_{\epsilon} {\cal A}_{\mu}{}^{M}
    &=&
    -i\,\sqrt{2}\,\Omega^{MN} {\cal V}_N{}^{AB}\,\Big(
    \bar\epsilon^{C}\,\gamma_{\mu}\,\chi_{ABC}
    +2\sqrt{2}\, \bar\epsilon_{A}\,\psi_{\mu B}\Big) + {\rm c.c.}
    \;, \nonumber \\[1ex]
    \delta_{\epsilon} {\cal B}_{\mu\nu\,\bfa} &=&-\frac{2}{3} \sqrt{2} \,
    (t_{\bfa})^{PQ}\, \Big( {\cal V}_{P\,AB} {\cal
    V}_{Q\,CD}\,
    \bar\epsilon^{[A}\,\gamma_{\mu\nu}\,\chi^{BCD]}
    + 2 \sqrt{2}\, {\cal V}_{P\,BC} {\cal V}_{Q}{}^{AC}\,
    \bar\epsilon_{A}\,\gamma_{[\mu}\,\psi_{\nu]}{}^{B}
    + {\rm c.c.}\Big)  \nonumber\\
    &&{}
    - (t_{\bfa})_{MN}\,{\cal A}_{[\mu}{}^{M}\,\delta_{\epsilon}
    {\cal A}_{\nu]}{}^{N} \;.
    \label{susybosons}
\eea
The supersymmetry variation of the constrained two-form ${\cal B}_{\mu\nu\,M}$ which is
invisible in the four-dimensional theory can be deduced from closure of the supersymmetry algebra
and yields
\bea
\delta_{\epsilon} {\cal B}_{\mu\nu\,M} &=&
\frac{16}3\,
{\cal V}^K{}^{AB}\, {\cal D}_M{\cal V}_K{}_{BC}
 \,\bar\epsilon^C\gamma_{[\mu} \psi_{\nu]A}  
 -   \frac{4\sqrt{2}}{3} \, {\cal V}^P{}_{AB} {\cal D}_M {\cal V}_{P\,CD}\,
    \bar\epsilon^{[A}\,\gamma_{\mu\nu}\,\chi^{BCD]}
\nonumber\\
&&{}
-8i \left(
   \bar\epsilon^{A}\,\gamma_{[\mu}  {\cal D}_M  \psi_{\nu] A}
 - {\cal D}_M \bar\epsilon^{A}\,\gamma_{[\mu}\, \psi_{\nu] A}\right)
+2 i \, e \varepsilon_{\mu\nu\rho\sigma}\,g^{\sigma\tau}\,
{\cal D}_M \left( \bar\epsilon^A \gamma^\rho \psi_{\tau\,A} \right)
~+~{\rm c.c.}
\nonumber\\[.7ex]
&&{}
 +  \Omega_{KL}\left(\cA_{[\mu}{}^K \partial_M \delta_{\epsilon} \cA_{\nu]}{}^L
-\partial_M \cA_{[\mu}{}^K \delta_{\epsilon} \cA_{\nu]}{}^L
\right)
\;,
\label{susyBM}
\eea
as we show explicitly in appendix~\ref{app:susy}. Note, that 
all ${\rm SU}(8)$ connections cancel in the variation~(\ref{susyBM}),
such that the external index is carried by $\partial_M$ and
this variation is indeed compatible with the constraint (\ref{sectionconditionB})
on ${\cal B}_{\mu\nu\,M}$. In particular, the variation (\ref{susyBM}) consistently vanishes
when $\partial_M = 0$\,.

In terms of the full spin connection (\ref{omegaMhat}), (\ref{full_spin}), 
introduced in the previous section,
the fermionic supersymmetry transformation rules 
take a very compact form given by
\bea
\delta_{\epsilon} \psi_{\mu}^A &=& 
2 \,{\cal D}_{\mu} \epsilon^A 
- 4i\,{\cal V}^{M\,AB} \widehat\nabla_M\left(\gamma_{\mu} \epsilon_B \right)
\;,\nonumber\\
\delta_{\epsilon} \chi^{ABC} &=& -2 \sqrt{2}\, {{\cal P}}_{\mu}{}^{ABCD} \gamma^{\mu} \epsilon_D 
-12\sqrt{2}i \, {\cal V}^M{}^{[AB}\,   \widehat\nabla_M \epsilon^{C]} 
\;.
\label{susyfermions_compact}
\eea
It is then straightforward to verify closure of the supersymmetry algebra.  
The algebra takes the same structural form as in the four-dimensional theory,  
\bea
  \label{susy-algebra}
  {}[\delta(\epsilon_1),\delta(\epsilon_2)] &=& \xi^\mu {\cal D}_\mu +
  \delta_{\rm Lorentz}(\Omega^{\alpha \beta}) + \delta_{\rm susy}(\epsilon_3) +
  \delta_{\rm SU(8)}(\Lambda^A{}_B) 
  +  \delta_{\rm gauge}(\Lambda^M) 
  \nonumber\\
  &&{}
  + \delta_{\rm gauge}(\Xi_{\mu \, \bfa}\,, \Xi_{\mu \,M}) 
  + \delta_{\rm gauge}(\Omega_{\mu\nu}{}^M{}_\bfa\,, \Omega_{\mu\nu}{}_M{}^N)  \,.
\eea
The first term refers to a covariantized general coordinate transformation
with diffeomorphism parameter
\be
\xi^\mu = 2\, \bar\epsilon_2{}^A \gamma^\mu \epsilon_{1\, A} + 2\, \bar\epsilon_{2\; A} \gamma^\mu \epsilon_1{}^A
\;.
\label{12diff}
\ee
The last three terms refer to generalized diffeomorphisms and
gauge transformations (\ref{gaugeLX}), (\ref{shiftB}), 
with parameters
\bea
\Lambda^N &=& 
-8i    \; \Omega^{NP} \left({\cal V}_P{}^{AB} \bar\epsilon_{2\, A} \epsilon_{1\, B} - {\cal V}_{P\;AB} \bar\epsilon_2^A \epsilon_1^B\right)
~\equiv~ {\cal V}^{-1}{}^{N}{}_{AB}\,\Lambda^{AB} + {\cal V}^{-1}{}^{N\,AB}\,\Lambda_{AB} 
\;,
\nonumber\\
  \Xi_{\mu\,\bfa} &=& \frac83 \,(t_\bfa)^{PQ}\,
  {\cal V}_{P\,AC} {\cal V}_Q{}^{BC} \left(\bar\epsilon_2{}^A \gamma_\mu \epsilon_{1B}
  + \bar\epsilon_{2B} \gamma_\mu \epsilon_1{}^A\right)
\;,
\label{12gauge}
\eea
again, as specified by the four-dimensional theory~\cite{deWit:2007mt}. The remaining (constrained) gauge parameters 
$\Xi_{\mu\,M}$, $\Omega_{\mu\nu}{}^M{}_\bfa$, $\Omega_{\mu\nu}{}_M{}^N$ are not present in the
four-dimensional theory and will be specified below.

Closure of the supersymmetry algebra on the vierbein $e_{\mu}{}^{\alpha}$ is confirmed by a standard
calculation:
\bea{}
[  \delta_{\epsilon_1}, \delta_{\epsilon_2}] \, e_{\mu}{}^{\alpha}&=&
 \Big( 2\,\bar\epsilon_{2\,A}\gamma^{\alpha}{\cal D}_\mu \epsilon_1^A  
  -4i\,{\cal V}^{M\,AB} \bar\epsilon_{2\,A}\gamma^{\alpha} \widehat\nabla_M\left(\gamma_\mu\epsilon_{1B}\right)
 ~+~\mbox{c.c.} \Big)~-~ ( 1 \leftrightarrow 2)
  \nonumber\\
   &=&
   2\,{\cal D}_\mu\left(\bar\epsilon_{2\,A}\gamma^{\alpha} \epsilon_1^A \right) 
 -4i\,\widehat\nabla_M \left({\cal V}^{M\,AB} \bar\epsilon_{2\,A} \, \epsilon_{1\,B}\right)\,e_\mu{}^{\alpha}
 -8i\,{\cal V}^{M\,AB} \bar\epsilon_{2\,A} \epsilon_{1\,B}\,\widehat\nabla_M e_{\mu}{}^{\alpha}
\nonumber\\
&&{}
 -4i\,e_{\mu\,\beta}\,{\cal V}^{M\,AB}
 \left( \bar\epsilon_{2\,A}\gamma^{\alpha \beta}  \,\widehat\nabla_M \epsilon_{1\,B}-
 \widehat\nabla_M \bar\epsilon_{2\,A} \gamma^{\alpha \beta} \epsilon_{1\,B}
 \right)
 ~+~\mbox{c.c.}
  \nonumber\\
   &=&
 {\cal D}_\mu\left( \xi^\nu e_\nu{}^{\alpha} \right) 
 +\Lambda^M\,\partial_M e_{\mu}{}^{\alpha} 
  +\frac12\,\partial_M \Lambda^M\,e_\mu{}^{\alpha}
+ \tilde\Omega^{\alpha \beta}\,e_{\mu\,\beta} 
\;,   \label{commE1}
  \eea
with parameters from (\ref{12diff}) and (\ref{12gauge}), and Lorentz transformation given by
\bea
\tilde\Omega^{\alpha \beta} &=& 
 -8i\,{\cal V}^{M\,AB} \bar\epsilon_{2\,A}\gamma^{\alpha \beta}  \,\widehat\nabla_M \epsilon_{1\,B}
  ~+~\mbox{c.c.}
  \;.
\eea
The $\Lambda^M$ terms in (\ref{commE1}) reproduce the transformation of $e_\mu{}^\alpha$ under generalized
diffeomorphisms as scalar densities of weight $\frac12$, cf.~table~\ref{tab:weights}.
Furthermore, the first term in (\ref{commE1}) can be rewritten in the standard way
\bea
{\cal D}_\mu \left( \xi^\nu e_\nu{}^\alpha \right) 
&=&  e_\nu{}^\alpha\, {\cal D}_\mu \xi^\nu + 
\xi^\nu {\cal D}_\nu e_\mu{}^\alpha + 2\, \xi^\nu {\cal D}_{[\mu} e_{\nu]}{}^\alpha  
\;,
\label{ddc}
\eea
into a sum of (covariantized) diffeomorphism and additional Lorentz transformation, upon
making use of the vanishing torsion condition (\ref{omega}) in the four-dimensional geometry.

An analogous calculation shows closure of the supersymmetry algebra on the 56-bein. 
We concentrate on the projection of the algebra-valued variation ${\cal V}^{-1} \delta {\cal V}$ 
onto the ${\bf 70}$ of ${\rm SU}(8)$, 
since the remaining part will entirely be absorbed into a local ${\rm SU}(8)$ transformation.
Using transformations~(\ref{susyfermions_compact}), we obtain
\begin{equation}
{\cal V}^{-1}{}^{M \, AB}\,[\delta_{\epsilon_1}, \delta_{\epsilon_2}] \, {\cal V}_M{}^{CD} =
\xi^\mu\,{\cal P}_\mu{}^{ABCD} 
+6i\, {\cal V}^{N\,[AB}
\nabla_N \Lambda^{CD]}
-\frac{i}4\,\epsilon^{ABCDEFGH}
{\cal V}^{N}{}_{EF}\nabla_N \Lambda_{GH}\;.
\nonumber
\end{equation}
While the first term is the action of the covariantized diffeomorphism,  the remaining terms can be 
rewritten in complete analogy to (\ref{ddc}) with the vanishing torsion condition in (\ref{ddc}) replaced
by the corresponding condition (\ref{inter1}) in the internal space.
Specifically,
\bea
{\cal V}^{-1}{}^{M \,AB}\,[\delta_{\epsilon_1}, \delta_{\epsilon_2}] \, {\cal V}_M{}^{CD} &=&
\xi^\mu\,{\cal P}_\mu{}^{ABCD} 
+
12\,{\cal V}_P{}^{[AB}  {\cal V}^{-1}{}^{CD]\,Q}\, 
\mathbb{P}^P{}_Q{}^N{}_{\underline{L}}\, 
\nabla_N \left({\cal V}_K{}^{\underline{L}} \Lambda^K\right)
\nonumber\\
&=&
\xi^\mu\,{\cal P}_\mu{}^{ABCD} 
+
12\,{\cal V}_P{}^{[AB}  {\cal V}^{-1}{}^{CD]\,M}\, 
\mathbb{P}^P{}_M{}^N{}_K\, 
\partial_N \Lambda^K \nonumber\\
&& {}+ 
 \Lambda^K \left(\nabla_K {\cal V}_M{}^{[AB} \right) {\cal V}^{-1}{}^{CD]\,M}
 \nonumber\\
 &=&
\xi^\mu\,{\cal P}_\mu{}^{ABCD} 
 + {\cal V}^{-1}{}^{M \, AB}\,\delta_\Lambda\,{\cal V}_M{}^{CD}
\;,
\eea
where we have used (\ref{inter1}) in the second equality. The second line 
of (\ref{inter1}) has been absorbed by the weight term associated with the non-trivial ${\rm E}_{7(7)}$
weight~$\frac12$ of~$\Lambda^K$\,.

Closure of the supersymmetry algebra on the vector and two-form fields can be verified by similar but 
more lengthy computations, which we relegate to appendix~\ref{app:susy}. 
Remarkably (and necessarily for consistency), 
closure on the two-forms ${\cal B}_{\mu\nu\,M}$ reproduces not only the 
action of generalized diffeomorphisms (\ref{gaugeLX}) but also the shift transformation (\ref{shiftB})
with parameter $\Omega_{\mu\nu\,M}{}^N$ and finally their rather
unconventional transformation behaviour (\ref{diffx}) under external diffeomorphisms.
Consistency of the algebra thus
confirms the above supersymmetry transformation rules and determines
the remaining gauge parameters on the right hand side of~(\ref{susy-algebra}):
\bea
\Xi_{\mu\,M} &=&
   8i \left(
   \bar\epsilon_2^{A}\,\gamma_{\mu}{\cal D}_M \epsilon_{1\,A}+
   {\cal D}_M \bar\epsilon_{2A}\,\gamma_{\mu}\, \epsilon_{1}^A\right) 
-\frac{16}3\,{\cal V}^K{}_{BC}\, {\cal D}_M{\cal V}_K{}^{AB}
\,\bar\epsilon_{2}^C\gamma_{\mu} \epsilon_{1\,A}~
+~{\rm c.c.}
\;,
 \nonumber\\
\Omega_{\mu\nu}{}^M{}_\bfa &=& 
 - \frac{32}{3}\,  i (t_{\bfa})^{PQ}  {\cal V}_{P}{}^{AB} {\cal V}_{Q}{}_{CB}{\cal V}^{M}{}_{AD}  \,
    \bar\epsilon_{2}^{(C}\,\gamma_{\mu\nu} \epsilon_{1}^{D)} ~+~ {\rm c.c.} 
    \;,
 \label{XXO}\\
    \Omega_{\mu\nu \, M}{}^N &=& -
 32\, {\cal V}^{N}{}_{AB}\bar\epsilon_{[2}^{A}\,\gamma_{[\mu} \nabla_{M} \left( \gamma_{\nu]} \epsilon_{1]}^{B}\right) 
 - \frac{32i}{3}\,   {\cal V}^{N}{}_{AC} {\cal V}^{P \,AB} {\cal D}_M {\cal V}_{P\, BD}  \,    
 \bar\epsilon_{2}^{(C}\,\gamma_{\mu\nu} \epsilon_{1}^{D)} 
  ~+~ {\rm c.c.} 
  \;.
  \nonumber
\eea
As required for consistency, the parameter $\Omega_{\mu\nu}{}^M{}_\bfa$ 
lives in the ${\bf 912}$, i.e.\ satisfies~(\ref{912Omega}).
Moreover, the parameters $\Xi_{\mu\,M}$ and $\Omega_{\mu\nu\,M}{}^N$ satisfy the required
algebraic constraints analogous to those given in (\ref{sectionconditionB}): one can verify that all ${\rm SU}(8)$ connection terms above (which would obstruct these constraints) mutually cancel.

\subsection{Supersymmetric field equations}
\label{subsec:susy_equations}

In this section we employ the formalism set up in the previous sections to 
spell out the fermionic field equations and sketch how under supersymmetry they transform 
 into the bosonic field equations of the E$_{7(7)}$ EFT (\ref{dualityM}), (\ref{finalaction}). 
The Rarita-Schwinger equation is of the form
\bea\nn
0~=~ ({\cal E}_{\psi})^\mu{}_A &\equiv&
-e^{-1} \varepsilon^{\mu \nu \rho \sigma} \gamma_{\nu} {\cal D}_\rho \psi_{\sigma\, A}
- \frac{\sqrt{2}}{6}  \gamma^\nu \gamma^\mu  \chi^{BCD} \, {\cal P}_{\nu \, BCDA}
\nonumber\\
&& {}
-2 \,i\,  e^{-1} \varepsilon^{\mu \nu \rho \sigma} \,  
{\cal V}^{M}{}_{AB} \, \gamma_{\nu} \widehat{\nabla}_M\left(\gamma_{\rho} \psi_\sigma^B \right)
- i\sqrt{2} \,  {\cal V}^N{}^{BC}\, \widehat{\nabla}_N \left(\gamma^\mu \chi_{ABC}\right)   
\;,\qquad
\label{eomgrav}
\eea
where the first two terms can be read off from the dimensionally reduced theory 
and the second line captures the dependence on the internal variables
and can be derived from verifying the supersymmetry transformation of (\ref{eomgrav}).
It is straightforward to check that the contractions of covariant derivatives
in (\ref{eomgrav}) are such that the undetermined part from the internal ${\rm SU}(8)$
connection ${\cal Q}_{M}$ precisely drops out, cf.~(\ref{covcom}) and~\cite{Coimbra:2012af}.
Hence, equation (\ref{eomgrav}) is fully defined via (\ref{covDmu}) and (\ref{QQUW}).

Under supersymmetry (\ref{susyfermions_compact}), and upon using the first order
duality equation (\ref{dualityM}), a somewhat lengthy computation confirms that
the Rarita-Schwinger equation (\ref{eomgrav}) transforms as 
\bea
\delta_\epsilon ({\cal E}_{\psi})^\mu{}_A &=&
({\cal E}_{\rm Einstein})^{\mu\nu}\,\gamma_\nu \epsilon_A
-2 \,({\cal E}_{\rm vector})^{\mu}{}_{AB} \, \epsilon^B
\;,
\label{var32}
\eea
into the Einstein and the second order vector field equations of motion
obtained from varying the action (\ref{dualityM}). It is instructive to give a few details of
this computation as it illustrates the embedding of the bosonic equations of motion
into the components of the curvature associated to the various blocks of the 
internal and external spin connections (\ref{full_spin}).

Let us first collect all terms in the variation (\ref{var32}) 
that contain an even number of $\gamma$-matrices acting on $\epsilon^A$,
which should combine into the second-order vector field equation.
These are the terms that carry precisely one internal derivative $\widehat\nabla_M$. After some
calculation, using in particular (\ref{dualityV}) and (\ref{vTo}), we find
\bea
\delta_\epsilon ({\cal E}_{\psi})^\mu{}_A\Big|_{{\rm even~}\#\gamma} 
&=&
4 \,i\,  e^{-1} \varepsilon^{\mu \nu \rho \sigma} \,  
{\cal V}^{M}{}_{AB} \, \gamma_{\nu} [\nabla_M, {\cal D}_\rho] \left( \gamma_{\sigma}  \epsilon^B \right)
+4 i  \,  {\cal V}^M{}^{CD}\, \gamma^{\mu\nu} \nabla_M   {\cal P}_\nu{\,}_{ABCD} \epsilon^B
\nonumber\\
&& {}
+4i \,  {\cal V}^M{}^{CD} \epsilon^B\,\nabla_M {\cal P}^\mu_{ABCD}  
+ 2\,{\cal P}_\nu{\,}_{ABCD}  \, {\cal F}^{\mu\nu}{\,}^{CD}\,
 \epsilon^B
\;.
 \label{varg1}
\eea
The commutator of covariant derivatives can be evaluated as 
\bea
{\cal V}^M{}^{AB}\,[\nabla_M, {\cal D}_\rho] X^C
&=&
-\frac12\,\nabla_M  {\cal P}_\rho{}^{ABDE}\,{\cal V}^M{}_{DE}\,X^C
+\frac14\, \cV^M{}^{AB} \widehat{R}_{M\rho}{}^{\alpha \beta}\,\gamma_{\alpha \beta}\,X^C
\;,
\label{DDmix1}
\eea
where the first term describes the mixed ${\rm SU}(8)$ curvature, and the second term refers to
the `mixed' curvature of the spin connections 
\bea
\widehat{R}_{M\rho}{}^{\alpha \beta} &\equiv& 
\partial_M \, \omega_\rho{}^{\alpha \beta}
-{\cal D}[\omega]_\rho \,\widehat\omega_M{}^{\alpha \beta}
\;.
\label{Rmix}
\eea
Evaluating this curvature in particular gives rise to the components
\bea
\widehat{R}_{M[\nu\,\rho\sigma]}
&=& 
\frac14\, 
{\cal D}_{[\nu} 
\left({\cal F}_{\rho\sigma]}{}^N\,{\cal M}_{NM} \right)
\;,
\nonumber\\
\widehat{R}_{M\nu}{}^{\mu\nu} &=&
-\frac12\, \widehat{J}^\mu{}_M +\frac14\,e_{\alpha}{}^\mu e_{\beta}{}^\nu\,
 {\cal D}_\nu \left( {\cal M}_{MN} {\cal F}^{\alpha \beta}{}^N \right)
\;,
\eea
with the current $\widehat{J}^\mu{}_M$ from (\ref{STep}).
Putting everything together,
we find for the variation (\ref{varg1})
\bea
\delta_\epsilon ({\cal E}_{\psi})^\mu{}_A\Big|_{{\rm even~}\#\gamma} 
 &=& 
 -2 \,   {\cal D}_\nu \left(
 {\cal F}^{\nu\mu\,+}{}_{AB}\right)   \epsilon^B 
-2\,
 {\cal P}_\nu{\,}_{ABCD}  \, {\cal F}^{\nu\mu-}{\,}^{CD}\,
 \epsilon^B
  +2i \,  \widehat{J}^\mu{}_M 
 {\cal V}^{M}{}_{AB} \, \epsilon^B 
\nonumber\\
&& {}
+4i \,  {\cal V}^M{}^{CD}\, \nabla_M \left(g^{\mu\nu}  {\cal P}_\nu{\,}_{ABCD} \right)  
 \epsilon^B
~~\equiv~
 -2\,({\cal E}_{\rm vector})^\mu{}_{AB} \,\epsilon^B
 \;,
 \eea
reproducing the second-order vector field equation obtained from varying
the action (\ref{finalaction}), cf.~(\ref{JMM}).

It remains to collect the remaining terms with
odd number of $\gamma$-matrices in the variation (\ref{var32}) 
which should combine into the Einstein field equations.
Many of these terms arrange precisely as in the dimensionally reduced theory.
Here we just focus on the additional terms carrying internal derivatives $\nabla_M$
and combining into
\bea
\delta_\epsilon ({\cal E}_{\psi})^\mu{}_A\Big|_{\nabla\nabla} &=&
16 \,  {\cal V}^M{}^{BC}{\cal V}^N{}_{AB}\,  \nabla_M \left( \gamma^\mu 
\nabla_N \epsilon_{C}
\right)   
+8 \,  {\cal V}^M{}^{BC}{\cal V}^N{}_{BC}\, \nabla_M \left( \gamma^\mu 
  \nabla_N \epsilon_{A}
\right)   
\nonumber\\
&&{}
-8\, e^{-1} \varepsilon^{\mu \nu \rho \sigma} \,  
{\cal V}^{M}{}_{AB}{\cal V}^{N\,BC}  \, \gamma_{\nu} \nabla_M\left(\gamma_{\rho} 
\nabla_N\left(\gamma_{\sigma} \epsilon_C \right)
\right)
\;.
\label{quadnabla}
\eea
Collecting all $\nabla_{M}  \nabla_{N} \epsilon_{A}$ terms in this variation gives rise to
\bea
&&
2\left(
8\,{\cal V}^{[M}{}_{AC}{\cal V}^{N]\,CB} + i\,\Omega^{MN} \delta^B_A \right) \,
  \gamma^\mu \,
 [ \nabla_M, \nabla_N] \, \epsilon_{B}
\nonumber\\
&&{}
\qquad\qquad+4\left(
16\, {\cal V}^{(M}{}_{AC}{\cal V}^{N)\,CB}  
+  {\cal M}^{MN} \delta_A^B \right)  \gamma^\mu 
   \nabla_{M}  \nabla_{N} \epsilon_{B}
\;,
\label{NabNab}
\eea
showing that all double derivatives $\partial_M \partial_N \epsilon_A$ vanish due
to the section condition (\ref{secSU8}). We evaluate the full expression (\ref{NabNab}) 
using the fact that the following combination of covariant derivatives~\cite{Coimbra:2012af} 
\bea
&& \Big(\,
6\, {\cal V}^{M}{}_{AC}{\cal V}^{N\,CB}  +2\,  {\cal V}^{N}{}_{AC}{\cal V}^{M\,CB}  
+{\cal V}^{M\,CD}\,{\cal V}^N{}_{CD}  \,\delta_A^B \,\Big) \;
  \nabla_{M}  \nabla_{N} \,\epsilon_{B}  \nonumber\\
  && \hspace{43mm}\equiv~ 
   \left(\frac1{16} {\cal R} \, \delta^{B}_{A} - \frac{1}{4} \cV^{M}{}_{AC} \cV^{N \, CB} \gamma^{\nu \rho} g^{\sigma \tau} \nabla_{M} g_{\nu \sigma} \nabla_{N} g_{\rho \tau} \right) \epsilon_{B},\qquad
  \label{curvK}
\eea
gives rise to the definition of the curvature ${\cal R}$
\bea
{\cal R} &\equiv& 
-4\,{\cal V}^{M}{}_{[AB} {\cal V}^N{}_{CD]}\,\left(
 \partial_M p_{N}{}^{ABCD} - \frac12\,q_{M\,E}{}^{A}\,p_{N}{}^{EBCD}\right)
 -\frac1{6}\,
 {\cal M}^{MN}\,p_{M}{}^{ABCD}p_{N\,ABCD} 
 \nonumber\\
 &&{}
 -4\,
 {\cal V}^{M}{}_{AB} {\cal V}^N{\,}^{CD}\,p_{M}{}^{ABEF}p_{N\,CDEF} 
  -\frac{3}{2}\, {\cal M}^{MN}\,e^{-1} \partial_M \partial_N e
 +\frac3{4}\, {\cal M}^{MN}\,e^{-2} \partial_Me\, \partial_N e
\nonumber\\
 &&{}
 -6\,{\cal V}^M{}_{AB} {\cal V}^N{}_{CD}\,e^{-1} \partial_M e\,p_N{}^{ABCD}
 \;,
 \label{defK}
\eea
which is invariant under generalized internal diffeomorphisms.
Comparing the explicit expression for the curvature to the scalar
potential $V$ (\ref{fullpotentialV}), we see that they are related by
\bea
e\,V &=&
-e\,{\cal R}
-\frac1{4}\, e\,{\cal M}^{MN} \nabla_Mg^{\mu\nu}\nabla_N g_{\mu\nu} + {\rm total\,derivative}
\;,
\label{VR}
\eea
in a form analogous to the ${\rm O}(d,d)$ DFT case discussed in Ref.~\cite{Hohm:2013nja}.
The operator on the left hand side of (\ref{curvK}) is such that the double derivatives $\partial_M\partial_N \epsilon_A$
as well as the single derivatives $\partial_M \epsilon_A$ disappear by virtue of the section constraint, 
and also all ambiguities drop out~\cite{Coimbra:2012af}.

The remaining terms in expression \eqref{quadnabla} can be written as 
\begin{align}
 & 4 \, \left(
16\, {\cal V}^{(M}{}_{AC}{\cal V}^{N)\,CB}  
+  {\cal M}^{MN} \delta_A^B \right) \nabla_{M} \gamma^{\mu} \nabla_{N} \epsilon_{B} - 8\, {\cal V}^{M}{}_{AC}{\cal V}^{N\,CB} \gamma^{\mu \nu \rho} \nabla_{M} \gamma_{\nu}  \nabla_{N} \gamma_{\rho} \epsilon_{B} \notag  \\ 
&\hspace{60mm} + 16 \, {\cal V}^{M}{}_{AC}{\cal V}^{N\,CB} \gamma^{\mu \nu} \nabla_{M}  \nabla_{N}\gamma_{\nu} \epsilon_{B}\;,
\end{align}
showing that $\partial_M \epsilon$ terms are also absent in these terms. These terms, which are independent of the ambiguities, can be further evaluated to give
\begin{align}
 & -\frac{1}{2} \, \partial_{M} g^{\mu \nu} \partial_{N} \cM^{MN} \gamma_{\nu} \epsilon_{A} - \frac{1}{4} e^{-1} \partial_{M} e \, \partial_{N} \cM^{MN} \gamma^{\mu} \epsilon_{A} + 2 \,\cV^{M}{}_{AC} \cV^{N \, CB} \gamma^{\mu \nu \rho} g^{\sigma \tau} \partial_{M} g_{\nu \sigma} \partial_{N} g_{\rho \tau} \epsilon_{B}  \notag \\
& + \frac{1}{8} \, \cM^{MN} \gamma^{\mu} \, \left(\, \partial_M g^{\rho \sigma}\, \partial_N g_{\rho \sigma} -2 \,\,e^{-1} \partial_M \partial_N e + e^{-2} \partial_Me\, \partial_N e  \right) \epsilon_{A} \notag \\
& + \frac{1}{2} \, \cM^{MN} g^{\mu \sigma} g^{\nu \rho} \left( \partial_M \partial_N g_{\rho \sigma} - g^{\tau \eta} \,\partial_{M} g_{\rho \tau}  \partial_{N} g_{\sigma \eta} + e^{-1} \partial_{M} e\, \partial_N g_{\rho \sigma} \right) \gamma_{\nu} \epsilon_{A}\;.
\end{align}
Together, using equation \eqref{curvK} and the expression above, the variation \eqref{quadnabla}  reduces to 
\bea
&&{}\frac12 \, {\cal R} \, \gamma^\mu\,\epsilon_{A} -\frac{1}{2} \, \partial_{M} g^{\mu \nu} \partial_{N} \cM^{MN} \gamma_{\nu} \epsilon_{A} - \frac{1}{4} e^{-1} \partial_{M} e \, \partial_{N} \cM^{MN} \gamma^{\mu} \epsilon_{A} 
\label{defT} \\
&&{} + \frac{1}{8} \, \cM^{MN} \gamma^{\mu} \, \left(\, \partial_M g^{\rho \sigma}\, \partial_N g_{\rho \sigma} - 2 \,\,e^{-1} \partial_M \partial_N e + e^{-2} \partial_Me\, \partial_N e  \right) \epsilon_{A} \notag \\
&&{} + \frac{1}{2} \, \cM^{MN} g^{\mu \sigma} g^{\nu \rho} \left( \partial_M \partial_N g_{\rho \sigma} - g^{\tau \eta} \,\partial_{M} g_{\rho \tau}  \partial_{N} g_{\sigma \eta} + e^{-1} \partial_{M} e\, \partial_N g_{\rho \sigma}\right) \gamma_{\nu} \epsilon_{A}
~\equiv~ {\cal T}^{\mu\nu}\,\gamma_\nu \epsilon_A
\;,
\nonumber
\eea
and gives part of the scalar matter contributions to the Einstein field equations, cf.~(\ref{var32}).
Indeed, ignoring the first term in the expression above, the remaining terms in ${\cal T}^{\mu\nu}$ precisely come 
from a variation of 
\begin{equation}
 \frac1{4}\, e\,{\cal M}^{MN} \nabla_Mg^{\mu\nu}\nabla_N g_{\mu\nu}
\end{equation}
with respect to the metric $g_{\mu\nu}$. Together with (\ref{VR}), and
noting that the variation
$$e \, \delta \cR = -\frac{3}{2} \, \partial_{M} \left( e \, \cM^{MN} \, \partial_{N} (e^{-1} \delta e) \right)\;$$ 
is a total derivative, we find that the variation of the potential (\ref{fullpotentialV}) with respect to the external metric
is given by
\begin{equation}
   \delta (-e \, V )  =  \cR \, \delta e + \frac1{4}\, 
   \delta \left(e\,{\cal M}^{MN} \nabla_Mg^{\mu\nu}\nabla_N g_{\mu\nu} \right)
   ~=~ {\cal T}^{\mu\nu}\,\delta g_{\mu\nu}
   \;,
\end{equation}
and precisely coincides with (\ref{defT}). In summary, the supersymmetry variation of the gravitino equation 
(\ref{eomgrav}) correctly reproduces the full Einstein equations from (\ref{finalaction}).

Finally, a similar discussion can be repeated for the field equation of the spin-1/2 fermions $\chi^{ABC}$,
which under supersymmetry transforms into vector and scalar field equations from (\ref{finalaction}).
Rather than going through the details of this computation, we present the final result in the compact form
of the full fermionic completion of the bosonic Lagrangian (\ref{finalaction}), given by
\begin{eqnarray}
  \label{Lferm}
   {\cal L}_{\rm ferm} &=&
   -\varepsilon^{\mu\nu\rho\sigma}\,
 \bar\psi_{\mu}{}^{A}\gamma_{\nu}{\cal D}_{\rho}\psi_{\sigma\,A}   
 -\frac1{6}e\,\bar\chi^{ABC}\gamma^{\mu}{\cal D}_{\mu}\chi_{ABC}
    -\frac1{3} \sqrt{2}\,e\,
   \bar\chi^{ABC}\gamma^{\nu}\gamma^{\mu}\psi_{\nu}^D\,
   {\cal P}_{\mu\,ABCD}  
   \nonumber\\
   &&{} 
-2i\,  \varepsilon^{\mu \nu \rho \sigma} \,  
{\cal V}^{M}{}_{AB} \, \bar\psi_\mu^A \gamma_{\nu} \,\widehat\nabla_M\left(\gamma_{\rho} \psi_\sigma^B \right)
-2\sqrt{2}i \, e {\cal V}^N{}^{AB}\,  \bar\psi_\mu^C \,\widehat\nabla_N \left(\gamma^\mu \chi_{ABC}\right)   
   \nonumber\\
   &&{} 
-\frac{i}{18} \,  e\, \epsilon_{ABCDEFGH}{\cal V}^M{}^{AB}\, \bar\chi^{CDE} \,\widehat\nabla_M \chi^{FGH}
~+~{\rm c.c.}
\;,
\end{eqnarray}  
up to terms quartic in the fermions. The latter can be directly lifted from the
dimensionally reduced theory \cite{deWit:1982ig}, for dimensional reasons
they are insensitive to $\nabla_M$ corrections.
We have thus obtained the complete supersymmetric extension of the bosonic E$_{7(7)}$ 
EFT (\ref{dualityM}), (\ref{finalaction}). In the rest of this paper, we shall discuss in detail
how this theory after the explicit solution (\ref{decIIA}) of the section constraint
relates to the reformulation \cite{deWit:1986mz,Godazgar:2013dma,Godazgar:2014sla}
of the full (untruncated) $D=11$ supergravity.


\section{Exceptional geometry from $D=11$ supergravity}
\label{sec:d11}


Independently of the construction of a field theory based on 
a particular duality group in Ref.~\cite{Hohm:2013uia} and other references 
alluded to earlier, and described in detail in the two foregoing sections, there is
the reciprocal (`ground up') approach of reformulating the higher-dimensional
theory in such a way that makes the role of duality groups directly manifest
in higher dimensions. This approach goes back to the early work of
Refs.~\cite{deWit:1986mz,Nicolai:1986jk}, and has been taken up again recently in a 
series of papers \cite{deWit:2013ija, Godazgar:2013dma, Godazgar:2014sla}, 
which have succeeded in providing an on-shell equivalent generalized geometric reformulation 
of the $D=11$ theory in which the bosonic degrees of freedom are assembled into 
E$_{7(7)}$ objects and where the supersymmetry transformations of the bosons 
assume a manifestly E$_{7(7)}\times $ SU(8) covariant form.\,\footnote{There exist 
   partial results along similar lines for the case of the 
   E$_{8(8)}$ duality group \cite{Nicolai:1986jk, Koepsell:2000xg, Godazgar:2013dma};
   the full bosonic  E$_{8(8)}$-covariant EFT is constructed in Ref.~\cite{HSE8}.} 
This reformulation is achieved by starting from the known supersymmetry variations of $D=11$ supergravity, and then rewriting the theory 
in such a way that the E$_{7(7)}$ and SU(8) structures become manifest (following 
the work of Cremmer and Julia \cite{Cremmer:1978ds}, where this strategy was applied 
first in the restricted context of the dimensionally reduced theory). One main
advantage of this procedure is that the on-shell equivalence of the reformulation with 
the original $D=11$ supergravity  is guaranteed at each step of the construction;
the detailed comparison between
the E$_{7(7)}$-covariant expressions and those originating from $D=11$ supergravity
is also an essential prerequisite for deriving non-linear Kaluza-Klein ans\"atze for 
all fields.\,\footnote{While the section constraint does admit a solution corresponding
   to IIB theory (with only six internal dimensions), the full consistency of the AdS$_5 \times S^5$
   reduction remains to be established; this would in fact require a detailed analysis
   of supersymmetric E$_{6(6)}$ theory similar to the one presented in this section.}
In this section, we briefly review these developments, and show how they tie 
up with the results of the two foregoing sections, eventually establishing the 
equivalence of the two approaches. As we will see, the full identification is subtle,
not only because it involves various redefinitions, but also because the ambiguities 
exhibited in the foregoing sections play a key role in establishing the precise relation.

\subsection{56-bein and GVP from eleven dimensions}

The first step is to identify an E$_{7(7)}$ 56-bein $\mathcal{V}_{M AB}$\,\footnote{The notations and conventions used here are slightly different to those used in \cite{deWit:1986mz, Godazgar:2013dma}.} {\em in
eleven dimensions} with the bosonic degrees of freedom that reduce to scalars under 
a reduction of the $D=11$ theory to four dimensions; this 56-bein will be eventually
identified  with the one introduced in the previous sections.
Decomposing the {\bf 56} of E$_{7(7)}$ under its SL(8) and GL(7) subgroups
\begin{equation}
{\bf 56} \; \rightarrow \; {\bf 28} \oplus \overline{\bf 28} \; \rightarrow \; {\bf 7} \oplus {\bf 21} \oplus \overline{\bf 21} \oplus \overline{\bf 7},
\end{equation}
we have the following decomposition of the 56-bein 
\begin{equation}
\cV_{M\, AB} \,\equiv\, 
\Big( \cV^m{}_{AB},  \cV_{mn\, AB} , \cV^{mn}{}_{AB} , \cV_{m\,AB} \Big),
\end{equation}
where we will often employ the simplifying notation $\cV^m{}_{AB} \equiv \cV^{m8}{}_{AB} = 
- \cV^{8m}{}_{AB}$, when considering the embedding of GL(7) into SL(8).
The main task is then to directly express this 56-bein in terms of components 
of eleven-dimensional fields along the seven-dimensional directions, {\it viz.}
\begin{equation}
\cV_{M\, AB}\, \equiv \, \cV_{M\, AB}\big( e_m{}^a, A_{mnp} , A_{mnpqrs}\big),
\end{equation}
where $e_m{}^a$ is the siebenbein, $A_{mnp}$ are the internal components of the three-form field, and $A_{mnpqrs}$ the internal components of the dual six-form field. In other words,
the 56-bein whose existence in eleven dimensions was postulated on the basis of symmetry considerations in the previous section is here given concretely in terms of certain components
of the $D=11$ fields and their duals. The calculation \cite{Godazgar:2013dma} yields 
the explicit formulae
\begin{align}
\cV^{m}{}_{AB} &=  \frac{1}8 \Delta^{-1/2} \Gamma^m_{AB}, \label{gv11} \\[3mm]
\cV_{mn}{}_{AB} &= \frac{1}{8}  \Delta^{-1/2} \left(
\Gamma_{mn}{}_{AB} + 6 \sqrt{2} A_{mnp} \Gamma^{p}_{AB} \right),
\label{gv12} \\[3mm]
\cV^{mn}{}_{AB} & = \frac1{4\cdot 5!} \, \eta^{mnp_1\cdots p_5}
   \Delta^{-1/2} \Bigg[
\Gamma_{p_1 \cdots p_5}{}_{AB} + 60 \sqrt{2} A_{p_1 p_2 p_3} \Gamma_{p_4
p_5}{}_{AB}  \notag \\
 & \hspace*{50mm}  - 6! \sqrt{2} \Big( A_{q p_1 \cdots p_5} - 
\frac{\sqrt{2}}{4} A_{qp_1 p_2} A_{p_3 p_4 p_5} \Big) \Gamma^{q}_{AB}
\Bigg], \label{gv13} \\[2mm]
\cV_{m\, AB} &=
\frac1{4\cdot 7!} \, \eta^{p_1\cdots p_7}
\Delta^{-1/2} \Bigg[
(\Gamma_{p_1 \cdots p_7} \Gamma_{m}{})_{AB} + 126 \sqrt{2}\ A_{m p_1
p_2} \Gamma_{p_3 \cdots p_7}{}_{AB} \notag \\
 & \hspace*{37mm}  + 3\sqrt{2} \times 7! \Big( A_{m p_1 \cdots p_5} +
\frac{\sqrt{2}}{4} A_{m p_1 p_2} A_{p_3 p_4 p_5} \Big) \Gamma_{p_6
p_7}{}_{AB} \notag \\[2mm]
  & \hspace*{42.5mm} + \frac{9!}{2} \Big(A_{m p_1 \cdots p_5} +
\frac{\sqrt{2}}{12} A_{m p_1 p_2} A_{p_3 p_4 p_5} \Big) A_{p_6 p_7 q }
\Gamma^{q}{}_{AB} \Bigg], \label{gv14}
\end{align}
where $\Delta$ is the determinant of the siebenbein $e_{m}{}^a.$ In particular, it can be 
explicitly verified that the 56-bein defined by the components above satisfies 
the identities \eqref{VOM}, and thus is indeed an element of the most general duality
group Sp(56,$\mathbb{R}$). To show that  that it is more specifically an 
E$_{7(7)}$-valued matrix one either verifies (\ref{varV}) directly, or invokes eqs. (14),(17) 
and (18) of Ref.~\cite{Godazgar:2014sla} where it is shown that $\cV$ transforms as a
generalized E$_{7(7)}$ covector.  From the point of view of 
Refs.~\cite{deWit:1986mz,Godazgar:2013dma}, this matrix corresponds to an element of the 
coset space E$_{7(7)}/$SU(8) in a specific gauge (where the local SU(8) is taken to act 
in the obvious way on the indices $A,B,...$), such that after a local SU(8) rotation the
direct identification as given above is lost. Note also the appearance of components
of the six-form potential in the expressions, as a consequence of whose presence
the identification of the EFT formulated in the previous section and the 
$D=11$ supergravity can only be achieved at the level of the equations of motion 
(which, of course, does not preclude 
the existence of suitable actions for either formulation).

In the same manner, one identifies a {\bf 56}-plet of E$_{7(7)}$ vector fields $\cA_{\mu}{}^{M}$ 
that incorporate the degrees of freedom corresponding to vectors under a reduction 
to four dimensions, combining the 28 electric and the 28 magnetic vectors of
maximal supergravity into a single representation that now live in eleven
dimensions. As before, the components in a GL(7) decomposition of the {\bf 56} 
of E$_{7(7)}$ can be explicitly written in terms of eleven-dimensional fields
\begin{align}
  {\cA_{\mu}}^{m} &= \frac{1}{2} {B_{\mu}}^m, \hspace{30.5mm}
\cA_{\mu\, mn} = 3\sqrt{2}\, \big(A_{\mu mn} - B_\mu{}^p A_{pmn}\big), \notag \\[4mm]
{\cA_{\mu}}^{m n} &= 6\sqrt{2}\, {\eta}^{mnp_1\dots p_5} 
\left( A_{\mu p_1 \cdots p_5} - B_\mu{}^q A_{qp_1\cdots p_5}  -
       \frac{\sqrt{2}}{4}\, \big(A_{\mu p_1p_2} - B_\mu{}^q A_{qp_1p_2}\big) A_{p_3p_4p_5} \right), \notag \\[4mm]
\cA_{\mu\, m} &= 36\, {\eta}^{n_1\dots n_7} 
\Bigg( A_{\mu n_1 \dots n_7, m} + (3 \tilde{c}-1)
\left( A_{\mu  n_1 \dots n_5} - B_{\mu}{}^{p} A_{p n_1 \dots n_5}
\right) A_{n_6 n_7 m} \notag \\[2mm]
& \quad \, + \tilde{c} A_{ n_1 \dots n_6} \left( A_{\mu n_7 m} -
B_{\mu}{}^{p} A_{pn_7m} \right) + \frac{\sqrt{2}}{12} \left( A_{\mu n_1 n_2} -
B_{\mu}{}^{p} A_{p n_1 n_2} \right) A_{n_3 n_4 n_5} A_{n_6 n_7 m} \Biggr).
\label{Bdef1}
\end{align}
The components of the six-form potential appear again in the expression above.  However, 
in the $\cA_{\mu\, m}$ component, there appears a new field $A_{\mu n_1 \dots n_7, m}$ 
(as well as an undetermined constant $\tilde{c}$), related to the {\em dual graviphoton}.

These E$_{7(7)}$ objects are found by analysing the $D=11$ supersymmetry transformations, 
which in the SU(8) invariant reformulation were found to take the precise form \cite{deWit:1986mz, deWit:2013ija, Godazgar:2013dma, Godazgar:2014sla}
\bea
  \delta e_{\mu}{}^{\alpha}&=&
  \bar\epsilon^{A}\gamma^{\alpha}\psi_{\mu A} ~+~
  \bar\epsilon_{A}\gamma^{\alpha}\psi_{\mu}{}^A \;, \nonumber\\[1ex]
  \delta {\cal V}_M{}^{AB} &=&  2\sqrt{2}\,{\cal V}_{M CD} \, \Big(
  \bar\epsilon^{[A}\chi^{BCD]}+\frac1{24}\varepsilon^{ABCDEFGH}\,
  \bar\epsilon_{D}\chi_{EFG}\Big)   \,,  \nonumber \\[1ex]
\delta {\cA}_{\mu}{}^{M}
    &=&
    -i\,\sqrt{2}\,\Omega^{MN} {\cal V}_N{}^{AB}\,\Big(
    \bar\epsilon^{C}\,\gamma_{\mu}\,\chi_{ABC}
    +2\sqrt{2}\, \bar\epsilon_{A}\,\psi_{\mu B}\Big)~+~ {\rm c.c.}
    \;, \label{susyva}
\eea
where a compensating SU(8) rotation has been discarded in the variation $\delta\cV_M{}^{AB}$,
as explained in Refs.~\cite{deWit:1986mz,Godazgar:2013dma}. 
Strictly speaking, the supersymmetry transformations  of the last seven components of the vectors cannot be derived from 
$D=11$ supergravity, due to the absence of a non-linear formulation of dual gravity, but 
are here obtained by `E$_{7(7)}$-covariantization'. 
The supersymmetry transformations of the last seven components of the vector field instead determine the supersymmetry transformation of the new field $A_{\mu n_1 \dots n_7, m}$ as discussed in Ref.~\cite{Godazgar:2013dma}.
While $A_{\mu n_1 \dots n_7, m}$, which is introduced to complete the {\bf 56} of E$_{7(7)}$ for the vectors, is clearly related to dual gravity degrees of freedom from a four-dimensional tensor hierarchy point of view, its direct relation to the eleven-dimensional fields cannot be determined. This is in stark contrast to the six-form potential that is related to the three-form potential via an explicit duality relation.  Nevertheless, our ignorance regarding this field is compensated by the fact that it does not appear in the GVPs (see below).

While the agreement in the supersymmetry variations of the boson fields as derived above 
and the exceptional field theory approach of the foregoing sections is thus manifest, the
agreement in the fermionic variations is much more subtle. This is because the latter 
depend on the connections, and a detailed comparison would thus require an analysis of the 
connection (\ref{internalQ}) in terms of the $D=11$ fields. Of course, ignoring the 
ambiguity (\ref{UU}) for the moment, we could simply try to work out the 
expressions (\ref{Cartan}) and (\ref{QQUW}) by substituting the explicit formulae 
(\ref{gv11})--(\ref{gv14}).  However, this would lead to extremely cumbersome expressions 
(but see appendix~\ref{app:gl7}
 for a simplified calculation), whose relation with the ones
given below would be far from obvious.
We will therefore proceed differently by starting `from the other end'.
The supersymmetry transformations of the fermions were already derived 
in \cite{deWit:1986mz}, {\it viz.} 
\begin{align}
 \delta \psi_{\mu}^{A} &= 2\left(\partial_{\mu} - B_{\mu}{}^{m} \partial_{m} -\frac{1}{4} \partial_m B_{\mu}{}^{m} \right) \epsilon^{A} + \frac{1}{2} \omega_{\mu}{}^{\alpha \beta} \gamma_{\alpha \beta} \epsilon^{A} + \cQ_{\mu}{}^{A}{}_{B} \epsilon^{B}  \nn\\[0mm]
 & \hspace{4mm}+2 \cG^{-}_{\alpha \beta}{}^{AB} \gamma^{\alpha \beta} \gamma_{\mu} \epsilon_{B}
- \frac{1}{4} e^{m \, AB} e_{\nu \, \beta} \partial_m e_{\rho}{}^{\beta} \gamma^{\nu \rho} \gamma_{\mu} \epsilon_B \nn\\[0mm]
& \hspace{4mm}+ e^{m \, AB} \partial_{m} \left( \gamma_{\mu} \epsilon_{B} \right) 
  + \frac{1}{2} e^{m \, AB} Q'_{m \, B}{}^{C} \gamma_{\mu} \epsilon_{C} 
 - \frac{1}{2} e^{m}{}_{CD} P'_{m}{}^{ABCD} \gamma_{\mu} \epsilon_{B}, \notag \\
\delta \chi^{ABC} &= - 2\sqrt{2} \cP_{\mu}{}^{ABCD} \gamma^{\mu} \epsilon_{D} + 6 \sqrt{2} \cG^{-}_{\alpha \beta}{}^{[AB|} \gamma^{\alpha \beta} \epsilon^{|C]} - \frac{3}{2 \sqrt{2}}  e_{\mu \, \beta} \partial_m e_{\nu}{}^{\beta} e^{m [AB} \gamma^{\mu \nu} \epsilon^{C]} \nn \\
& \hspace{3mm}+ 3 \sqrt{2} e^{m [AB} \partial_{m}\epsilon^{C]} 
  - \frac{3\sqrt{2}}{2} e^{m [AB} Q'_{m \, D}{}^{C]} \epsilon^{D} - \frac{3\sqrt{2}}{2} e^{m}{}_{DE} P'_{m}{}^{DE[AB} \epsilon^{C]}  \nn \\
& \hspace{3mm}-2\sqrt{2} e^{m}{}_{DE} P'_{m}{}^{ABCD} \epsilon^{E},
\label{spsichi}
\end{align}
where
\begin{equation}
 e^{m}{}_{AB}= e^{m AB} = i \Delta^{-1/2} \Gamma^{m}_{AB}
\end{equation}
is just part of the 56-bein $\cV^M{}_{AB}$ given above in \eqref{gv11}, and 
\begin{gather}
 \cG_{\alpha \beta AB} \equiv - \frac{i}{8} \Delta^{1/2} e_{[\alpha}{}^{\mu} e_{\beta]}{}^{\nu} (\partial_{\mu} - B_{\mu}{}^{m} \partial_{m}) B_{\nu}{}^{n} \Gamma_{n AB} + \frac{\sqrt{2}}{32} i \Delta^{-1/2}
  F_{\alpha \beta mn} \Gamma^{mn}_{AB} 
\end{gather}
comprises the contribution from the spin one degrees of freedom. The link of the particular
expressions involving the Kaluza-Klein vectors $B_\mu{}^m$ with those of the previous
two sections is easily seen by noting that
\be
\partial_\mu - B_\mu{}^m\partial_m \equiv
\partial_\mu - \cA_\mu{}^M\partial_M
\ee
upon taking the canonical solution of the section constraint. Furthermore,
the direct comparison with the fermion transformations of $D=11$ supergravity
yields the expressions
\begin{align}
{Q'}_{m \,A}{}^B &= \frac12 \, q_{m\,ab} \, \Gamma^{ab}_{AB} \, + \,
  \frac{\sqrt{2}}{48}\,  F_{mabc} \, \Gamma^{abc}_{AB} \, + \,
  \frac{\sqrt{2}}{14\cdot 6!} F_{mabcdef} \, \Gamma^{abcdef}_{AB}, \nn \\\label{QPdWN}
{P'}_{mABCD} &= - \frac34 \, p_{m\,ab} \, \Gamma^a_{[AB} \Gamma^b_{CD]} \, + \,
   \frac{\sqrt{2}}{32} \, F_{mabc} \Gamma^a_{[AB} \Gamma^{bc}_{CD]}  \, - \,
   \frac{\sqrt{2}}{56\cdot 5!} F_{mabcdef} \, \Gamma^a_{[AB} \Gamma^{bcdef}_{CD]},  
\end{align}
where 
\be\label{SPQ}
q_{m\,ab} \equiv e_{[a}{}^n \partial_{|m} e_{n|b]} \;\; , 
\quad p_{m\,ab} \equiv e_{(a}{}^n \partial_{|m} e_{n|b)}
\ee
are the components of the GL(7) Cartan form, with analogous notation as in 
the previous section. These objects transform properly under local SU(8): 
$Q'_{m \,A}{}^B$ is the SU(8) connection, while $P'_{m\,ABCD}$ transforms covariantly 
in the complex self-dual $\bf{35}$ representation of SU(8). However, as written, these 
connections are not fully covariant under internal diffeomorphisms, because 
$q_{m\,ab}$ and $p_{m\,ab}$ do {\em not} transform as proper vectors under internal 
diffeomorphisms. For this reason we will switch to a slightly different choice below, 
see (\ref{QGGN}) and (\ref{PGGN}), which satisfies all covariance requirements.

The other important feature of the reformulation \cite{deWit:1986mz,Godazgar:2013dma,Godazgar:2014sla} is the 
so-called generalized vielbein postulate (GVP). When evaluated on the different components
of $\cV^M{}_{AB}$ this consists of certain differential equations satisfied by the 56-bein 
which are analogous to the usual vielbein postulate in differential geometry.  The GVPs are equations satisfied by the 56-bein and in the approach of \cite{deWit:1986mz,Godazgar:2013dma,Godazgar:2014sla}
they can be checked explicitly on a component by component basis, while they
appear as genuine postulates in the approach of the previous section.
Moreover, the direct comparison with $D=11$ supergravity allows 
for a direct understanding of four-dimensional maximal gauged theories and the embedding tensor \cite{Godazgar:2013pfa, Godazgar:2013oba, Godazgar:2014sla} that defines them from a higher-dimensional perspective as well as providing generalized geometric structures that can be interpreted as generalized connections and used to construct a generalized curvature tensor.   

The external GVP, which gives the dependence of the 56-bein on the four-dimensional coordinates is given by equation \eqref{PQ} (see Refs.~\cite{Godazgar:2013dma, Godazgar:2014sla}), where the explicit expressions for $\cQ_{\mu}$ 
and $\cP_{\mu}$ in terms of the $D=11$ fields were already given in Ref.~\cite{deWit:1986mz}.
Here we concentrate on the internal part of the GVP which was given in \cite{Godazgar:2013dma,Godazgar:2014sla} in the form
\begin{equation} \label{intgvp}
 \partial_m \cV_{M\, AB} \, - \, {\bf{\Gamma}}_m{}_{M}{}^{N} \cV_{N\, AB} 
 \,+\,  Q_{m}^{C}{}_{[A} \cV_{M\, B]C} \,=\,  P_{m\, ABCD} \cV_{M}{}^{CD},
\end{equation}
where\,\footnote{Note that in this paper our conventions are such that Cartan's first structure equation takes the form $T^{a} = \textup{d} e^{a} + \omega^{a}{}_{b} \wedge e^{b}.$}
\bea\label{QGGN}
Q_{m \,A}{}^B &=&  \frac12 \, \omega_{m\,ab} \, \Gamma^{ab}_{AB} \, + \,
  \frac{\sqrt{2}}{48}\,  F_{mabc} \, \Gamma^{abc}_{AB} \, + \,
  \frac{\sqrt{2}}{14\cdot 6!} F_{mabcdef} \, \Gamma^{abcdef}_{AB}, \\
P_{m \,ABCD} &=& 
   \frac{\sqrt{2}}{32} \, F_{mabc} \Gamma^a_{[AB} \Gamma^{bc}_{CD]}  \, - \,
   \frac{\sqrt{2}}{56\cdot 5!} F_{mabcdef} \, \Gamma^a_{[AB} \Gamma^{bcdef}_{CD]}. \label{PGGN}
\eea 
Notice that ${Q'}_{m \,A}{}^B$ and ${P'}_{m \,ABCD}$ defined in equations \eqref{QPdWN} 
and ${Q}_{m \,A}{}^B$ and ${P}_{m \,ABCD}$ defined above, \eqref{PGGN}, differ in their
components relating to the siebenbein since we have replaced $q_{m\, ab}$ by the
spin connection $\omega_{m\,ab}$ and $p_{m\,ab}$ by zero.
As explained in Ref.~\cite{Godazgar:2014sla} this change is required if the connections 
are to satisfy all the requisite covariance properties, as is indeed the case
for (\ref{QGGN}) and (\ref{PGGN}). However, there appears to be 
no way to reproduce these covariant expressions in terms of the 56-bein $\cV$ and
its internal derivatives $\partial_m\cV$ without `breaking up' the matrix $\cV$,
and this is one of the main difficulties in 
establishing agreement between the above expressions and the ones obtained 
in the previous section. Fortunately, the apparent discrepancy turns out to 
reside in the $\bf{1280}$ part of the SU(8) connection (see (\ref{UU})) and the hook ambiguity described in section \ref{sec:hook} and will 
thus drop out in all relevant expressions.

The internal GVP  as given in (\ref{VP2}) and (\ref{intgvp}) (and also \eqref{intgvpe7}, 
see below) differ in two respects. First of all, and prior to imposing the section constraint,
(\ref{VP2}) involves all 56 components, whereas (\ref{intgvp}) involves only the
seven internal dimensions with index $M=m$. The second distinctive feature is 
the appearance of a non-zero term proportional to $P_m$ on the right-hand side
of the GVP.  As we will explain in more detail below, this term  corresponds to 
a {\em generalized non-metricity}.\,\footnote{We would like to thank Malcolm 
Perry for pointing this out to us.} We will show below how to absorb this
non-metricity, and thereby bring the GVP into the same form as \eqref{VP2}.
Finally, the connection coefficients ${\bf{ \Gamma}}_m$ can appear in the supersymmetry transformations of the fermions only via their traces, because the fermions, while transforming
as densities, are otherwise only sensitive to the local SU(8).

Given the coefficients $Q_{m \,A}{}^B$ and $P_m{}^{ABCD}$ we can solve for the affine connection
coefficients ${\bf\Gamma}_{mM}{}^N$ in terms of the fields of  $D=11$ supergravity;
we use boldface letters here to indicate that these coefficients are {\em different
from the ones identified in} (\ref{AffCon1}) of the previous section. 
With \eqref{QGGN} and \eqref{PGGN}, ${\bf{\Gamma}}_m{}_{M}{}^{N}$ takes values 
in the Lie algebra of E$_{7(7)}$
\begin{equation} \label{gammdef}
{\bf{\Gamma}}_m{}_{M}{}^{N} = {\bf{\Gamma}}_m{}^\bfa (t_\bfa)_{M}{}^{N}.
\end{equation} 
The comparison with $D=11$ supergravity allows to solve 
for the components of ${\bf{\Gamma}}_m{}^\bfa$ directly in 
terms of $D=11$ fields; the non-vanishing components are
\begin{gather} 
 ({\bf{\Gamma}}_m)_n{}^p \equiv - \Gamma_{mn}^p + 
 \textstyle\frac14 \delta_n^p \Gamma_{mq}^q, \qquad ({\bf{\Gamma}}_m)_8{}^8 = 
 - \textstyle \frac34\, \Gamma_{mn}^n, \notag \\[1mm]
 ({\bf{\Gamma}}_{m})_{8}{}^{n} = \sqrt{2} \eta^{n p_1 \cdots p_6}\, \Xi_{m|p_1 \cdots p_6}, \qquad
 ({\bf{\Gamma}}_{m})^{n_1 \cdots n_4} =  
 \textstyle\frac{1}{\sqrt{2}} \eta^{n_1 \cdots n_4 p_1 p_2 p_3}\, \Xi_{m| p_1 p_2 p_3},
 \label{affcon}
\end{gather}
where $\Gamma_{mn}{}^p$ is the usual Christoffel symbol, and where
\begin{align} \label{Xi1}
 \Xi_{p|mnq} &\equiv\,  D_{p} A_{mnq} - \frac{1}{4!} F_{pmnq}, \\[1mm]
 \Xi_{p|m_1 \cdots m_6} &\equiv \, D_{p} A_{m_1 \cdots m_6} + \frac{\sqrt{2}}{48} F_{p[m_1 m_2 m_3} A_{m_4 m_5 m_6]} \notag \\[3pt]
 & \qquad - \frac{\sqrt{2}}{2} \left( D_{p} A_{[m_1 m_2 m_3} - \frac{1}{4!} F_{p[m_1 m_2 m_3} \right) A_{m_4 m_5 m_6]} - \frac{1}{7!} F_{p m_1 \dots m_6}. \label{Xi2}
 \end{align}
One notices that these objects, like the usual Christoffel symbol,  indeed transform with second derivatives of the tensor gauge parameters, as would be expected for a generalized affine connection (see Ref.~\cite{Godazgar:2014sla} for details).  Another noteworthy feature is that they vanish under full antisymmetrization:
 \begin{equation} \label{symmxi}
 \Xi_{[p|mnq]}=0, \qquad \qquad \Xi_{[p|m_1 \dots m_6]} = 0. 
 \end{equation}
Therefore, they correspond to hook-type Young tableaux diagrams, and thus 
encapsulate the non-gauge invariant part of the derivatives of 
the three-form and the six-form fields. In terms of SL(7) these $\Xi$'s correspond 
to the $\bf{210}$ and $\bf{48}$ representations, respectively; when further decomposed 
into SO(7) representations, these will become the $\bf{21} \oplus \bf{189}$ and 
$\bf{21}\oplus\bf{27}$ of SO(7), all of which appear in the $\bf{1280}$ of  SU(8).
We will also see below that the irreducibility property 
\eqref{symmxi} is crucial for the absence of torsion in the sense of generalized geometry.

As given above, the connection coefficients $Q_{m \,A}{}^B$, $P_m{}^{ABCD}$ and 
${\bf \Gamma}_{mN}{}^P$ have all the desired transformation properties 
with respect to local SU(8) and generalized diffeomorphisms, 
as can be verified explicitly from their definitions (see Ref.~\cite{Godazgar:2014sla}). That is, 
$Q_{m \,A}{}^B$ transforms as an SU(8) connection (as is obvious from the way the local 
SU(8) has been introduced in Ref.~\cite{deWit:1986mz} as a St\"uckelberg-type symmetry), while
$P_m{}^{ABCD}$ transforms covariantly under SU(8) transformations. Both $Q_{m \,A}{}^B$ 
and $P_m{}^{ABCD}$ transform as generalized vectors under generalized diffeomorphisms (for the natural truncation of generalized Lie
derivatives to vectors with only seven vector indices). 
Furthermore, the generalized affine connection ${\bf \Gamma}$ is invariant  under SU(8) transformations, and transforms as a generalized connection (with a second derivative of the gauge parameters).

A distinctive feature of the internal GVP as given here, to be contrasted with the one given 
in (\ref{VP2}), is that, at this point, the connections have non-zero components only along the seven internal dimensions, but vanish otherwise -- just like the partial derivative $\partial_M$ after
imposition of the section constraint. Nevertheless, we can formally write the internal GVP as
\begin{equation} \label{intgvpe7}
 \partial_{M} \cV_{N\, AB} \, - \, {\bf{\Gamma}}_{M}{}_{N}{}^{P} \cV_{P\, AB} 
 \,+\,  Q_{M}^{C}{}_{[A} \cV_{N\, B]C} \,=\,  P_{M\, ABCD} \cV_{N}{}^{CD}
\end{equation}
by trivially promoting the GL(7) index $m$ to part of a {\bf 56} of E$_{7(7)}$.  Hence, taking
\begin{equation}
 \partial_{M} = \begin{cases}
                   \partial_{m}  \quad & \text{if $M = m8$}, \\
		    0 & \text{otherwise}
                  \end{cases}
\label{secconsol}
\end{equation}
and identifying the $m$ components of the connection coefficients with those that appear in equation \eqref{intgvp}, with all other components vanishing, gives back \eqref{intgvp}. In this form the internal GVP can be compared to equation \eqref{VP2}, {\em with the proviso} that the section constraint also applies to the connections. However, in view of the derivation
given in the foregoing section, a natural question that arises at this point is why all 
other components of the connection coefficients should vanish. Would it not be more ``natural'' from a generalized geometric point of view if the connection coefficients had non-trivial components in the other directions of the {\bf 56} representation, as has been assumed in section \ref{sec:e7covgeo} and, for example, Ref.~\cite{Coimbra:2011ky}? Indeed, we will see below that the introduction of non-vanishing connection components along the other directions will actually be required if we want to recast the supersymmetry variations of the fermions in order to
achieve full agreement with the formalism of the preceding section.

We now proceed to reformulate these structures in order to exhibit their precise
relationship to those constructed in section \ref{sec:e7covgeo}.  However, given that vanishing torsion is taken to be an important ingredient for defining generalized connections in section \ref{sec:e7covgeo}, we will first consider the generalized 
torsion associated to the generalized affine connection ${\bf \Gamma}$.

\subsection{Generalized torsion}

In Ref. \cite{Godazgar:2014sla}, the generalized torsion ${\bf T}_{M N}{}^{P}$ is defined as follows 
\begin{equation}  \label{torgen}
 [\nabla_M, \nabla_{N}] S = {\bf T}_{M N}{}^{P} \partial_P S
\end{equation}
for some scalar $S$ and where $\nabla_{M}$ is defined using the connection 
${\bf \Gamma}_{MN}{}^P.$ The generalized torsion as defined above vanishes \cite{Godazgar:2014sla}.   An alternative (and {\it a priori}
independent) definition of the torsion is given in equation \eqref{GENtorsion} of section 
\ref{sec:e7covgeo}, which leads to the formula \eqref{FULLtau}.
While the above definition of torsion and that defined in \eqref{GENtorsion} are equivalent in usual differential geometry, this is not the case in generalized geometry.  Here we 
will evaluate the generalized torsion \eqref{FULLtau} explicitly in terms of the connection 
coefficients ${\bf \Gamma}_{mN}{}^P$ given in Ref.~\cite{Godazgar:2014sla} and above.  
A simple component-wise calculation using the components of ${\bf \Gamma}_{mN}{}^P$ 
identified above now shows that the generalized torsion does indeed vanish.  
For example, consider
\begin{equation}
 \cT_{m8\,  n8}{}^{p8} = {\bf \Gamma}_{m8\,  n8}{}^{p8}-48\,\mathbb{P}^{p8}{}_{n8}{}^{q8}{}_{r8}\, {\bf \Gamma}_{q8 \, m8}{}^{r8}
  +16\,\mathbb{P}^{p8}{}_{n8}{}^{q8}{}_{m8} {\bf \Gamma}_{r8\, q8}{}^{r8}.
\end{equation}
Using the fact that
\begin{equation}
 \mathbb{P}^{p8}{}_{r8}{}^{q8}{}_{s8} = \frac{1}{96} \left( 2 \delta^{p}_{s} \delta^{q}_{r} + \delta^{p}_{r} \delta^{q}_{s} \right),
\end{equation}
the above equation reduces to
\begin{equation}
 \cT_{m8\,  n8}{}^{p8} = 2 {\bf \Gamma}_{[m\,  n]}{}^{p} - \frac{2}{3} {\bf \Gamma}_{r \, [m}{}^{r} \delta^p_{n]}.
\end{equation}
However, the right hand side of the above equation vanishes by substituting the relevant components of ${\bf \Gamma}$ from \eqref{affcon}. Hence,
\begin{equation}
 \cT_{m8\,  n8}{}^{p8} = 0.
\end{equation}
Next consider, for example,
\begin{equation}
  \cT_{m8\,  pq \, r8} = {\bf \Gamma}_{m8\,  pq \, r8}  - 24 \mathbb{P}_{pq\,  r8}{}^{st\, u8} {\bf \Gamma}_{u8 \, m8\, st}.
\end{equation}
Using the fact that
\begin{equation}
 \mathbb{P}_{pq\,  r8}{}^{st\, u8} = \frac{1}{8} \delta^{stu}_{pqr},
\end{equation}
the above equation reduces to
\begin{equation}
  \cT_{m8\,  pq \, r8} = 4 {\bf \Gamma}_{[m\,  pq \, r]}.
\end{equation}
However,
\begin{equation}
 {\bf \Gamma}_{[m\,  pq \, r]} \sim \Xi_{[m|pqr]} = 0
\end{equation}
by equation \eqref{symmxi}.  Finally, consider the following components
\begin{equation}
   \cT_{m8\,  n8}{}^{pq} = {\bf \Gamma}_{m8\,  n8}{}^{pq}  - 24 \mathbb{P}^{pq}{}_{n8}{}^{r8}{}_{st} {\bf \Gamma}_{r8 \, m8}{}^{st}.
\end{equation}
Using the fact that
\begin{equation}
 \mathbb{P}^{pq}{}_{n8}{}^{r8}{}_{st} = -\frac{1}{12} \delta^{pq}_{n[s} \delta^r_{t]},
\end{equation}
we obtain
\begin{align}
   \cT_{m8\,  n8}{}^{pq} &= {\bf \Gamma}_{m8\,  n8}{}^{pq}  + 2 {\bf \Gamma}_{r8\,  m8}{}^{r[p} \delta^{q]}_n \notag \\
   &= 3 \sqrt{2} \eta^{pq t_1 \dots t_5} \left( \Xi_{m|nt_1 \dots t_5} - \Xi_{n|mt_1 \dots t_5} + 5 \Xi_{t_1|mnt_2 \dots t_5} \right) \notag \\
   &= 21 \sqrt{2} \eta^{pq t_1 \dots t_5} \Xi_{[m|nt_1 \dots t_5]} \, = \, 0,
\end{align}
where we have used the expression for ${\bf \Gamma}_{m8\,  n8}{}^{pq}$ in the second equality and equation \eqref{symmxi} in the final equality.  All other components of the generalized torsion can be similarly shown to be zero.  It should be emphasized that the fact that the full antisymmetrization of the $\Xi$ quantities is zero, equation \eqref{symmxi}, is crucial for this argument.

In summary, the generalized torsion, as defined by equation \eqref{FULLtau} is zero
\begin{equation}
 \cT_{MN}{}^{P} = 0.
\end{equation}
Let us emphasize again the remarkable feature that the vanishing of the generalized
torsion, as originally defined on the basis of very different considerations based on generalized
geometry, here follows from the direct comparison with $D=11$ supergravity.

\subsection{Hook ambiguity} \label{sec:hook}

As we have already mentioned, the supersymmetry transformations are insensitive 
to the generalized affine connection, modulo density
contributions involving the trace of the affine connection, because the fermions 
transform only under the chiral SU(8). With the connections as
originally given in Ref.~\cite{deWit:1986mz}, or equivalently from equations \eqref{spsichi}, 
the supersymmetry variations of the eight gravitini and the 56 dilatini contain 
the following combinations of $Q'_m$ and $P'_m$
\begin{eqnarray}\label{SUSY}
\delta\psi_\mu^A &\propto&...+ \, \Big( e^{m \, AC} {Q'}_{m \,C}{}^{B}
  - e^m{}_{CD} {P'}_m{}^{ABCD}\Big) \gamma_\mu  \varepsilon_B, \nn\\
\delta\chi^{ABC} &\propto& ...+  \, \Big( 3\, e^{m \, [AB} {Q'}_m{}^{C]}{}_D  +
3 \, e^m{}_{EF} {P'}_m{}^{EF[AB} \delta^{C]}_D +
4\, {P'}_m{}^{ABCE} e^m{}_{ED} \Big) \varepsilon^D.  \hspace{3mm}
\end{eqnarray}  
An important property of the expressions appearing here on the right hand side, is that they
are actually insensitive to certain modifications of the connections. We first recognize
that these are exactly the same combinations that appear in the two first 
equations of \eqref{covcom}. Secondly, the expressions on the right hand side 
of \eqref{SUSY} admit a non-trivial kernel  which is found by looking for solutions of
\bea\label{Aeqs}
0 &=&  \Gamma^{m}_{AC} \, \delta {Q'}_{m \,C}{}^{B} - 
   \Gamma^m_{CD} \, \delta {P'}_m^{ABCD}, \nn\\
0 &=&  3\, \Gamma^{m[AB} \, \delta {Q'}_m{}^{C]}{}_D  -
3\,  \Gamma^m_{EF} \, \delta {P'}_m{}^{EF[AB} \delta^{C]}_D -
4\, \delta {P'}_m{}^{ABCE} \, \Gamma^m_{ED}. 
\eea
Let us proceed with the following ans\"atze 
\bea
\delta {Q'}_{m \,A}{}^B &=& X^{(3)}_{m | ab} \, \Gamma^{ab}_{AB} 
                + X^{(4)}_{m | abc} \, \Gamma^{abc}_{AB} 
                + X^{(7)}_{m | abcdef} \, \Gamma^{abcdef}_{AB}, \nn\\
\delta {P'}_m^{ABCD} &=& Y^{(3)}_{m | ab} \, \Gamma^a_{[AB} \Gamma^b_{CD]}      +          
                Y^{(4)}_{m | abc} \, \Gamma^{[a}_{[AB} \Gamma^{bc]}_{CD]}                +
                Y^{(7)}_{m | abcdef} \, \Gamma^{[a}_{[AB} \Gamma^{bcdef]}_{CD]},                
\eea   
where the slash $| $ simply indicates that no {\it a priori} symmetry conditions are
imposed on the $X$'s and $Y$'s other than the obvious ones (to wit, anti-symmetry
in $[ab]$, $[abc]$ and $[abcdef]$, respectively).  For the form field contributions 
it was already shown in Ref.~\cite{Nicolai:2011cy} that the GVP remains valid if
\be
Y^{(4)}_{m| abc} = \frac32 \,X^{(4)}_{m| abc} \; ,\quad
Y^{(7)}_{m| abcdef} = - \frac32 \, X^{(7)}_{m|  abcdef}
\ee
with no further restrictions on the $X$'s and $Y$'s.
Notice that both $X^{(4)}$ and $X^{(7)}$ have {\em two} irreducible parts: besides the fully antisymmetric pieces appearing in (\ref{QPdWN}) there are the hook diagram
contributions. Furthermore, it was shown in Ref.~\cite{Nicolai:2011cy} that $X^{(4)}, Y^{(4)}$ 
and $X^{(7)}, Y^{(7)}$ are in the kernel of the supersymmetry variations (\ref{Aeqs}) provided that
\be
X^{(4)}_{[m |  abc]} = 0 \;\; , \quad   X^{(7)}_{[m| abcdef]} = 0.
\ee
That is, the fully antisymmetric parts (the four-form and seven-form field strengths)
are determined, but the hook diagram contributions can be chosen freely, as they drop
out in the supersymmetry variations of the fermions in (\ref{SUSY}).
Note that $\Xi_{m|npq}$ and
$\Xi_{m|npqrst}$ that appear in the generalized affine connection in \eqref{Xi1} and \eqref{Xi2}
are precisely of the hook-type, hence providing a {\em geometrical} explanation 
for the ambiguities found in \cite{Nicolai:2011cy}.

As for the remaining SO(7) part $X^{(3)}_{m|ab}$, which was not considered
in Ref.~\cite{Nicolai:2011cy}, the first expression in equations (\ref{Aeqs}) reduces to 
\be\label{X30}
X^{(3)}_{a|bc} \,\Gamma^{abc}_{AB}  \,+
\left( 2 X^{(3)}_{a|ab} + \frac43 Y^{(3)}_{a|ab}\right) \Gamma^b_{AB}
\, - \, Y^{(3)}_{a|bb} \Gamma^a_{AB}=0 \, .
\ee
Whence we read off the condition
\be
Y^{(3)}_{m|ab} = -\frac32\, X^{(3)}_{m|ab}.
\ee
With this identification the second line in (\ref{Aeqs}) becomes
\be\label{X31}
X^{(3)}_{a | bc} \, \Big( 2 \Gamma^{[a}_{[AB} \Gamma^{b]c}_{C]D}
             - \Gamma^c_{[AB} \Gamma^{ab}_{C]D}\Big)  \, - \,
             X^{(3)}_{a | bb} \Gamma^a_{[AB} \delta_{C]D}=0.
\ee
We now see that all terms in (\ref{X30}) and (\ref{X31}) except the last ones 
involving $X^{(3)}_{a|bb}$ cancel, provided we demand that
\be\label{X32}
X^{(3)}_{[a | b]c} = 0.
\ee

To interpret the remaining term let us check the difference between the expressions for the connection coefficients given 
in Ref.~\cite{deWit:1986mz}, equation \eqref{QPdWN}, and in Ref.~\cite{Godazgar:2014sla}, equations \eqref{QGGN} and \eqref{PGGN}.  These connections are fully covariant under internal diffeomorphisms.  The difference is thus
\be
X^{(3)}_{m | ab} \; = \;  \frac12 \Big(e^{n}{}_{b} \partial_m e_{n\, a} + \omega_{m\, ab} \Big) \, = \,
    \frac12 \, e^n{}_a e_{p\, b} \Gamma_{mn}^p,
\ee
where we have used the usual vielbein postulate satisfied by the siebenbein and 
$\Gamma_{mn}^p$ is the usual Christoffel symbol.  Hence (\ref{X32}) is indeed
satisfied for a torsion-free affine connection. The only extra term in the supersymmetry
variations then comes from the `leftover' term in (\ref{X31}) which is just a 
density term proportional to $\Gamma_{km}{}^k$, which is required here because 
the supersymmetry parameter is a density. This is the same term that was obtained above
with the connections (\ref{QPdWN}) just from $Q'_{m\, ab}$ and $P'_{m\, ab}$ alone.
We thus see that the switch from \eqref{QPdWN} to \eqref{QGGN} and \eqref{PGGN}
reintroduces the density term proportional to $\Gamma_{km}{}^k$ 
that was absent in Ref.~\cite{deWit:1986mz}. In other words, 
even the density term which is there with the correct weight if the GVP 
is formulated with the usual affine connection as in Ref.~\cite{Godazgar:2014sla} can 
be absorbed into a redefinition of $Q_{m \,A}{}^B$ and $P_m{}^{ABCD}$, as they were 
originally given in Ref.~\cite{deWit:1986mz}. In fact we are free to also
choose any {\em interpolating} solution where the coefficient of the density 
term changes, as part of it is absorbed into $Q_{m \,A}{}^B$, while the other into $P_m{}^{ABCD}$. 

Let us also point  out how the apparent discrepancy between (\ref{traceconn}), where
$\Gamma_{km}{}^k \, \propto \, e^{-1}\partial_m e$ (with $e$ the usual vierbein determinant), and
the above result, where $\Gamma_{km}{}^k \, \propto\, \Delta^{-1} \partial_m \Delta$, 
is resolved: while in (\ref{covcom1})  the contribution proportional to $\Gamma_{KM}{}^K$ cancels 
with the  weight assignments given there, the contribution proportional to $\Gamma_{km}{}^k$ 
here can be eliminated by shifting back to the non-covariant connections $Q'_m$ and $P'_m$,
and only then the two pictures can be made to agree.
Otherwise the two sets of connections (both of which
are consistent) simply reflect the unavoidable ambiguities identified in section \ref{subsec:e7covgeo}.

Let us emphasize once again that the connections given in equations \eqref{QGGN} and
 \eqref{PGGN} satisfy {\em all required covariance properties} of generalized or 
exceptional geometry provided we break up $\cV$ by choosing the specific `frame'
as derived from D=11 supergravity. First of all, the covariance under local SU(8) follows by 
the same arguments as in Ref.~\cite{deWit:1986mz}: as given, these expressions 
correspond to objects in a special SU(8) gauge (namely the one that accords with the D=11 theory), such that $Q_{m \,A}{}^B$ transforms as a proper SU(8) connection (for the SO(7) 
subgroup this is anyhow obvious). Secondly,  $P_m{}^{ABCD}$ transforms covariantly when 
we apply an SU(8) rotation that moves us out of the given gauge. Furthermore, these 
objects are also covariant under generalized diffeomorphisms: for the 7-dimensional
internal diffeomorphisms this is manifestly true, while the fact that they do not transform 
at all under the remaining generalized diffeomorphisms with parameters $\xi_{mn}\, ,\,
\xi^{mn}$ and $\xi_m$ is consistent with the formulae (17) and (18) of Ref.~\cite{Godazgar:2014sla} 
because $Q_M = P_M = 0$ for $M\neq m$. Of course, these statements apply only
to the specific `frame' as derived from $D=11$ supergravity, that we have adopted 
here, where the connections have non-vanishing coefficients only along the seven internal 
dimensions. However, it is straightforward to see that the manipulations we are now 
going to perform on these specific connections to bring them in line with the constructions described in the two foregoing sections are themselves fully covariant and 
therefore {\em preserve these covariance properties}.

Let us point out once more that the existence of
covariant connections is possible here because we have given the connections explicitly in terms of
$D=11$ fields. It is not possible to achieve if all quantities are to be expressed only 
in terms of the generalized vielbein $\cV$ and its derivatives in an E$_{7(7)}$-covariant manner, as we already saw in the foregoing section (and will explain
again for a simplified example in appendix \ref{app:gl7}).

\subsection{Non-metricity and redefinition of the generalized connection}

In order to understand how the appearance of $P_{M}$ on the right-hand side of the GVP
(\ref{intgvpe7}) can be reconciled with the absence in the corresponding 
relation given previously in equation \eqref{VP2},  it is useful to recall that similar ambiguities 
arise in standard  differential geometry.  While the vielbein postulate is usually quoted as
\begin{equation}
 \partial_m e_{n}{}^a + \omega_{m}{}^{a}{}_{b} e_n{}^b - \Gamma_{mn}^{p} e_p{}^a = 0
\end{equation}
with $\Gamma_{mn}^p$ the Christoffel symbols, there is a more general expression
\begin{equation}
 \partial_m e_{n}{}^a + \omega_{m}{}^{a}{}_{b} e_n{}^b - \Gamma_{mn}^{p} e_p{}^a = T_{mn}{}^p e_p{}^a + P_{m}{}^a{}_b\, e_n{}^b,
\end{equation}
where $\Gamma_{mn}^{p}$ is no longer given by the Christoffel symbols, $T_{mn}{}^p=T_{[mn]}{}^p$ is referred to as the torsion and $P_{m\, ab}=P_{m\, (ab)}$ is referred to as the non-metricity,
as it `measures' the failure of the metric to be covariantly constant (see for example Ref.~\cite{Hehl:1976kj}).  Notice that there is quite a lot of freedom in the definition of the various objects in the equation above.  For example, the antisymmetric part of the affine connection $\Gamma_{[mn]}^p$ can be absorbed into a redefinition of $T_{mn}^p$ so that $\Gamma_{mn}^p=\Gamma_{(mn)}^p$.  Similarly, the non-metricity can be absorbed into a redefinition of the affine connection and the torsion:
\begin{gather}
 \Gamma_{mn}^{p} \longrightarrow \Gamma_{mn}^{p} - P_{(m}{}^c{}_{|d|}\, e_{n)}{}^d e^p{}_c, \notag \\
 T_{mn}{}^p \longrightarrow T_{mn}{}^p - P_{[m}{}^c{}_{|d|}\, e_{n]}{}^d e^p{}_c.
\end{gather}
Furthermore, the fully anti-symmetric part of the torsion can be absorbed into a redefinition of the spin connection
\begin{equation}
 \omega_{m\, ab} \longrightarrow \omega_{m\, ab} - T_{mnp}\, e^{n}{}_a \, e^p{}_{b}.
\end{equation}
Hence, in differential geometry there is a great deal of freedom in how one defines various structures such as non-metricity, torsion and the affine and spin connections.

In complete analogy with this discussion, connection coefficient $P_{M}$ can be absorbed into a redefinition of ${\bf \Gamma}_{M}$ in the internal GVP, equation \eqref{intgvpe7}:
\be\label{Ga1}
{\bf \Gamma}_{M N}{}^{P} \longrightarrow \tilde{\bf \Gamma}_{M N}{}^{P} \, = \, {\bf \Gamma}_{M N}{}^{P} \, + \,
i \Big( \cV_{N}{}^{AB} P_{M\, ABCD} \cV^{P\,CD} 
         - \cV_{N\,AB} P_{M}{}^{ABCD} \cV^{P}{}_{CD} \Big)  
\ee
so that the internal GVP becomes
\begin{equation} \label{intgvpnop}
 \partial_{M} \cV_{N\, AB} \, - \, \tilde{\bf{\Gamma}}_{M}{}_{N}{}^{P} \cV_{P\, AB} 
 \,+\,  Q_{M}{}^{C}{}_{[A} \cV_{N\, B]C} \,=\,  0.
\end{equation}
We note that this shift only changes the affine connection, but {\em does not affect
the} SU(8) {\em connection} $Q_{M \, A}{}^B$. The GVP is now of the form of \eqref{VP2} 
in section \ref{sec:e7covgeo}, but the connections are still different. In particular,
the $Q_{M \, A}{}^{B}$ and $\tilde{\bf\Gamma}_{MN}{}^P$ are still non-zero only
for the first seven components given by equations \eqref{QGGN}. However, by removing the non-metricity in the affine connection we have 
reintroduced torsion in $\tilde{\bf\Gamma}$ where there was none before, in analogy to ordinary 
differential geometry. Therefore, in order to recover a torsion-free affine connection
we follow the same procedure as in section \ref{subsec:e7covgeo}, and accordingly redefine the 
affine connection once more, as follows:
\begin{align}
 Q_{M\, A}{}^{B} &\longrightarrow \widehat\cQ_{M \, A}{}^B \equiv Q_{M\, A}{}^{B} + \mathbb{Q}_{M \, A}{}^{B}, \\
\tilde{\bf\Gamma}_{M N}{}^P &\longrightarrow \widehat{\bf\Gamma}_{M N}{}^P 
\equiv \tilde{\bf\Gamma}_{M N}{}^P 
+ i \Big( \cV^P{}_{AB} \MQ_{M}{}^A{}_C \cV_N{}^{BC}  -
          \cV^{P AB} \MQ_{M\, A}{}^C \cV_{N BC} \Big),
\label{torfreeredef}
\end{align}
where, modulo the remaining ambiguity $U_{M \, A}{}^B$,  the modification $\MQ_M$ is 
now chosen to obtain precisely the connection $\cQ$ in section \ref{sec:e7covgeo}, namely
\begin{equation} \label{bbQdef}
\mathbb{Q}_{M \, A}{}^B = R_{M \, \, A}{}^{B} + U_{M \, \, A}{}^{B} + W_{M \, \, A}{}^{B} + \frac{2i}{3} \, {\bf{\Gamma}}_{MN}{}^{P} \cV_{P \, AC} \cV^{N \, CB}.
\end{equation}
With the redefinitions \eqref{torfreeredef}, we have now brought the GVP into 
the standard form
\begin{equation} \label{GVPfinal}
 \partial_{M} \cV_{N\, AB} \, - \, \widehat{\bf{\Gamma}}_{M}{}_{N}{}^{P} \cV_{P\, AB} 
 \,+\,  \cQ_{M}^{C}{}_{[A} \cV_{N\, B]C} \,=\,  0,
\end{equation}
with the following properties:
\begin{itemize}
\item The affine connection $\widehat{\bf\Gamma}_{MN}{}^P$ is torsion-free, an SU(8) 
          singlet and transforms properly under generalized diffeomorphisms.
\item The SU(8) connection $\cQ_{M \, A}{}^B$ transforms as a connection under SU(8), and
          as a generalized vector under generalized diffeomorphisms.
\item The connections have non-vanishing components for {\em all} 56 components,
          and this is necessary for the supersymmetry variations of the fermions to be 
          expressible in terms of the SU(8) connection $\cQ_{M \, A}{}^B$ alone (see the previous section).           
\item The remaining differences between the above connections and the ones obtained
          in the previous section are all contained in the hook-type ambiguity.
\end{itemize}

Modulo the ambiguity, these connections are now equivalent to the connections defined 
in section \ref{sec:e7covgeo}, namely ${\widehat{\bf{\Gamma}} \cong \Gamma}$. 
We should point out that, with the formulae at hand, we could in principle proceed to work
out explicit expressions for $\cQ_{M \, A}{}^B$ and $\Gamma_{MN}{}^P$ in 
terms of the $D=11$ fields. However, after the redefinitions these expressions will be very complicated, and by themselves not very illuminating.

The trace of the affine connection $\bf\Gamma$ is given by the determinant of the siebenbein \cite{Godazgar:2014sla},
 \begin{equation}
   {\bf{\Gamma}}_{K M}{}^{K} = \frac{3}{2} \partial_{M} \textup{log} \Delta.
 \end{equation}
The connection used to construct the exceptional geometry in section \ref{sec:e7covgeo} is required to be compatible with the vierbein density, \eqref{viercomp}, which implies equation \eqref{traceconn}. This condition can be satisfied by the torsion-free connection by choosing $W$ in equation \eqref{bbQdef} appropriately. In particular the trace of $\bf\Gamma$ drops out of $ \Gamma_{KM}{}^{K}$:      
\begin{align*}
 \Gamma_{KM}{}^{K}
 &= \tilde{\bf\Gamma}_{KM}{}^{K} + i \Big( \cV^K{}_{AB} \MQ_{K}{}^A{}_C \cV_M{}^{BC}  -
          \cV^{K AB} \MQ_{K\, A}{}^C \cV_{M BC} \Big) \\
&= i \left( \cV^{K}{}_{AB} W_{K}{}^{A}{}_{C} \cV_{M}{}^{BC} - \cV^{K \, AB} W_{K \, A}{}^{C} \cV_{M \, BC} \right).
\end{align*}
The $W$ given in equation \eqref{QQUW} ensures that the affine connection $\Gamma$ satisfies the condition \eqref{traceconn}. Note that the part of the fermion supersymmetry transformations given by the internal connection are independent of the vierbein determinant. This remains so despite the contribution from $W$, which is cancelled by the density contributions in the covariant derivative $\nabla_{M}$ of weighted tensors in the supersymmetry transformations.

\subsection{Connections and fermion supersymmetry transformations}

In section \ref{subsec:susy}, we give the fermion supersymmetry transformations \eqref{susyfermions_compact} in terms of the torsion-free connection constructed in section \ref{subsec:e7covgeo}. Solving the section condition to obtain the ${D=11}$ supergravity, the fermion supersymmetry transformation should yield those of the SU(8) invariant reformulation \cite{deWit:1986mz}, \eqref{spsichi}. Using the definition of the covariant derivative \eqref{full_nabla} and equations \eqref{omegaMhat} and \eqref{internalQ}, transformations \eqref{susyfermions_compact} become
\bea
\delta_{\epsilon} \psi_{\mu}^A &=& 2\, {\cal D}_{\mu} \epsilon^A
+ \frac{1}{4} {{\cal F}}^-_{\rho \sigma}{}^{AB} \gamma^{\rho \sigma} \gamma_{\mu} \epsilon_B 
+ \,i\, e_{\nu \, \beta} \partial_{M} e_{\rho}{}^{\beta} \cV^{M \, AB} \gamma^{\nu \rho} \gamma_{\mu} \epsilon_{B} 
- 4 \,i\, \,{\cal V}^{M\,AB} \partial_M\left(\gamma_{\mu} \epsilon_B \right)
\nonumber\\
&&{}
- 2 \,i\, \, \cV^{M \, AB} q_{M \, B}{}^{C} \gamma_{\mu} \epsilon_{C} 
- 2 \,i\, \, \cV^{M}{}_{CD} p_{M}{}^{ABCD} \gamma_{\mu} \epsilon_{B}  
\;,\nonumber\\
\delta_{\epsilon} \chi^{ABC} &=& -2 \sqrt{2}\, {{\cal P}}_{\mu}{}^{ABCD} \gamma^{\mu} \epsilon_D 
+ \frac{3 \sqrt{2}}{4}\, {{\cal F}}^-_{\mu \nu}{}^{[AB} \gamma^{\mu \nu} \epsilon^{C]} 
+ 3 \sqrt{2} \,i\, e_{\mu \, \beta} \partial_{M} e_{\nu}{}^{\beta} \cV^{M \, [AB} \gamma^{\mu \nu} \epsilon^{C]}  \nonumber\\
 && -12\sqrt{2} i \, {\cal V}^M{}^{[AB}\,   \partial_M \epsilon^{C]} 
+ 6 \sqrt{2} \,i\, \, \cV^{M \, [AB} q_{M \, D}{}^{C]} \epsilon^{D} 
- 8 \sqrt{2} \,i\, \, \cV^{M}{}_{DE} p_{M}{}^{ABCD}  \epsilon^{E}  \nonumber\\
&& -6 \sqrt{2} \,i\, \, \cV^{M}{}_{DE} p_{M}{}{}^{DE[AB} \epsilon^{C]}  
\;,
\label{susyfermions_complete}
\eea
In this form, the supersymmetry transformations \eqref{susyfermions_compact} reduce to the following expressions upon use of the canonical solution of the section condition
\begin{align}
 \delta \psi_{\mu}^{A} &= 2\left(\partial_{\mu} - B_{\mu}{}^{m} \partial_{m} -\frac{1}{4} \partial_m B_{\mu}{}^{m} \right) \epsilon^{A} + \frac{1}{2} \omega_{\mu}{}^{\alpha \beta} \gamma_{\alpha \beta} \epsilon^{A} + \cQ_{\mu}{}^{A}{}_{B} \epsilon^{B}  \nn\\[0mm]
 & \hspace{4mm}+\frac{1}{4} \cF^{-}_{\alpha \beta}{}^{AB} \gamma^{\alpha \beta} \gamma_{\mu} \epsilon_{B}
- \frac{1}{4} e^{m \, AB} e_{\nu \, \beta} \partial_m e_{\rho}{}^{\beta} \gamma^{\nu \rho} \gamma_{\mu} \epsilon_B \nn\\[0mm]
& \hspace{4mm}+ e^{m \, AB} \partial_{m} \left( \gamma_{\mu} \epsilon_{B} \right) 
  + \frac{1}{2} e^{m \, AB} q_{m \, B}{}^{C} \gamma_{\mu} \epsilon_{C} 
 - \frac{1}{2} e^{m}{}_{CD}\, p_{m}{}^{ABCD} \gamma_{\mu} \epsilon_{B}, \notag \\
\delta \chi^{ABC} &= - 2\sqrt{2} \cP_{\mu}{}^{ABCD} \gamma^{\mu} \epsilon_{D} + \frac{3\sqrt{2}}{4} \cF^{-}_{\alpha \beta}{}^{[AB|} \gamma^{\alpha \beta} \epsilon^{|C]} - \frac{3\sqrt{2}}{4}  e_{\mu \, \beta} \partial_m e_{\nu}{}^{\beta} e^{m [AB} \gamma^{\mu \nu} \epsilon^{C]} \nn \\
& \hspace{4mm}+ 3 \sqrt{2} e^{m [AB} \partial_{m}\epsilon^{C]} 
  - \frac{3\sqrt{2}}{2} e^{m [AB} q_{m \, D}{}^{C]} \epsilon^{D} - \frac{3\sqrt{2}}{2} e^{m}{}_{DE}\, p_{m}{}^{DE[AB} \epsilon^{C]} \nn \\[2mm]
& \hspace{4mm} -2\sqrt{2} e^{m}{}_{DE} \, p_{m}{}^{ABCD} \epsilon^{E}.
\end{align}
Comparing the supersymmetry transformations above that come from the supersymmetric EFT with the canonical solution of the section condition with those of the $D=11$ theory as written in Ref.~\cite{deWit:1986mz}, transformation \eqref{spsichi}, we find that they are identical upon identifying $\frac{1}{8} \cF_{\alpha \beta}{}^{AB}$ with $\cG_{\alpha \beta}{}^{AB}$ and $Q'$, $P'$ with $q$, $p$, respectively.

First, let us consider the relation between $\cF_{\alpha \beta}{}^{AB}$ and $\cG_{\alpha \beta}{}^{AB}.$ Note that $\cF_{\alpha \beta}{}^{AB}$ satisfies a twisted self-duality condition, which means that on-shell $$ \cF^{-}_{\alpha \beta}{}^{AB} = \cF_{\alpha \beta}{}^{AB}.$$
The $\cG_{\alpha \beta}{}^{AB}$, however, does not satisfy a twisted self-duality condition and in order to modify it so that it does, we need to add to it the Hodge dual of the field strengths, \emph{viz.}
\begin{gather}
 \cG_{\alpha \beta AB} \equiv - \frac{i}{16} \Delta^{1/2} e_{[\alpha}{}^{\mu} e_{\beta]}{}^{\nu} (\partial_{\mu} - B_{\mu}{}^{m} \partial_{m}) B_{\nu}{}^{n} \Gamma_{n AB} + \frac{\sqrt{2}}{64} i \Delta^{-1/2}
  F_{\alpha \beta mn} \Gamma^{mn}_{AB} \notag \\
 \hspace{23mm} +\frac{\sqrt{2}}{64 \cdot 5!} \Delta^{-1/2} \epsilon_{\alpha \beta \gamma \delta}
  F^{\gamma \delta m_1 \dots m_5} \Gamma_{m_1 \dots m_5 \, AB} + i \Delta^{1/2} \epsilon_{\alpha \beta \gamma \delta} X^{\gamma \delta|n} \Gamma_{n AB},
\end{gather}
 where $X_{\alpha \beta|m}$ would correspond to the field strength of the field dual to $B_{\mu}{}^{m}.$ However, since the first term in the expression above is not exact, $B_{\mu}{}^{m}$ cannot be dualized in the usual way. This is why the new field $B_{\mu \nu M}$ is necessary in the definition of $\cF_{\mu \nu}{}^{M}$, \eqref{YM}, schematically ``eating up'' the non-exact terms to allow dualization.       

Regarding the relation between $Q'$, $P'$ and $q$, $p$: as explained in section \eqref{sec:hook}, the $Q'$ and $P'$ are related to $Q$ and $P$ by the usual Christoffel symbol associated with the siebenbein. Moreover, the $Q$ and $P$ are related to $q$ and $p$ by the generalized affine connection $\bf\Gamma,$
\begin{align} 
 Q_{m \, A}{}^{B} &= q_{m \, A}{}^{B} -
 \frac{2i}{3} \, {\bf{\Gamma}}_{mN}{}^{P} \cV_{P \, AC} \cV^{N \, CB}, \nn\\
 P_{m ABCD} &= p_{m ABCD} + 
 i \, {\bf{\Gamma}}_{mN}{}^{P} \cV_{P AB} \cV^{N}{}_{CD}. \label{Qqreln}
\end{align} 
In both cases, the redefinitions correspond to hook-type redefinitions to which the supersymmetry transformations are insensitive, as explained in section \ref{sec:hook}.  Therefore, at the level of the supersymmetry transformations, the two sets of connection coefficients are equivalent.

The fermion supersymmetry transformations of a truncation of the $D=11$ theory have been studied in Ref.~\cite{Coimbra:2012af}, where they are also given in terms of a generalized SU(8) connection constructed in Ref.~\cite{Coimbra:2011ky}. In this paper, we use a connection that allows us to express the fermion supersymmetry transformations covariantly in terms of the 56-bein, rather than its components. This is done by using some of the components in the \textbf{1280} representation, to which supersymmetry transformations are insensitive to \cite{Coimbra:2011ky} (see also section \ref{sec:e7covgeo}). Therefore, the connection ${\cQ - U}$ still contains terms, not expressible in terms of the 56-bein and its derivatives, that are in the \textbf{1280} representation.  These terms are precisely the difference between the ${\cQ - U}$ and the unambiguous part of the connection of Ref.~\cite{Coimbra:2011ky}.  In practice, an explicit expression of this difference is rather complicated.

\newpage

\section*{Acknowledgments}
H.G.\ and M.G.\ would like to thank the AEI, in particular H.N., ENS Lyon, in particular H.S., for hospitality, as well as the Mitchell foundation for hospitality at Great Brampton House. H.G.\ and H.N.\ would also like to thank KITPC in Beijing for hospitality
during the final stages of this work. H.G.\ and M.G.\ are supported by King's College, Cambridge. H.G.\ acknowledges funding from the European Research Council under the European Community's Seventh Framework Programme (FP7/2007-2013) / ERC grant agreement no. [247252].
The work of O.H.\ is supported by the 
U.S. Department of Energy (DoE) under the cooperative 
research agreement DE-FG02-05ER41360 and a DFG Heisenberg fellowship. 
The work of H.N.\ is supported by a Gay-Lussac-Humboldt prize.
We would like to thank B. de Wit, E. Musaev, M. Perry, and D. Waldram
for useful comments and discussions.   

\newpage

\section*{Appendix}

\begin{appendix}

\section{Notations and conventions}  The index notation used in this paper is as follows:
\label{app:notations}

\begin{itemize}
 \item $\mu, \nu, \ldots$ and $\alpha, \beta, \ldots$ denote $D=4$ spacetime and tangent space indices, respectively.
 \item $m, n, \ldots$ and $a, b, \ldots$ denote $D=7$ spacetime and tangent space indices, respectively.
 \item $M, N, \ldots$ label the fundamental (\textbf{56}) of E$_{7(7)}$.
 \item $\bfa$ labels the adjoint (\textbf{133}) of E$_{7(7)}$.
\item $ A,B, \dots $ denote SU(8) indices.
\end{itemize}
Furthermore, the following notations are used for covariant derivatives:
\begin{itemize}
 \item ${D}_\mu=\partial_\mu - \mathbb{L}_{\cA_\mu}$ denotes the E$_{7(7)}$-covariant derivative.
\item $\cD_{\mu} = {D}_\mu + \omega_{\mu}{}^{\alpha}{}_{\beta} + \cQ_{\mu}{}^{A}{}_{B}$ denotes the E$_{7(7)}$-covariant derivative that is also covariant with respect to the local SO(1,3) and SU(8) symmetries.
\item $\nabla_{\mu} = {\cD}_\mu + \Gamma_{\mu \nu}^{\rho}$ is the fully covariant derivative.
\end{itemize}
Analogously,
\begin{itemize}
\item $\cD_{M} = {\partial}_M + \omega_{M}{}^{\alpha}{}_{\beta} + \cQ_{M}{}^{A}{}_{B}$ denotes derivative that is also covariant with respect to the local SO(1,3) and SU(8) symmetries.
\item $\nabla_{M} = {\cD}_M + \Gamma_{M N}^{P}$ is the fully covariant derivative,
\end{itemize}
and $\widehat\cD_{M}$ and $\widehat\nabla_{M}$ are defined with the modified spin connection $\widehat\omega_{M}.$

\section{Useful identities}
\label{app:identities}

In this appendix we collect a handful of useful relations and identities in order
to deal with the ${\rm E}_{7(7)}$ projectors (\ref{adjproj}) and the section constraint (\ref{sectioncondition})
upon contractions with the 56-bein. Let us first note the projector identity
\bea
{\mathbb{P}}^M{}_N{}^P{}_Q\,{\cal V}_P{}_{AB}\,{\cal V}^Q{}^{CD}
&=&
\frac13\,{\cal V}_N{}_{E[A}\,{\cal V}^M{}^{E[C}\,\delta_{B]}{}^{D]}
+\frac13\,{\cal V}^M{}_{E[A}\,{\cal V}_N{}^{E[C}\,\delta_{B]}{}^{D]}
\nonumber\\
&&{}
-\frac{1}{24}\left({\cal V}_N{}_{EF}\,{\cal V}^M{}^{EF}+{\cal V}^M{}_{EF}\,{\cal V}_N{}^{EF}\right)
\delta_{AB}^{CD}
\;.
\eea
As a consistency check, we may calculate the trace of this relation
\bea
{\mathbb{P}}^M{}_N{}^P{}_Q\,{\cal V}_P{}_{AB}\,{\cal V}^Q{}^{CB}
&=&
\frac{1}{2}\,{\cal V}_N{}_{AB}\,{\cal V}^M{}^{CB}\,
+\frac{1}{2}\,{\cal V}^M{}_{AB}\,{\cal V}_N{}^{CB}\,
\nonumber\\
&&{}
-\frac{1}{16}
\left({\cal V}_N{}_{EF}\,{\cal V}^M{}^{EF}+{\cal V}^M{}_{EF}\,{\cal V}_N{}^{EF}\right)
\delta_{A}{}^{C}
\;,
\eea
confirming that ${\mathbb{P}}^M{}_N{}^P{}_Q$ acts as an identity on the 
right hand side.
Similarly, one finds that
\bea
{\mathbb{P}}^M{}_N{}^P{}_Q\,{\cal V}_P{}_{AB}\,{\cal V}^Q{}_{CD}
&=&
\frac12\,{\cal V}_N{}_{[AB}\,{\cal V}^M{}_{CD]}
-\frac1{48}\,\epsilon_{ABCDEFGH} {\cal V}_N{}^{EF}\,{\cal V}^M{}^{GH}
\;.
\eea

The section constraint (\ref{sectioncondition}) states that
\bea
({\mathbb P}_{{\bf 1}+{\bf 133}})_{PQ}{}^{MN} \;\partial_M \otimes \partial_N &=& 0
\;.
\eea
where ${\bf133}$ and ${\bf 1}$ are in the symmetric and antisymmetric tensor product, respectively.
Contracting this equation with the 56-bein, we obtain explicitly
\bea
{\cal V}^{(M}{}_{AC} {\cal V}^{N)}{}^{BC} \;\partial_M \otimes \partial_N &=& 
\frac18\,\delta_A^B\, {\cal V}^{(M}{}_{CD} {\cal V}^{N)}{}^{CD} \;\partial_M \otimes \partial_N
\;,\nonumber\\
{\cal V}^{M}{}_{[AB} {\cal V}^{N}{}_{CD]} \;\partial_M \otimes \partial_N &=& 
\frac1{24}\,
\epsilon_{ABCDEFGH}\,
{\cal V}^{M}{}^{EF} {\cal V}^{N}{}^{GH} \;\partial_M \otimes \partial_N
\;.
\label{secSU8}
\eea


\section{The supersymmetry algebra}
\label{app:susy}


In this appendix, we show that the commutator of supersymmetry transformations 
(\ref{susybosons})--(\ref{susyfermions_compact}) closes into the supersymmetry 
algebra~(\ref{susy-algebra}). For the commutator on the 
external and internal vielbeine $e_\mu{}^\alpha$ and 
${\cal V}_M{}^{AB}$ we have seen in section~\ref{subsec:susy} above that closure of the algebra 
is a direct consequence of the vanishing torsion conditions (\ref{omega}) and (\ref{GENtorsion}),
respectively. Here, we complete the algebra on the vectors ${\cal A}_\mu{}^M$ and two-forms
${\cal B}_{\mu\nu\,\bfa}$ and ${\cal B}_{\mu\nu\,M}$\,.

We start with the vector fields, 
for which the commutator of two supersymmetry transformations yields
\bea
  [\delta_{\epsilon_1}, \delta_{\epsilon_2}]\,
  {\cal A}_{\mu}{}^{M}
     &=&
     -8 i \,{\cal D}_\mu\left({\cal V}^M{}^{AB} \bar\epsilon_{2\, A} \epsilon_{1\, B}\right)
  +16\, {\cal V}^N{}_{AB}\,   {\cal V}^M{}^{AB}\,\bar\epsilon_2^{C}\,\gamma_{\mu}\,\widehat\nabla_N \epsilon_{1\,C} 
\nonumber\\
&&{}
 +32\, {\cal V}^N{}_{CA}\,   {\cal V}^M{}^{AB}\,\bar\epsilon_2^{C}\,\gamma_{\mu}\,\nabla_N \epsilon_{1\,B} 
 +32\,{\cal V}^M{}^{AB}{\cal V}^{K}{}_{BC} \,  \bar\epsilon_{2\,A}\,\widehat\nabla_K\left(\gamma_{\mu} \epsilon_1^C \right)
~~+~ {\rm c.c.}
\nonumber\\
&=&
 {\cal D}_\mu \Lambda^M 
  +4\, g_{\mu\nu} {\cal M}^{MN} \partial_N \left(\bar\epsilon_2^{A}\,\gamma^{\nu} \epsilon_{1\,A}\right)
  +8\,  {\cal M}^{MN}  \left(\bar\epsilon_2^{A}\,\gamma_{\alpha} \epsilon_{1\,A}\right)  e_{\mu \beta}  
  \, e^{\nu [\alpha} \widehat\nabla_N e_\nu{}^{\beta]}    
  \nonumber\\
&&{}
   +8i \,\Omega^{MN}\left(
   \bar\epsilon_2^{A}\,\gamma_{\mu}\widehat\nabla_N \epsilon_{1\,A}-
   \widehat\nabla_N \bar\epsilon_2^{A}\,\gamma_{\mu}\, \epsilon_{1\,A}\right)
   \nonumber\\
&&{}
 +32  \left(
 {\cal V}^M{}^{AB}{\cal V}^{K}{}_{BC}  
 + {\cal V}^M{}_{BC}{\cal V}^{K}{}^{AB} +\frac18\,\delta_C^A\,{\cal M}^{MK}\right)\;
\nabla_K \left(\bar\epsilon_{2}^C\gamma_{\mu} \epsilon_{1\,A} \right). \qquad
\label{ddA1}
\eea
In the first line, we recognize the action of a gauge transformation together with the
non-covariant contribution $g_{\mu\nu}{\cal M}^{MN} \partial_N \xi^\nu$ 
of the diffeomorphism action~(\ref{diffx}).
The third term can be reduced using (\ref{vTo}).
Let us rewrite the last term of (\ref{ddA1}) as
\bea
&&{}\hspace*{-2cm}
32 \,\nabla_K \left\{\left(
 {\cal V}^M{}^{AB}{\cal V}^{K}{}_{BC}  
 + {\cal V}^M{}_{BC}{\cal V}^{K}{}^{AB} +\frac18\,\delta_C^A\,{\cal M}^{MK}\right)
\left(\bar\epsilon_{2}^C\gamma_{\mu} \epsilon_{1\,A} \right)\right\}
\nonumber\\
&=&
32 \,\partial_K \left\{\left(
 {\cal V}^M{}^{AB}{\cal V}^{K}{}_{BC}  
 + {\cal V}^M{}_{BC}{\cal V}^{K}{}^{AB} +\frac18\,\delta_C^A\,{\cal M}^{MK}\right)
\left(\bar\epsilon_{2}^C\gamma_{\mu} \epsilon_{1\,A} \right)\right\}
\nonumber\\
&&{} 
-32\, \left(
 {\cal V}^{K}{}_{BC} {\cal D}_K{\cal V}^M{}^{AB}  
 + {\cal V}^{K}{}^{AB} {\cal D}_K{\cal V}^M{}_{BC}    -\frac18\,\delta_C^A \, ({\rm trace})\right)
\left(\bar\epsilon_{2}^C\gamma_{\mu} \epsilon_{1\,A} \right)
\nonumber\\
&&{} 
+8 \left(e^{-1}\partial_K e\right) \left(
 {\cal V}^{K}{}_{BC} {\cal V}^M{}^{AB}  
 + {\cal V}^{K}{}^{AB} {\cal V}^M{}_{BC}    +\frac18\,\delta_C^A \,{\cal M}^{MK} \right)
\left(\bar\epsilon_{2}^C\gamma_{\mu} \epsilon_{1\,A} \right)
\nonumber\\
&=&
12\,(t^\bfa)^{MN} \, \partial_N \Xi_{\mu \, \bfa}
-\frac83\,\Omega^{MN} 
\left({\cal V}^K{\,}_{BC}\, {\cal D}_N{\cal V}_K{}^{AB}
+{\cal V}^K{\,}^{AB}\, {\cal D}_N{\cal V}_K{}_{BC}
\right)
\left(\bar\epsilon_{2}^C\gamma_{\mu} \epsilon_{1\,A} \right)
\;,
\nonumber
\eea
reproducing the parameter $\Xi_{\mu\,\bfa}$ from (\ref{12gauge}), and
where we have used (\ref{traceconn}) in the first equality and the vanishing torsion condition (\ref{inter2})
in the second. 
Together, we obtain
\bea
  [\delta_{\epsilon_1}, \delta_{\epsilon_2}]\,
  {\cal A}_{\mu}{}^{M}
     &=&
 {\cal D}_\mu \Lambda^M 
  +\, g_{\mu\nu} {\cal M}^{MN} \partial_N\xi^\nu
  - \frac{1}{2}\,    \xi^\nu    {\cal F}_{\mu\nu}{}^M
-12\,(t^\bfa)^{MN} \, \partial_N \Xi_{\mu \, \bfa}
  \nonumber\\
&&{}
   +8i \,\Omega^{MN}\left(
   \bar\epsilon_2^{A}\,\gamma_{\mu}\widehat{\cal D}_N \epsilon_{1\,A}-
   \widehat{\cal D}_N \bar\epsilon_2^{A}\,\gamma_{\mu}\, \epsilon_{1\,A}\right)
\nonumber\\
&&{} 
-\frac83\, \,\Omega^{MN} 
\left({\cal V}^K{}_{BC}\, {\cal D}_N{\cal V}_K{}^{AB}
+{\cal V}^K{}^{AB}\, {\cal D}_N{\cal V}_K{}_{BC}
\right)
\left(\bar\epsilon_{2}^C\gamma_{\mu} \epsilon_{1\,A} \right)
   \;.
\label{ddA2}
\eea
We observe, that we can simultaneously drop the ${\rm SU}(8)$ connection part in the last 
two lines since they mutually cancel. 
The spin connection $\widehat{\omega}_M{}^{\alpha \beta}$ in the second line yields 
additional contributions which explicitly carry the field strength ${\cal F}_{\mu\nu}{}^M$
and can be simplified using the
twisted self-duality equation (\ref{dualityM}):
\bea
-\,i\,\Omega^{MN}\,\varepsilon_{\mu\nu\rho\sigma} \,
\bar\epsilon_2^{A}\,\gamma^\nu\, \epsilon_{1A}\,{\cal M}_{NK} \,{\cal F}^{\rho\sigma}{}^K &=&
-\frac{1}{2}\,\xi^\nu\,{\cal F}_{\mu\nu}{}^M
\;.
\eea
In total, the commutator (\ref{ddA2}) then takes the expected form 
\bea
  [\delta_{\epsilon_1}, \delta_{\epsilon_2}]\,
  {\cal A}_{\mu}{}^{M}
     &=&
   \xi^\nu    {\cal F}_{\nu\mu}{}^M  
+g_{\mu\nu} {\cal M}^{MN} \partial_N\xi^\nu
+ {\cal D}_\mu \Lambda^M    
+12\,(t^\bfa)^{MN} \, \partial_N \Xi_{\mu \, \bfa}
\nonumber\\
&&{}
+\frac12\,\Omega^{MN}\,\Xi_{\mu\,N}
\;,
\eea
with the last term corresponding to the action of a tensor gauge transformation (\ref{gaugeLX})
with parameter $\Xi_{\mu\,M}$ from (\ref{XXO}).

Next, let us check the commutator of supersymmetry transformations
on the two-forms ${\cal B}_{\mu\nu\,\bfa}$. First, we note that to lowest order in the fermions
the terms descending from variation of the 
$(t_{\bfa})_{MN}\,{\cal A}_{[\mu}{}^{M}\,\delta_\epsilon {\cal A}_{\nu]}{}^{N}$  
contribution in (\ref{susybosons}) simply reproduce the corresponding terms of type
$(t_{\bfa})_{MN}\,{\cal A}_{[\mu}{}^{M}\,[\delta_1, \delta_2] {\cal A}_{\nu]}{}^{N}$ 
in the action of gauge transformations (\ref{gaugeLX}) and diffeomorphisms (\ref{diffx}),
by virtue of the closure of the algebra (\ref{ddA2}) on the vector fields.
We can thus in the following ignore all terms that carry explicit gauge fields ${\cal A}_\mu{}^M$.
With some calculation the various remaining terms 
organize into
\bea
[\delta_1 ,\delta_2 ] \,{\cal B}_{\mu\nu\,\bfa}
&=&
-\frac{8}{3}  \,
    (t_{\bfa})^{PQ}\, \Big( -  {\cal V}_{P\,AB} {\cal
    V}_{Q\,CD}\,
    \bar\epsilon_2^{[A}\,\gamma_{\mu\nu}\,{\cal P}_\rho{}^{BCD]E}\gamma^\rho\epsilon_{1\,E}
    \nonumber\\
    &&{}\qquad\qquad\quad
    + 2 \, {\cal V}_{P\,BC} {\cal V}_{Q}{}^{AC}\,
    \bar\epsilon_{2\,A}\,\gamma_{[\mu}\,{\cal D}_{\nu]}{}\epsilon_1^{B}
      \nonumber\\
    &&{}\qquad\qquad\quad
     -6 i {\cal V}_{P\,AB} {\cal
    V}_{Q\,CD}\,
    \bar\epsilon_2^{[A}\,\gamma_{\mu\nu}\,{\cal V}^{M}{}_{\vphantom{1}}^{\vphantom{[}BC} \,\widehat\nabla_M \epsilon_1^{D]}
    \nonumber\\
    &&{}\qquad\qquad\quad
    - 4 i \, {\cal V}_{P\,BC} {\cal V}_{Q}{}^{AC}\,
    \bar\epsilon_{2\,A}\gamma_{[\mu} {\cal V}^{M}{}^{BD} \,\widehat\nabla_M( \gamma_{\nu]} \epsilon_{1\,D})
    ~+~ {\rm c.c.}\Big)    \nonumber\\
    &&{}~-~ (1 \leftrightarrow 2)
\nonumber\\[1ex]
&=&
    2 {\cal D}_{[\mu} \Xi_{\nu]\bfa}
  + \frac{1}{3} \,
    (t_{\bfa})^{PQ}\,  {\cal V}_{P\,AB} {\cal
    V}_{Q\,CD}\,
   {\cal P}^\sigma{}^{ABCD}\,e\varepsilon_{\mu\nu\rho\sigma}\, \xi^\rho 
  +
    (t_{\bfa})_{MN}\, \Lambda^M
    {\cal F}_{\mu\nu}{}^{N}
\nonumber\\
    &&{}
  - \frac{32}{3} (t_{\bfa})^{P Q}\, \partial_{M} 
    \left(i{\cal V}_{PAC} {\cal V}_{Q}{}^{BC} {\cal V}^{M}{}_{BD} 
    \bar\epsilon_{2}^A \gamma_{\mu\nu} \epsilon_{1}^D  ~+~ {\rm c.c.}\right)
    \nonumber\\
    &&{}
    - \frac{4}{3}
    (t_{\bfa})^{PQ}  \Big(-12 {\cal V}_{P\, CD} \, \bar\epsilon_{2}^{C} \gamma_{\mu \nu} \partial_{Q} \epsilon_{1}^{D} + 4 i {\cal V}_{P AC} {\cal V}^{N \,CD} \partial_{Q} {\cal V}_{N\, DB} \,
    \bar\epsilon_{2}^{A} \gamma_{\mu \nu} \epsilon_{1}^{B}     
\nonumber\\
    &&\hspace{33mm}
    + 3 {\cal V}_{P \, CD} \Omega_{Q \rho \sigma} \, \bar\epsilon_{2}^{C} \gamma_{\mu \nu} \gamma^{\rho \sigma} \epsilon_{1}^{D}  ~+~ {\rm c.c.} ~~ -~ (1 \leftrightarrow 2) \Big)
\nonumber\\[1ex]
&=&
    2 {\cal D}_{[\mu} \Xi_{\nu]\bfa}
  + \xi^\rho\,  {\cal H}_{\rho\mu\nu\,\bfa}
     +
    (t_{\bfa})_{MN}\, \Lambda^M
    {\cal F}_{\mu\nu}{}^{N}
    +\partial_M\Omega_{\mu\nu}{}^M{}_\bfa + (t_{\bfa})_{M}{}^{N} \Omega_{\mu \nu N}{}^{M} \, , \nn \\
     \label{ddB}
\eea
with the gauge parameters
$\Lambda^M$ and $\Xi_{\mu \,\bfa}$ defined in (\ref{12gauge}) above,
and the shift parameters $\Omega_{\mu\nu}{}^M{}_\bfa$, $\Omega_{\mu \nu N}{}^{M}$
given in (\ref{XXO}). Finally, we have used the first-order duality equations (\ref{dualH})
for the last equality in (\ref{ddB}) in order to reproduce on-shell the transformation (\ref{diffx})
under external diffeomorphisms.
Together, we confirm the supersymmetry algebra (\ref{susy-algebra}) 
on the two-forms ${\cal B}_{\mu\nu\,\bfa}$.

Closure of the supersymmetry algebra on the vector fields and two-forms ${\cal B}_{\mu\nu\,\bfa}$
thus has not only determined the supersymmetry transformation rules but also uniquely fixed all
the gauge parameters appearing on the right hand side of (\ref{susy-algebra}). 
The remaining commutator for the constrained two-forms ${\cal B}_{\mu\nu\,M}$ 
thus becomes a consistency check of the entire construction with no more 
free or adjustable parameters to be determined. 
Indeed, closure of two supersymmetry transformations on  ${\cal B}_{\mu\nu\,M}$ 
into (\ref{susy-algebra}) can be shown by a rather lengthy calculation 
of which we will give only a few essential ingredients here.

As for ${\cal B}_{\mu\nu\,\bfa}$, we can consistently 
ignore all terms that carry explicit gauge fields ${\cal A}_\mu{}^M$ which separately organize into
the correct contributions due to closure (\ref{ddA2}) on the vector fields.
After some calculation, we then find for the remaining commutator
\bea
{}[\delta_{\epsilon_1},\delta_{\epsilon_2}]\,{\cal B}_{\mu\nu\,M} &=&
 2\,{\cal D}_{[\mu}\Xi_{\nu]M} 
-4i \, \xi^\rho\,e\varepsilon_{\mu\nu\rho\sigma} 
   \,  R_{M\tau}{}^{\sigma\tau}
\nonumber\\
&&{}
-2i\, e \varepsilon_{\mu\nu\rho}{}^{\sigma}\,
{\cal D}^{\rho} \left(g_{\sigma\lambda} \partial_M  \xi^\lambda \right)
 -\frac{2}{3}\,e\varepsilon_{\mu\nu\rho\sigma} \,{\cal P}^\rho{}^{ABCD}\,
  {\cal V}^P{}_{AB} {\cal D}_M {\cal V}_{P\,CD}\,
   \xi^\sigma  
\nonumber\\
&&{}
+\frac{128}3\,i \,
 \,\bar\epsilon_2^{(A}\gamma_{[\mu} \widehat\nabla_L (\gamma_{\nu]}\epsilon_1^{B)} )   
   {\cal D}_M{\cal V}_K{}_{CA} {\cal V}^{L}{}_{BD} {\cal V}^K{}^{CD}
 ~+~{\rm c.c.}
\nonumber\\
&&{}
-64i  \,
   \bar\epsilon_2^{C}\, \epsilon_1^D \,e_{[\mu}{}^{\alpha} \widehat\nabla_K e_{\nu]}{}_{\alpha}
   {\cal V}^K{}^{AB} {\cal V}^N{}_{AB}  {\cal D}_M {\cal V}_N{}_{CD}
   ~+~{\rm c.c.}
\nonumber\\
&&{}
+64 \,{\cal V}^{K}{}_{CD}\,
   \bar\epsilon_2^{C}\,\gamma_{[\mu}    \widehat{\cal D}_M \widehat\nabla_K(\gamma_{\nu]} \epsilon_1^D))
   ~+~{\rm c.c.}
\nonumber\\
&&{}
 - 64\,
 {\cal D}_M \bar\epsilon_2^{C}\,\gamma_{[\mu}\, {\cal V}^{K}{}_{CD} \widehat\nabla_K(\gamma_{\nu]} \epsilon_1^D)
 ~+~{\rm c.c.}
\nonumber\\
&&{}
-16 i \,\Omega_{MN}\,{\cal F}_{\rho[\mu}{}^N \bar\epsilon_2^{C}\,\gamma^\rho \widehat\nabla_K(\gamma_{\nu]} \epsilon_1^D) {\cal V}^K{}_{CD}\,
~+~{\rm c.c.}
\nonumber\\
&&{}
-16 \, e \varepsilon_{\mu\nu\rho}{}^{\sigma}\,{\nabla}_M \left( \bar\epsilon_2^C \gamma^\rho 
{\cal V}^{N}{}_{CD} \widehat\nabla_N( \gamma_{\sigma} \epsilon_{1}^D) \right)
~+~{\rm c.c.}\;.
\label{ddBM1}
\eea   
Here the curvature in the second term
refers to the curvature of the corresponding spin connections
\bea
R_{M\tau}{}^{\sigma\tau} &\equiv&
e_{\alpha}{}^\sigma e_{\beta}{}^\rho 
\left(
\partial_M \, \omega_\rho{}^{\alpha \beta}
-{\cal D}[A,\omega]_\rho \,\omega_M{}^{\alpha \beta}\right)
\nonumber\\
&=&
e_{\alpha}{}^\sigma e_{\beta}{}^\rho  \left(
\partial_M \omega_\rho{}^{\alpha \beta}
- {\cal D}_\rho \left(e^{\tau[\alpha} \partial_M e_\tau{}^{\beta]} \right)
 \right)
\;.
\eea
In the calculation of (\ref{ddBM1}), we have made use of
\bea
{\cal D}_M {\cal V}^L{}_{AB} &=&
\frac{2i}3\,  {\cal V}^K{}^{CD}  {\cal D}_M {\cal V}_K{}_{C[B}{\cal V}^L{}_{A]D}
-i {\cal V}^L{}^{CD} {\cal V}^K{}_{CD}  {\cal D}_M {\cal V}_K{}_{AB}\;,
\\
\Longleftrightarrow\qquad
{\cal V}^P{}_{AB}\,\Gamma_{MP}{}^L  &=&
\frac{2i}3\,  {\cal V}^K{}^{CD}  {\cal V}_P{}_{C[A}{\cal V}^L{}_{B]D}\, \Gamma_{MK}{}^P
+i {\cal V}^L{}^{CD} {\cal V}^K{}_{CD} {\cal V}_P{}_{AB} \,\Gamma_{MK}{}^P \;, \nn
\eea
as well as
\bea
8 i \, e \varepsilon_{\mu\nu\rho}{}^{\sigma}\,
\partial_M \left( \bar\epsilon_2^A \gamma^\rho {\cal D}_{\sigma} \epsilon_{1\,A} \right)
~+~{\rm c.c.}
&=&
8 i \, e \varepsilon_{\mu\nu\rho}{}^{\sigma}\,
\partial_M {\nabla}_{\sigma} \left( \bar\epsilon_2^A \gamma^\rho  \epsilon_{1\,A} \right)
\\
&=&
8 i \, e \varepsilon_{\mu\nu\rho}{}^{\sigma}\,
{\nabla}_{\sigma} \partial_M  \left( \bar\epsilon_2^A \gamma^\rho  \epsilon_{1\,A} \right)
+2 i \, e \varepsilon_{\mu\nu\rho}{}^{\sigma}\,
\partial_M \Gamma_{\sigma\tau}{}^\rho \xi^\tau  
\nonumber\\
&=&
-2i\, e \varepsilon_{\mu\nu\rho}{}^{\sigma}\,
{\cal D}^{\rho} \left(g_{\sigma\lambda} \partial_M  \xi^\lambda \right)
+2i\, e \varepsilon_{\mu\nu\sigma\tau}\,
R_{M\rho}{}^{\sigma\tau} \, \xi^\rho 
\;,\nonumber
\eea
and
\bea
32i
   \left(\bar\epsilon_2^{A}\,\gamma_{[\mu} [{\cal D}_{\nu]}, \widehat{\cal D}_M]_{\rm spin} \, \epsilon_{1\,A}
   ~-~{\rm c.c.}\right)
   &=&
-2i e\varepsilon_{\alpha \beta[\mu|\rho} 
  \xi^\rho\,
   \widehat R_{M|\nu]}{}^{\alpha \beta}
   ~+~{\rm c.c.}
   \nonumber\\
      &=&
-4i e\varepsilon_{\alpha \beta[\mu|\rho} 
  \xi^\rho\,
   R{}_{M|\nu]}{}^{\alpha \beta}
  - 2\Omega_{MN}   
  \xi^\rho\,
   {\cal D}_{[\mu}  {\cal F}_{\nu]\rho} {}^N 
   \nonumber\\
      &=&
-4i e\varepsilon_{\mu\nu\tau\rho} 
  \xi^\rho\,
   R{}_{M\sigma}{}^{\sigma\tau}
-2i e\varepsilon_{\mu\nu\sigma\tau} 
  \xi^\rho\,
   R{}_{M\rho}{}^{\sigma\tau}
   \nonumber\\
   &&{}
  - 2\Omega_{MN}   
  \xi^\rho\,
   {\cal D}_{[\mu}  {\cal F}_{\nu]\rho} {}^N 
\;.
   \eea

Let us start by considering the first five terms of (\ref{ddBM1}). After some further calculation
and upon using the first-order duality equation (\ref{dualH}),
they reduce to
\bea
&\longrightarrow&
 2\,{\cal D}_{[\mu}\Xi_{\nu]M} 
-4i \, \xi^\rho\,e\varepsilon_{\rho\mu\nu\sigma} 
    R_{M\tau}{}^{\sigma\tau}
\nonumber\\
&&{}
-2i\, e \varepsilon_{\mu\nu}{}^{\rho\sigma}\,
{\cal D}_{\rho} \left(g_{\sigma\lambda} \partial_M  \xi^\lambda \right)
 -\frac{2}{3}\,e\varepsilon_{\mu\nu\rho\sigma} \,{\cal P}^\rho{}^{ABCD}\,
  {\cal V}^P{}_{AB} {\cal D}_M {\cal V}_{P\,CD}\,
   \xi^\sigma  
   \nonumber\\
   &=&
 2\,{\cal D}_{[\mu}\Xi_{\nu]M} 
+2i \xi^\rho\,e\varepsilon_{\rho\mu\nu\sigma} 
    \left(\widehat{J}_{M}{}^{\sigma}
     +\frac{1}{3}\, {\cal P}^\sigma{}^{ABCD}\,p_{M\,ABCD} 
 \right)
 \nonumber\\
&&{}
-2i\, e \varepsilon_{\mu\nu\rho}{}^{\sigma}\,
{\cal D}^{\rho} \left(g_{\sigma\lambda} \partial_M  \xi^\lambda \right)
\nonumber\\
&=&
 2\,{\cal D}_{[\mu}\Xi_{\nu]M} 
+ \xi^\rho\,{\cal H}_{\rho\mu\nu\, M}
-2i\, e \varepsilon_{\mu\nu\rho\sigma}{} g^{\sigma\tau}\,
{\cal D}^{\rho} \left(g_{\tau\lambda}\, \partial_M  \xi^\lambda \right)
\;.
\label{ddBM2}
\eea
This exactly reproduces the expected transformation 
of ${\cal B}_{\mu\nu\,M}$ under external diffeomorphisms (\ref{diffx}).
Next, we collect all ${\cal F}^M$ terms on the right hand side of (\ref{ddBM1}).
This yields
\bea
[\delta_1,\delta_2] \,{\cal B}_{\mu\nu\,M}\Big|_{{\cal F}} &=&
\frac{8}{3}\,
{\cal V}_N{}_{CB} {\cal V}^K{}^{CD}\, {\cal D}_M{\cal V}_K{}_{DA}
 \,\bar\epsilon_2^A\gamma_{[\mu} \gamma^{\rho\sigma} \gamma_{\nu]}\epsilon_1{}^B\,
 {\cal F}_{\rho\sigma}{}^N   
\nonumber\\
&&{}
  -   4 \, {\cal V}^P{}_{AB} {\cal D}_M {\cal V}_{P\,CD}\,
    \bar\epsilon_2^{A}\,\gamma_{\mu\nu}\,\gamma^{\rho\sigma} \epsilon_1^B {\cal V}_N{}^{CD} {\cal F}_{\rho\sigma}{}^N 
\nonumber\\
&&{}
-4 i \,
   \bar\epsilon_2^{A}\,\gamma_{[\mu} {\cal D}_M \left(\gamma^{\rho\sigma} \gamma_{\nu]}\epsilon_1{}^B\,
 {\cal F}_{\rho\sigma}{}^N {\cal V}_N{}_{AB}  
\right)  
\nonumber\\
&&{}
   +
4 i \,  {\cal D}_M \bar\epsilon_2^{A}\,\gamma_{[\mu}\, 
\gamma^{\rho\sigma} \gamma_{\nu]}\epsilon_1{}^B\,
 {\cal F}_{\rho\sigma}{}^N {\cal V}_N{}_{AB}  
\nonumber\\
&&{}
+i \, e \varepsilon_{\mu\nu\rho}{}^{\sigma}\,
{\cal D}_M \left( \bar\epsilon_2^A \gamma^\rho \gamma^{\lambda\tau} \gamma_{\sigma}\epsilon_1{}^B\,
 {\cal F}_{\lambda\tau}{}^N {\cal V}_N{}_{AB}   \right)
~+~{\rm c.c.}
\;.
\eea
After some further calculation, these terms may be brought into the form
\bea
&=&
-\frac{32}3 {\cal V}^K{}^{CD}  {\cal D}_M {\cal V}_K{}_{AD} \,{\cal F}_{\mu\nu}{}_{CB}\,
\bar\epsilon_2^A\epsilon_1{}^B
+16 {\cal V}^K{}_{CD} {\cal D}_M {\cal V}_K{}_{AB}  {\cal F}_{\mu\nu}{}^{CD}\,
\bar\epsilon_2^A\epsilon_1{}^B
\nonumber\\
&&{}
+8 i \,{\cal F}_{\mu\nu}{}_{AB}  {\cal D}_M ( \bar\epsilon_2^A\epsilon_1{}^B)
-8 i\, {\cal V}^K{}^{AB} {\cV}_{K\, CD} {\cal D}_M ( {\cal F}_{\mu\nu}{}_{AB}) \bar\epsilon_2^C\epsilon_1{}^D
\nonumber\\
&=&
-8 {\cal V}^K{}^{CD} {\cal F}_{\mu\nu}{}_{CD} {\cal D}_M ({\cal V}_K{}_{AB} \bar\epsilon_2^A\epsilon_1{}^B)
+8 {\cal V}^K{}_{CD} {\cal F}_{\mu\nu}{}^{CD} {\cal D}_M ({\cal V}_K{}_{AB} \bar\epsilon_2^A\epsilon_1{}^B)
\nonumber\\
&&{}
+8 {\cal D}_M ( {\cal V}^K{}^{CD} {\cal F}_{\mu\nu}{}_{CD}) {\cal V}_K{}_{AB} \bar\epsilon_2^A\epsilon_1{}^B
-8 {\cal D}_M ({\cal V}^K{}_{CD} {\cal F}_{\mu\nu}{}^{CD}) {\cal V}_K{}_{AB} \bar\epsilon_2^A\epsilon_1{}^B
\nonumber\\
&=&
{\cal F}_{\mu\nu}{}^K\partial_M \Lambda_K-
     \Lambda_K\partial_M {\cal F}_{\mu\nu}{}^K      
\;,
\eea
and precisely reproduce the gauge transformation (\ref{gaugeLX}) of the two-form ${\cal B}_{\mu\nu\,M}$\,.

It remains to show that all the remaining terms in (\ref{ddBM1}) combine into the $\Omega$ transformations
of (\ref{shiftB}) with parameter $\Omega_{\mu\nu\,M}{}^N$ from (\ref{XXO}). This can be verified by a lengthy
but direct computation.
In the course of this computation, it is useful to explicitly develop the curvature
\bea
R_{MN}{}^{\alpha\beta } &\equiv&
2\partial_{[M} \omega_{N]}{}^{\alpha\beta } + 2 \omega_{[M}{}^{\alpha\gamma } \omega_{N]}{}_{\gamma }{}^\beta 
\nonumber\\
&=&
e^\nu{}_\gamma  e^\rho{}^{[\alpha}  \partial_{[M} e_{\nu}{}^{\beta]}  \partial_{N]}  e_\rho{}^\gamma     
-\frac12\, g^{\mu\nu} \partial_{[M} e_{\nu}{}^{\alpha}   \partial_{N]} e_{\mu}{}^{\beta} 
-\frac12\, e^{\nu \alpha} e^{\mu \beta } \partial_{[M} e_{\nu}{}^{\gamma }  \partial_{N]} e_{\mu}{}_{\gamma } 
\;, \qquad
\eea
from which one obtains
\bea
R_{MN}{}_{\mu\nu} &\equiv& R_{MN}{}^{\alpha\beta }\,e_{\mu \alpha } e_{\nu \beta }
\nonumber\\
&=& -\frac12\,g^{\lambda\kappa}  \partial_{[M} g_{\mu\lambda} \partial_{N]} g_{\nu\kappa}
~=~ -\frac12\,g^{\lambda\kappa}  \nabla_{[M} g_{\mu\lambda} \nabla_{N]} g_{\nu\kappa}
\;.
\eea
We conclude that the supersymmetry algebra consistently closes also on the field
${\cal B}_{\mu\nu\,M}$.

\section{Non-exceptional gravity} 
\label{app:gl7}

In this appendix we will illustrate in terms of a simple example (taken from standard
differential geometry) how the difficulties encountered in constructing a fully covariant 
connection can be understood and resolved in our framework. The main point will be 
that fully covariant 
expressions can be obtained in terms of the $D=11$ connections, but that these cannot 
be written just in terms of the generalized vielbein and its ordinary derivatives --
unlike in ordinary differential geometry.

In standard differential geometry and in the absence of torsion,
the spin connection is defined as 
$$
\omega_{mab} = - \frac12 e_m{}^c\big( \Omega_{ab\,c} - \Omega_{bc\, a} - \Omega_{ca\,b}\big)
$$
with coefficients of anholonomy 
$$
\Omega_{ab\,c} \equiv e_a{}^p e_b{}^q \partial_p e_{qc} - e_b{}^p e_a{}^q \partial_p e_{qc}.
$$
Now define the Cartan form
$$
S_{m\,ab} \equiv e_a{}^n \partial_m e_{nb},
$$
which is the analogue of $\cV^{-1}\partial\cV$ in (\ref{Cartan}),
and decompose this into a symmetric and an antisymmetric part
$$
q_{m\,ab} \equiv S_{m\,[ab]} \;\; , \quad p_{m\,ab} \equiv S_{m\,(ab)}.
$$
These are the same as the $q_{m\,ab}$ and $p_{m\,ab}$ in \eqref{QPdWN}.
Now a quick calculation shows that
$$
\omega_{m\,ab} =  q_{m\,ab} \,- \, 
\big( e_a{}^p e_m{}^c p_{p\,bc} - e_b{}^p e_m{}^c p_{p\,ac}\big)
\equiv  q_{m\,ab} - 2 p_{[a\, b]m}.
$$
Under an arbitrary diffeomorphism, the non-covariant contributions are
$$
\Delta^{nc} q_{m\,ab} = e_{[a}{}^r e_{q|b]} \partial_m \partial_r \xi^q \;\; , \quad
\Delta^{nc} p_{m\,ab} = e_{(a}{}^r e_{q|b)} \partial_m \partial_r \xi^q
$$
and these two contributions cancel in the variation of $ \omega_{m\,ab}$, as expected.
So the spin connection is indeed a covariant object under diffeomorphisms, and 
we also know that it is the {\em only} such object that can be built from the vielbein and
its derivative. Under local SO(1,3) we have
$$
\delta q_{m\,ab} = \partial_m \Lambda_{ab} + 
    \Lambda_a{}^c q_{m\,cb}  + \Lambda_b{}^c q_{m\,ac} \;\; , \quad 
\delta p_{m\,ab} = 
 \Lambda_a{}^c p_{m\,cb}  + \Lambda_b{}^c p_{m\,ac} \;\; , \quad 
$$
so $q_{m\,ab}$ and hence $\omega_{m\,ab}$ transform non-covariantly
as SO(1,3) gauge fields, while $p_{m\,ab}$ is covariant under local SO(1,3).

Next we repeat this calculation in the E$_{7(7)}$ formalism, replacing the siebenbein
by the 56-bein $\cV^M{}_{AB}$ of exceptional geometry. To simplify things
we set $A^{(3)} = A^{(6)} = 0$, and this will suffice to make clear our
main point. Then the E$_{7(7)}$ 56-bein (whose components are explicitly
given in \eqref{gv11}--\eqref{gv14}) simplifies to
\bea
\cV^{m8}{}_{AB} &=& \frac{1}8 \Delta^{-1/2} \Gamma^m_{AB}  \; , \quad
\cV_{mnAB} = \frac{1}8 \Delta^{-1/2} \Gamma_{mnAB},   \nn\\
\cV^{mn}{}_{AB} &=& \frac{i}4 \Delta^{1/2} \Gamma^{mn}_{AB} \; ,\quad  
\cV_{m8 AB} = - \frac{i}4 \Delta^{1/2} \Gamma_{mAB}. 
\eea 
Note that $\cV^{m8}{}_{AB}$ and $\cV^{mn}{}_{AB}$ are imaginary, while 
$\cV_{m8AB}$ and $\cV_{mnAB}$ are real (this is true only in this particular
SU(8) gauge).  By direct computation we find
\bea\label{QmPm}
q_{m \,A}{}^B  &=& \frac{2i}3 \cV^{N \,BC} \partial_m \cV_{N\,CA} \,=\, 
 \frac12 q_{m\,ab} \Gamma^{ab}_{AB},  \nn\\
p_{m \, ABCD} &=& - i\cV^N{}_{AB} \partial_m \cV_{N\,CD} \, = \, 
- \frac34 p_{m\,ab} \Gamma^a_{[AB} \Gamma^b_{CD]}.
\eea
As a check on the coefficients we compute (this is the combination appearing
in the variation of the gravitino)
\be
e^{m \, AC} q_{m \,C}{}^B - e^m{}_{CD} \, p_m{}^{ABCD} =
-\frac12 \omega_{m\,ab} (\Gamma^m \Gamma^{ab})_{AB} 
- \frac12 P_{m\,aa} \Gamma^m_{AB},
\ee
which is indeed the correct result. The last term proportional to
$ -\frac12 \Delta^{-1}\partial_m\Delta$ is just the density contribution proportional to
$\Gamma_{mp}^p$ that is required because the supersymmetry 
parameter $\varepsilon$ is a density, showing again how the density contribution
was absorbed into the connections given in Ref.~\cite{deWit:1986mz}.

With this information we can now compute
\bea\label{QM}
R_{M \, A}{}^B &=& 
\frac{4i}3\Big(\cV^{N \, BC} \cV_M{}^{DE} p_{N \,ACDE} + \cV^N{}_{AC} \cV_{M\, DE} p_N{}^{B CDE}\Big)
\nn\\ && + \,
\frac{20i}{27}\Big(\cV^{N \,DE} \cV_M{}^{BC} p_{N \,ACDE} + \cV^N{}_{DE} \cV_{M \,AC} p_N{}^{BCDE}\Big)
\nn\\ && - \, 
\frac{7i}{27} \delta_A^B
\Big(\cV^{N CD} \cV_M{}^{EF} p_{N \,CDEF} + \cV^N{}_{CD} \cV_{M \,EF} p_N{}^{CDEF}\Big).
\eea
This gives
\bea
R_{m \,A}{}^B &=& -\frac1{6} p_{a\,bm} \Gamma^{ab}_{AB} 
   + \frac{5}{54} p_{a\,ab}\Gamma_{bm AB} + \frac1{27} p_{a\, bb} \Gamma_{maAB}, \nn\\
R_{pq  \,A}{}^B &=& - \frac{4i}{27} \Delta^{-1} p_{a\,a[p} \Gamma_{q]AB} -
\frac{i}{3} \Delta^{-1} p_{[p\,q]a} \Gamma^a_{AB}  +
\frac{7i}{27} p_{[p\, aa} \Gamma_{q]AB}, \nn\\
R^{pq}{}_A{}^B &=& \frac13 p_{a\,b}{}^{[p} \Gamma^{q]ab}_{AB} 
     + \frac5{54} p_{a\,ab} \Gamma^{bpq}_{AB} 
     - \frac1{27} p_{a\,bb} \Gamma^{apq}_{AB}, \nn\\
R^m{}_A{}^B &=& 0.
\eea     
The last component drops out because for this term the first two lines in (\ref{QM}) give
something proportional to $\delta_A^B$, and hence are cancelled by the third term 
in the definition of $R_{M \, A}{}^B$. This shows very explicitly, that no matter how we
combine expressions depending only on $\cV$ and its derivative, there is no way
of getting rid of $p_{m\,ab}$ and replacing $q_{m\,ab} \rightarrow \omega_{m\,ab}$
by such manipulations, without `breaking up' the 56-bein $\cV$.
In other words, full covariance cannot be achieved in this way, 
but requires the explicit introduction `by hand' of the spin connection.

In principle we could extend the above calculation to non-vanishing form fields;
but this will be far more tedious than the calculation just presented (and 
the resulting expressions will not be any prettier). Perhaps the only interesting aspect
here is that, again, there appears to be {\em no} combination of $\cV$'s and $\partial\cV$'s 
that would produce the fully anti-symmetrized (exterior) derivatives on the 3-form and the 
6-form field, and this is the reason why the hook-like contributions in the affine
connection are needed. It is therefore very remarkable that the supersymmetric
theory avoids this problem by picking precisely the combinations \eqref{covcom}
where these terms drop out.

\section{Covariant ${\rm SU}(8)$ connection}
\label{app:su8connection}

In this appendix we provide yet more evidence that  an SU(8) connection satisfying
all desired covariance properties {\em cannot} be constructed in terms of only
$\cV$ and its derivative $\partial\cV$. Namely, we will show by explicit computation 
how the SU(8) connection of section~3 can be made to transform as a generalized
vector under generalized diffeomorphisms, which implies a unique expression
for $U_{M \, A}{}^B$ in terms of $\cV$ and its derivatives. However, the modifications required
to achieve this come at the price of destroying the covariance under SU(8).

Let the SU(8) connection be
\bea \label{su8con}
{\cal Q}_M{}_A{}^B &=&
q_M{}_A{}^B + R_M{}_A{}^B + U_M{}_A{}^B + W_M{}_A{}^B \;,
\eea
with $q_M{}_A{}^B$, $R_M{}_A{}^B$, and $W_M{}_A{}^B$
given by (\ref{Cartan}) and (\ref{QQUW}), and we make the 
following choice for the undetermined part $U_M{}_A{}^B$
\bea
U_M{}_A{}^B &\equiv& 
-\frac23\,q_M{}_A{}^B  +\frac{2i}3\, \left(
 {\cal V}_{M\,CD} {\cal V}^N{}^{BC} q_N{}_A{}^{D} -
 {\cal V}_{M}{}^{CD} {\cal V}^N{}_{AC} q_N{}_D{}^{B}\right)
\nonumber\\
&&{}
-\frac{34i}{189}
\left({\cal V}_{M\,AC}\,  {\cal V}^N{}^{CD} q_N{}_D{}^{B} - {\cal V}_{M}{}^{BC}\,  {\cal V}^N{}_{CD} q_N{}_A{}^{D} \right)
\nonumber\\
&&{}
-\frac{20i}{189}\left(
{\cal V}_{M\,AD}\,{\cal V}^N{}^{BC}  \,  q_N{}_C{}^{D}- {\cal V}_{M}{}^{BD}\,{\cal V}^N{}_{AC}\,  q_N{}_D{}^{C} \right)
\nonumber\\
&&{}
-\frac{2i}{27}\,\delta_A^B 
\left({\cal V}_{M\,CD}\,{\cal V}^N{}^{EC} q_N{}_E{}^{D}-{\cal V}_{M}{}^{CD}\,{\cal V}^N{}_{EC} q_N{}_D{}^{E}\right)
\;.
\label{QQUW_app}
\eea
These are indeed all the objects that one can construct in terms of $\cV$ and
its derivative $\partial\cV$. However, while the first term $q_{M \, A}{}^B$,
$R_M{}_A{}^B$, and $W_M{}_A{}^B$  have indeed the required covariance properties of an SU(8) 
connection, the expression (\ref{QQUW_app}) for $U_{M \, A}{}^B$ does not, and will therefore violate 
SU(8) covariance if general covariance requires such a contribution.

To see that the full connection can be made to transform covariantly under generalized diffeomorphisms, consider the non-covariant contributions in the transformation of $q_{M \, A}{}^{B}$ and $p_{M ABCD}$
\begin{gather}
 \Delta^{\rm nc} q_{M \, A}{}^{B} = 8i \, \cV^{N BC} \, \mathbb{P}^K{}_N{}^S{}_R \, \partial_{M} \partial_{S} \Lambda^{R} \, \cV_{K \, CA}, \\
\Delta^{\rm nc} p_{M}{}^{ABCD} = 12i \, \cV^{N AB} \, \mathbb{P}^K{}_N{}^S{}_R \, \partial_{M} \partial_{S} \Lambda^{R} \, \cV_{K}{}^{CD},  \label{ncP}
\end{gather}
where we have used
\begin{equation}
 \mathbb{P}^M{}_N{}^P{}_Q = \frac{1}{24} \left(2 \delta^M_Q \delta^P_N + \delta^M_N \delta^P_Q - \Omega_{NQ} \Omega^{MP} \right) + (t_{\bfa})_{NQ} (t^{\bfa})^{MP}
\end{equation}
and the section condition. Note that the covariant part of the transformations of $q_{M}$ and 
$p_{M}$ contain a weight term. So in fact they transform as generalized tensor densities of weight $-1/2.$

Furthermore, 
\begin{align}
  \Delta^{\rm nc} R_{M \, A}{}^{B} =  \cV_{M \, CD}& \left(  -8 \, \cV^{N}_{AE} \cV_{R}{}^{[BE|} \cV^{S \,|CD]} + 10 \, \delta^{[B|}_{A}  \cV^{N| CD|} \cV^{S\, |EF]} \cV_{R \, EF} \right. \notag \\[2mm]
&  \quad - \frac{40}{9} \delta^{C}_{A} \cV^{N}{}_{EF} \cV_{R}{}^{[BD|} \cV^{S \, |EF]} + \frac{40}{9} \delta^{C}_{A} \cV^{N \,[BD|} \cV^{S |EF]} \cV_{R \, EF}  \notag \\[3mm]
&\quad \left. + \frac{14}{9}  \delta^{A}_{B} \cV^{N}_{EF} \cV_{R}{}^{[EF|} \cV^{S |CD]} - \frac{14}{9}  \delta^{A}_{B} \cV^{N[ EF|} \cV^{S |CD]} \cV_{R\,EF} \right) \partial_{N} \partial_{S} \Lambda^{R} + {\rm c.c.} \, , \label{ncbbQ1}
\end{align}
where we have used equations \eqref{ncP}, (A.3) and 
$$ \cV^{M}{}_{[AB} \cV^{N}{}_{CD]} \partial_{M} \partial_{N} \cdot = \frac{1}{24}
\epsilon_{ABCDEFGH} \cV^{M \,EF} \cV^{N \,GH} \partial_{M} \partial_{N}
\cdot $$ 
which can be proved using identity (A.3) and the section condition. Now using, 
\begin{equation}
  \cV^{M AC} \cV^{N}_{BC} \partial_{M} \partial_{N} \cdot = \frac{1}{8}
\delta^{A}_{B}\, \cV^{M CD} \cV^{N}_{CD} \partial_{M} \partial_{N} \cdot\, , \label{VVppiden2}
\end{equation}
 which holds by identity (A.2) and the section condition, equation \eqref{ncbbQ1} can be simplified to:
\begin{align}
  \Delta^{\rm nc} R_{M \, A}{}^{B} = \nn \\
- \frac{1}{3} \cV_{M \,CD}& \left( 4 \, \cV^{N \,CD} \left[ \cV_{R \, AE} \cV^{S \,BE} + \cV^{S}{}_{AE} \cV_{R}{}^{BE}  - \frac{1}{36} \, \delta_{A}^{B} \left( 4 \cV_{R \,EF} \cV^{S \,EF} + 7 \cV^{S}{}_{EF} \cV_{R}{}^{EF} \right) \right]   \right. \notag \\[2mm]
&  \quad + 8 \, \cV^{N BC} \left[ \cV_{R \,AE} \cV^{S \,DE} +  \cV^{S}{}_{AE} \cV_{R}{}^{DE} \right]
+ \frac{1}{9} \, \delta^{C}_{A} \cV^{N EF} \cV^{S}{}_{EF} \cV_{R}{}^{BD} \notag \\[3mm]
& \quad - \frac{8}{9}\, \delta^{C}_{A} \cV^{N \,DE} \cV^{S\, BF} \cV_{R \,EF} - \frac{4}{9} \, \delta^{C}_{A} \cV^{N\, BD} \left[ \cV_{R \,EF}\cV^{S \, EF} - 5 \, \cV^{S}{}_{EF} \cV_{R}{}^{EF} \right]   \notag \\[3mm]
&\quad \left.  + \frac{1}{9}\, \delta^{B}_{A} \cV^{N \,EF} \cV^{S}{}_{EF} \cV_{R}{}^{CD} + \frac{8}{9}\, \delta^{B}_{A} \cV^{N \, EC} \cV^{S \,FD} \cV_{R \, EF}  \right) \partial_{N} \partial_{S} \Lambda^{R} + {\rm c.c.}\, .
\end{align}
Similarly, using identities (A.2) and \eqref{VVppiden2} 

\begin{align}
  \Delta^{\rm nc} U_{M \, A}{}^{B} = \nn \\
- \frac{1}{3} \cV_{M \, CD}& \left( 8 \, \cV^{N \,CD} \left[ \cV_{R \, AE} \cV^{S \,BE} + \cV^{S}{}_{AE} \cV_{R}{}^{BE}  - \frac{1}{9} \, \delta_{A}^{B} \left( \cV_{R \,EF} \cV^{S \,EF} + \cV^{S}{}_{EF} \cV_{R}{}^{EF} \right) \right]   \right. \notag \\[2mm]
&  \quad - 8 \, \cV^{N \,BC} \left[ \cV_{R \,AE} \cV^{S \,DE} +  \cV^{S}{}_{AE} \cV_{R}{}^{DE} \right]
- \frac{1}{9} \, \delta^{C}_{A} \cV^{N \,EF} \cV^{S}{}_{EF} \cV_{R}{}^{BD}  \notag \\[3mm]
& \quad + \frac{8}{9}\, \delta^{C}_{A} \cV^{N \,DE} \cV^{S \, BF} \cV_{R \, EF} - \frac{8}{9} \, \delta^{C}_{A} \cV^{N \, BD} \left[ \cV_{R \,EF}\cV^{S \,EF} + \cV^{S}{}_{EF} \cV_{R}{}^{EF} \right]  \notag \\[3mm]
&\quad \left. - \frac{1}{9}\, \delta^{B}_{A} \cV^{N\,EF} \cV^{S}{}_{EF} \cV_{R}{}^{CD} - \frac{8}{9}\, \delta^{B}_{A} \cV^{N \,EC} \cV^{S \,FD} \cV_{R \,EF}  \right) \partial_{N} \partial_{S} \Lambda^{R} + {\rm c.c.} \, .
\end{align}
It is straightforward to verify that 
\begin{align}
  \Delta^{\rm nc} &\left(R_{M \, A}{}^{B} + U_{M \, A}{}^{B} \right) = \nn \\
&- 4 \, \cV_{M \,CD} \left( \, \cV^{N \, CD} \left[ \cV_{R \,AE} \cV^{S \,BE} + \cV^{S}{}_{AE} \cV_{R}{}^{BE}  - \frac{1}{36} \, \delta_{A}^{B} \left( 4 \cV_{R \,EF} \cV^{S \,EF} + 5 \cV^{S}{}_{EF} \cV_{R}{}^{EF} \right) \right]   \right. \notag \\[2mm]
& \left. \hspace{26mm} - \frac{1}{9} \, \delta^{C}_{A} \cV^{N \,BD} \left[ \cV_{R \,EF}\cV^{S \,EF} - \cV^{S}{}_{EF} \cV_{R}{}^{EF} \right]   \right) \partial_{N} \partial_{S} \Lambda^{R} + {\rm c.c.}\, , \notag \\[3mm]
&= 8 \left( \cV_{M \,CD} \cV^{N \,CD} - \cV_{M}{}^{CD} \cV^{N}{}_{CD} \right)   \, \mathbb{P}^P{}_Q{}^S{}_R  \, \cV_{P AE} \cV^{Q BE}\, \partial_{N} \partial_{S}  \Lambda^{R} \notag \\[3mm]
& \quad -\frac{4i}{9} \left[ \cV_{M AC} \cV^{N BC} + \cV_{M}{}^{AC} \cV^{N}{}_{BC} - \frac{1}{8} \delta^{B}_{A} \left( \cV_{M CD} \cV^{N CD} + \cV_{M}{}^{CD} \cV^{N}{}_{CD} \right)  \right] \partial_{N} \partial_{S}  \Lambda^{S}, \notag \\[3mm]
&= - \Delta^{\rm nc} q_{m \, A}{}^{B} - \Delta^{\rm nc} W_{m \, A}{}^{B}.
\end{align}
Therefore, ${\cal Q}_{M \, A}{}^{B}$ defined in equation \eqref{su8con} is a 
generalized tensor density of weight $-1/2.$ However, as the term $U_{M \, A}{}^B$ 
itself depends on $q_{M \, A}{}^B$ in a definite manner, the total SU(8) connection 
no longer transforms properly under SU(8). As we explained, this conclusion can only
be evaded if one drops the assumption that all parts of $\cQ_M$ should be expressible
in terms of $\cV$ and $\partial_M\cV$.

\end{appendix}

\newpage


\bibliographystyle{JHEP2}
\bibliography{refs2}
\end{document}